\documentclass[11pt,a4paper]{article}
\pdfoutput=1
\usepackage{jheppub}
\usepackage{latexsym,amsfonts,amsmath,amssymb}
\usepackage{bbm}
\usepackage{empheq}
\usepackage{graphicx}
\usepackage{color}
\usepackage[normalem]{ulem}
\usepackage{comment}
\usepackage{url}
\usepackage{slashed}
\usepackage{tabu}
\usepackage{multirow}
\usepackage{extarrows}
\usepackage{amsmath}
\usepackage{mathrsfs}
\usepackage{bm}
\usepackage{stmaryrd}
\usepackage[dvipsnames]{xcolor}
\usepackage[utf8]{inputenc}
\usepackage[T1]{fontenc}

\usepackage[super]{nth}
\usepackage{mathbbol}

\usepackage{mathtools}
\DeclarePairedDelimiter{\ceil}{\lceil}{\rceil}
\DeclarePairedDelimiter\floor{\lfloor}{\rfloor}

\usepackage{amsxtra,graphics,epsfig,bm,tikz,xfrac,lscape}
\usetikzlibrary{decorations.pathmorphing}
\usetikzlibrary{decorations.markings}
\usetikzlibrary{arrows, decorations.markings, calc, fadings, decorations.pathreplacing, patterns, decorations.pathmorphing, positioning}

\usetikzlibrary{positioning,shapes}
\usetikzlibrary{chains}
\usetikzlibrary{arrows,fit,decorations.pathreplacing}
\tikzstyle{every picture}+=[remember picture]
\tikzstyle{na} = [baseline=-.5ex]

\definecolor{islamicgreen}{rgb}{0.0, 0.56, 0.0}

\newcommand{\np}{\textrm{N.P.}}
\newcommand{\ssp}{\textrm{S.P.}}

\newcommand\cC{{\cal C}}

\newcommand\cG{{\cal G}}

\newcommand\cL{{\cal L}}

\newcommand\cN{{\cal N}}

\newcommand\cV{{\cal V}}

\newcommand{\Tr}{\mbox{Tr}}

\newcommand{\half}{\frac{1}{2}}

\newcommand{\lhs}{\textrm{L.H.S.}}

\newcommand{\nn}{\nonumber}

\def\i{\mathrm{i}}

\def\beqa{\begin{eqnarray}}
\def\eeqa{\end{eqnarray}}
\def\be{\begin{equation}}
\def\ee{\end{equation}}
\def\bse{\begin{subequations}}
\def\ese{\end{subequations}}

\newcommand{\bem}{\begin{pmatrix}}
\newcommand{\eem}{\end{pmatrix}}

\renewcommand{\=}{\;  = \;}
\def\+{\, + \,}

\def\wt{\widetilde}

\def\bar{\overline}
\def\ads2{$AdS_2$}
\def\ss2{S$^2$}

\def\rt2{\sqrt{2}}

\newcommand\qeq{Q_{eq}}

\def\a{\alpha}

\def\eps{\epsilon}

\def\l{{\lambda}}

\newcommand\myeq{\stackrel{\mathclap{\normalfont\mbox{reg.}}}{=}}

\def\r{ r}

\newcommand{\etaN}{{\eta^{\text{N}}}}
\newcommand{\etaS}{{\eta^{\text{S}}}}
\def\S{{\sigma}}
\def\P{{\rho}}

\newcommand{\ba}{\begin{array}}
\newcommand{\ea}{\end{array}}

\newcommand\redsout{\bgroup\markoverwith{\textcolor{red}{\rule[0.5ex]{2pt}{0.4pt}}}\ULon}

\newcommand\Imtak[1]{\textcolor{red}{(IJ: #1)}}
\newcommand\Aaron[1]{\textcolor{blue}{(AP: #1)}}
\newcommand\augniva[1]{\textcolor{teal}{(AR: #1)}}

\newcommand\HK[1]{\textcolor{orange}{(HK: #1)}}

\title{Supersymmetric localisation of $\mathcal{N}=(2,2)$ theories on a spindle
}

\author{ Imtak Jeon$^{a}$,} 
\author{Hyojoong Kim$^b$,}
\author{Nakwoo Kim$^c$,}
\author{Aaron Poole$^d$}
\author{and Augniva Ray$^b$}
\affiliation{$^a$ School of Science, Huzhou University, Huzhou 313000, Zhejiang, China }
\affiliation{$^b$ Department of Physics and Astronomy \& Center for Theoretical Physics, Seoul National University, Seoul 08826, Korea }
\affiliation{$^c$ Department of Physics and Research Institute of Basic Science, Kyung Hee University, 26 Kyungheedae-ro, Seoul 02447, Korea}
\affiliation{$^d$ Department of Physics and Center for Theoretical Physics, National Taiwan University, No. 1, Sec. 4, Roosevelt Road, Taipei 106319, Taiwan}
\emailAdd{imtakjeon@gmail.com}
\emailAdd{hjkim1996@gmail.com}
\emailAdd{nkim@khu.ac.kr}
\emailAdd{apoole@phys.ntu.edu.tw}
\emailAdd{augniva@gmail.com}

\abstract{We consider two-dimensional $\mathcal{N}=(2,2)$ supersymmetric field theories living on a weighted projective space $\mathbb{WCP}_{[n_1,n_2]}^1$, often referred to as a spindle. Starting from the spindle solution of five-dimensional minimal gauged supergravity, we construct a theory on a spindle which preserves supersymmetry via the anti-twist mechanism and admits two Killing spinors of opposite $R$-charge. We apply the technique of supersymmetric localisation to compute the exact partition function for a theory consisting of an abelian vector multiplet and a chiral multiplet, finding that the path integral localises to a real moduli space of vector multiplet fluctuations. We compute the one-loop determinants via the equivariant index, using both the method of unpaired eigenvalues and the fixed point theorem, finding agreement between the two approaches. We conclude with  the explicit partition function for an example of a charged chiral multiplet in the presence of a Fayet-Iliopoulos term and comment on its dependence on the overall length scale of the geometry. This work paves the way towards uncovering two-dimensional dualities, such as mirror symmetry, for field theories defined on orbifold backgrounds.}

\begin{document}

\maketitle

\section{Introduction}
\label{sec:introduction}

Supersymmetric localisation \cite{Duistermaat:1982vw, Atiyah:1984px, Witten:1988ze, Witten:1988xj, Nekrasov:2002qd, Pestun:2007rz} is a powerful tool to compute exact observables in supersymmetric quantum field theories. Such techniques have been applied to stu\textrm{d}y in great detail the properties of supersymmetric field theories defined on curved manifolds \cite{Festuccia:2011ws}, both in two dimensions \cite{Benini:2012ui, Doroud:2012xw, GonzalezLezcano:2023cuh} and higher \cite{Kapustin:2009kz, Jafferis:2010un, Hama:2010av, Hosomichi:2012ek, Closset:2012ru, Closset:2013sxa}.  These results have further allowed for the discovery of dualities between seemingly unrelated models \cite{Hori:2000kt, Hori:2006dk, Beem:2012mb, Benini:2015noa}, as well as the realisation that gauge theory observables provide efficient methods of computing invariants of the manifold upon which the theory lives \cite{Gromov, Jockers:2012dk, Gomis:2012wy, Manschot:2023rdh}.

One of the most active areas of work involving localisation is that of  holographic (AdS/CFT) duality \cite{Maldacena:1997re, Witten:1998qj, Aharony:1999ti}, where localisation techniques are important in verifying precise checks between gravity and the dual field theory. In particular, localisation in the field theory can be used to derive supersymmetric indices whose large-$N$ limit provides a quantum origin of supersymmetric AdS black hole entropy \cite{Benini:2015eyy, Cabo-Bizet:2018ehj, Choi:2018hmj, Benini:2018ywd}. It is an active area of interest to extend the application of localisation to the supergravity side of the correspondence \cite{Dabholkar:2010uh,Dabholkar:2011ec, Gupta:2015gga,Jeon:2018kec, Iliesiu:2022kny}, a direction which is hoped to provide deeper and more stringent tests of quantum gravity.

A new avenue of holographic stu\textrm{d}y has been driven by the discovery of several new supergravity solutions often referred to as ``spindle solutions'' \cite{Ferrero:2020laf, Ferrero:2020twa, Hosseini:2021fge, Boido:2021szx, Ferrero:2021wvk, Ferrero:2021ovq, Couzens:2021rlk, Faedo:2021nub, Ferrero:2021etw, Suh:2022olh, Arav:2022lzo, Couzens:2022yiv, Couzens:2022lvg, Faedo:2022rqx, Suh:2022pkg, Suh:2023xse, Amariti:2023mpg, Hristov:2023rel, Amariti:2023gcx, Hristov:2024qiy, Suh:2024fru, Crisafio:2024fyc}. These solutions correspond to the wrapping of various branes on the spindle \textit{orbifold} $\mathbb{\Sigma} = \mathbb{WCP}^1_{[n_1,n_2]}$, a weighted complex projective space characterised by two co-prime integers $n_{1,2}$.\footnote{The co-prime condition $\text{gcd}(n_1,n_2) =1$ is related to the smooth uplifiting of certain spindle solutions into $D=10, 11$, see \cite{Ferrero:2020laf, Ferrero:2020twa}.} The four-dimensional solutions \cite{Ferrero:2020twa, Ferrero:2021ovq} correspond to the near-horizon region of accelerating, asymptotically locally AdS$_4$ black hole solutions \cite{Plebanski:1976gy, Dias:2002mi, Griffiths:2005qp, Podolsky:2022xxd} with spindle horizon topology. From the gravity side, the thermo\textrm{d}ynamics \cite{Cassani:2021dwa, Kim:2023ncn} and entropy functions \cite{Boido:2022iye, Boido:2022mbe} have been studied directly, and it is in the interest of precision holography to recover such results from the dual field theory at the conformal boundary $\mathscr{I} \cong \mathbb{\Sigma} \times S^1$. This setup led the authors of \cite{Inglese:2023wky} to consider the problem of localisation of $\mathcal{N}=2$ theories defined on $\mathbb{\Sigma} \times S^1$, a calculation which is carried out via a special case of the orbifold index \cite{10.1215/S0012-7094-96-08226-5, MEINRENKEN1998240} dubbed the \textit{spindle index}. Moreover, the large-$N$ limit of the spindle index \cite{Colombo:2024mts} has been shown to provide a microscopic origin  of the entropy of accelerating black hole solutions, extending the applications of holographic duality into the new territory of orbifold backgrounds.

This example has led to recent interest in extending the application of localisation techniques to field theories defined on various orbifold backgrounds \cite{Inglese:2023wky, Inglese:2023tyc, Pittelli:2024ugf, Mauch:2024uyt, Ruggeri:2025kmk, Mauch:2025irx} (see also \cite{Hosomichi:2015pia} for earlier work). The localisation techniques developed to stu\textrm{d}y the partition functions of $\mathcal{N}=2$ theories on $\mathbb{\Sigma} \times S^1$ are also applicable to  various other cases of supersymmetric theories on orbifolds.  For example $\mathcal{N}=2$ theories on the squashed-branched Lens space $\mathbb{\Lambda}$ \cite{Inglese:2023tyc}, one-loop determinants for 4d $\mathcal{N}=1$ theories on $\mathbb{\Sigma} \times T^2$ and $\mathbb{\Lambda} \times S^1$ \cite{Pittelli:2024ugf}, and 4d $\mathcal{N}=2$ theories on $\mathbb{\Sigma} \times  \mathbb{\Sigma}$ \cite{Ruggeri:2025kmk} have all been considered.  

A class of field theories which has been comparatively less explored for orbifold backgrounds are those in two dimensions (see however Section 3.3 of \cite{Inglese:2023tyc}). Two-dimensional $\mathcal{N}=(2,2)$ theories \cite{Closset:2014pda, Benini:2016qnm} are a topic of great interest in their own right, both through their acting as a good playground to stu\textrm{d}y four-dimensional gauge theories, as well as through their connection with string theory, often appearing as the worldsheet theories of strings compactified down to four dimensions. Of particular interest in mathematical physics are two-dimensional non-linear sigma models whose target space is Calabi-Yau \cite{Witten:1993yc}. Such models play an important role in understanding dualities such as mirror symmetry \cite{Hori:2000kt, Hori:2002fa}, and Gromov-Witten invariants \cite{Gromov, Witten:1988xj}, the latter of which are computable using localisation techniques \cite{Jockers:2012dk, Gomis:2012wy} applied to $\mathcal{N}=(2,2)$ theories on $S^2$.

It is thus of interest to examine whether such a variety of rich applications exists when one places the theory on an orbifold such as the spindle -  a question which we begin to explore concretely in this work. We will consider two-dimensional $\mathcal{N}=(2,2)$ abelian gauge theory defined on the spindle $\mathbb{\Sigma} = \mathbb{WCP}^1_{[n_1,n_2]}$ and compute the partition function using supersymmetric localisation. In order to do this, we will begin by considering the original spindle solution of $D=5$ minimal gauged supergravity found in \cite{Ferrero:2020laf}. This will lead us to consider the so called \textit{anti-twisted} spindle. Starting from this solution, we will then use the technique of \cite{Festuccia:2011ws} as described in detail in \cite{Closset:2014pda} to construct a two-dimensional $\mathcal{N}=(2,2)$ theory living on $\mathbb{\Sigma}$. Of particular importance in that work are the Killing spinor equations 
\begin{align}
\begin{split}
    D_\mu\epsilon = - \frac{1}{2} H \gamma_\mu \epsilon - \frac{1}{2} G \gamma_\mu \gamma_{12} \epsilon\, , \qquad   D_\mu\widetilde{\epsilon} = - \frac{1}{2} H \gamma_\mu \widetilde{\epsilon} + \frac{1}{2} G \gamma_\mu \gamma_{12} \widetilde{\epsilon}\,,
    \end{split}
\end{align}
where $\epsilon, \widetilde{\epsilon}$ are Dirac spinors of $R$-charge $\mp1$. These equations are characterised by the supergravity background fields $H,G$, which we will read off from the spindle components of the five-dimensional Killing spinor equations originally solved in \cite{Ferrero:2020laf}. 

This setup will allow us to build supersymmetric actions on $\mathbb{\Sigma}$. We will consider a simple theory consisting of one abelian vector multiplet (Yang-Mills action), one charged chiral multiplet (matter action) and Fayet-Iliopoulos term, the last of which can be understood as a special choice of a twisted superpotential action. We will then employ localisation on the Coulomb branch of this theory in order to compute the exact partition function 
\begin{equation} 
    Z_{\mathbb{\Sigma}} = \int_{\mathcal{M}} \mathcal{D \varphi} \, e^{-S[\varphi]}\, ,  
\end{equation}
giving a detailed discussion of the method and subtleties involved when the theory is placed on a spindle. As a consistency check, we will see that the full partition function indeed reproduces the $S^2$ partition function in the appropriate limit of $n_1 = n_2 = 1$.

This paper is organised as follows: in Section \ref{sec: minimal_spindle} we introduce the spindle geometry via a discussion of the $D=5$ minimal gauged supergravity solution found in \cite{Ferrero:2020laf}. We solve the five-dimensional Killing spinor equations and comment on various properties of the spinors on a spindle, consistent with the very general analysis of \cite{Ferrero:2021etw}. In Section \ref{sec: D=2N=2,2_on_spindle} we use the spindle components of the Killing spinor equation of the 5d theory in order to construct a two-dimensional $\mathcal{N}=(2,2)$ theory on $\mathbb{\Sigma}$. We give the definitions of the vector and chiral multiplets on $\mathbb{\Sigma}$ and discuss the supersymmetry transformations and algebra of such transformations, including the approach to supersymmetry given by writing the components of the chiral multiplet in ``cohomological variables'' \cite{GonzalezLezcano:2023cuh}. We conclude the section with a discussion of the reality properties of the fields in the multiplets as well as a construction of the supersymmetric action for our theory under consideration. In Section \ref{sec: localisation} we discuss the general procedure of supersymmetric localisation. We begin with a brief overview of the general localisation argument, breaking down the path integral into classical and one-loop contributions, before applying it explicitly to our theory on $\mathbb{\Sigma}$. We compute the BPS locus to which the path integral localises and evaluate the action on the locus. A point of note here is that our locus will be entirely real as opposed to the complex loci found in \cite{Inglese:2023tyc}. In Section \ref{sec: one-loop} we compute the one-loop determinant part of the path integral. We compute the determinants via both the method of unpaired eigenvalues and the Atiyah-Bott fixed point theorem \cite{atiyah1966lefschetz,10.2307/1970694,Atiyah:1974obx}, allowing us to confirm our results using two separate avenues, as well as relating to the predictions for the $\mathcal{N} =(2,2)$ one-loop results via dimensional reduction of the related determinants on $\mathbb{\Sigma} \times S^1$ as considered in \cite{Inglese:2023tyc}. In Section \ref{sec: Mirror Symmetry} we compute the partition function explicitly for a charged chiral multiplet, performing the integral and sum over the moduli space of the BPS locus. We discuss the physical interpretation of the partition function as counting vacua in the limit of vanishing Fayet-Iliopoulos term, as well as comment on the structure of the general result in relation to factorisation properties and Higgs branch localisation \cite{Benini:2012ui,Doroud:2012xw}. Finally we conclude and discuss some interesting further directions in Section \ref{sec: conclusions}. We also include several appendices where our conventions are established and several additional technical steps are highlighted. 

\section{Spindle from minimal \texorpdfstring{$D=5$}{D=5} gauged supergravity} \label{sec: minimal_spindle}

\subsection{The spindle solution}
  We consider the spindle solution of minimal gauged supergravity in $D=5$ \cite{Gunaydin:1983bi}. Let us first consider the solution of the type studied in \cite{Ferrero:2020laf}, consisting of a warped product of $AdS_3$ and spindle $\mathbb{\Sigma}$  given by 
\begin{subequations}\label{5Dbackground}
\begin{align}
    \textrm{d}s^2 &= {\color{black}L^2} \frac{4y}{9} \textrm{d}s^2_{AdS_3} + \textrm{d}s^2_{\mathbb{\Sigma}} \,,\\
    \textrm{d}s^2_{\mathbb{\Sigma}} &= {\color{black}L^2} \left( \frac{y}{q} \textrm{d}y^2 + \frac{q}{36y^2} \textrm{d}z^2 \right) \label{eq: spindle_metric}\,,\\
    q&= 4y^3 - 9 y^2 +6ay - a^2\,,
    \end{align}
\end{subequations}
together with an abelian $R$-symmetry gauge field 
\begin{equation}
    A^{5d} =  \frac{L}{4}\left( 1 - \frac{a}{y}\right) \textrm{d}z\,. \label{eq: singular_gauge_field}
\end{equation}
The parameter $a$ is related to the order of orbifolding. Assuming $a \in (0,1)$, the function $q(y)$ has three real positive roots, which we denote by $y_i\,, i = 1,2,3$. Without losing generality, we choose the ordering of the roots to be $y_1 < y_2 < y_3$, and we take the range of $y$ to be $y \in [y_1, y_2]$. This gives us $q(y)\geq 0$ and thus the metric on $\mathbb{\Sigma}$ is positive definite.\footnote{One can also obtain a positive definite metric by choosing the range $y \in [y_3, \infty)$. Such a case is examined in \cite{Bomans:2024vii}. However, this choice is not analysed here.} 
In terms of two co-prime positive integers $\{ n_1, n_2\}$ with $n_1 < n_2$, the parameter $a$ and the roots  $y_1$ and $y_2$ are given by
\begin{align}
\begin{split}
	a  =  \frac{(n_1-n_2)^2(2n_1 +n_2)^2(n_1 +2n_2)^2}{4(n_1^2 +n_1 n_2 + n_2^2)^3}\, , \qquad \qquad 
\\
	y_1 = \frac{(n_1 -n_2)^2(2 n_1 +n_2)^2}{4(n_1^2 +n_1 n_2 + n_2^2)^2} \, , \qquad 	y_2 = \frac{(n_1-n_2)^2(n_1+ 2n_2)^2}{4(n_1^2 +n_1 n_2 + n_2^2)^2} \,. \label{eq: ys}
    \end{split}
\end{align}
Then with the choice of periodicity for $z$ given by
\begin{equation}  \label{eq: deltaz}
    \Delta z = 4\pi \frac{n_1^2 + n_1 n_2 +n_2^2}{3 n_1 n_2 (n_1 + n_2)}\,,
\end{equation}
as shown in Appendix~\ref{app: coordinates}, we have a weighted projective space 
$\mathbb{\Sigma}= \mathbb{WCP}^1_{[n_1,n_2]}$ having deficit angles $2\pi(1-1/n_1)$ and $2\pi(1-1/n_2)$ at $y_1$ and $y_2$ respectively. Despite the appearance of conical singularities in this solution, it is rendered completely smooth when uplifted to ten-dimensional type IIB supergravity \cite{Ferrero:2020laf}. 

Evaluation of the gauge field \eqref{eq: singular_gauge_field} at the poles of the spindle gives
\begin{equation}
     A^{5d}|_{y_{1,2}} = \frac{L}{4}\left( 1 - \frac{a}{y_{1,2}}\right) \textrm{d}z = -  \frac{L}{2n_{1,2}}\cdot \frac{2\pi}{\Delta z}\textrm{d}z 
     \neq 0\,,
\end{equation}
where the non-vanishing of the gauge field at the poles indicates that the gauge field is \textit{singular} at these points. In later sections we will employ gauge transformations in order to construct gauge fields which are \textit{regular} at one of the poles.

We note that the Ricci scalar curvature of $\mathbb{\Sigma}$ is found to be
\begin{align} \label{eq: Ricci_scalar_spindle}
    R_{\mathbb{\Sigma}} = \frac{5a^2 - 9a y -2 y^3}{{\color{black}L^2}y^3} \,,
\end{align}
which gives the Euler number as 
\begin{equation} \label{eq: Euler_number}
   \chi =  \frac{1}{4\pi}\int_{\mathbb{\Sigma}} R_{\mathbb{\Sigma}} \, {\rm vol}_{\mathbb{\Sigma}} = {\frac{1}{n_1}+\frac{1}{n_2}}   \,,
\end{equation}
and the magnetic flux through $\mathbb{\Sigma}$ is 
\be \label{eq: R-flux}
 \frac{1}{2\pi {\color{black}L}} \int_{\mathbb{\Sigma}} \textrm{d} A^{5d} =  \frac{1}{2}\cdot\frac{n_2 -n_1}{n_1 n_2} \,:= \,\frac{1}{2} \chi_- \,.
\ee
This configuration is called the `anti-twist' configuration. This is contrasted with the other way of preserving supersymmetry through the twist configuration, where the flux is given by $\chi/2$. While the minimal $D=5$ supergravity only admits an anti-twist solution, more general supergravity theories such as STU supergravity have both twisted and anti-twisted spindle solutions \cite{Ferrero:2021etw}. 
\subsection{Killing spinor equation} The background admits a Killing spinor $\varepsilon$ satisfying the Killing spinor equation 
\begin{equation} 
   \left[ \nabla_M -\frac{\i}{12}\left(\Gamma_M{}^{N P} - 4 \delta_M^N \Gamma^P \right)F^{5d}_{NP} -\frac{1}{2  {\color{black}L}}\Gamma_M -  {\color{black}\frac{\i}{L}} A^{5d}_M \right]\varepsilon =0\,,
   \label{5dkse}
\end{equation}
which we will now analyse in some detail. Let us set the five-dimensional gamma matrix as 
\begin{equation}\label{GammaSplit}
        \Gamma^a = \rho^a \otimes \gamma_{3}\,,\quad a= \{ 0,1,2\}\,,\qquad\Gamma^3 = 1 \otimes s\, \i \gamma^1 \gamma^3 \,,\qquad~~\Gamma^4 = 1\otimes s\, \i \gamma^2 \gamma^3\,,
\end{equation}
where $\rho^a$ is the 3-dimensional gamma matrix along the $AdS_3$ directions and $\gamma^m$ the 2-dimensional gamma matrix along the spindle directions with $\gamma_3 \equiv -\i \gamma_{12}$. See Appendix~\ref{app: Defs} for our conventions for gamma matrices and indices. Note that we also introduce the sign factor $s= \pm 1$. We decompose the Killing spinor as $\varepsilon =  \vartheta \otimes \chi$, where $\vartheta$ is a Killing spinor for $AdS_3$. Using this decomposition, we show in Appendix~\ref{KSEAdS3} that the full Killing spinor equation \eqref{5dkse} is solved as long as the spindle directions of \eqref{5dkse} are solved, i.e. $\chi$ is a Killing spinor on the spindle. Since the background \eqref{5Dbackground} is given as warped product, the derivation for $AdS_3$ is more involved. In the main text, we will focus entirely on the Killing spinors on the spindle and will not use any details of the $AdS_3$ part.

The Killing spinor equation on the spindle is given by 
\begin{eqnarray}\label{KSSpindle1}
  \left(  \nabla_\mu - {\color{black}\frac{\i}{L}} A^{5d}_\mu \right)\chi = s \frac{\i}{3} F^{5d}\gamma_\mu  \chi+ s\frac{1}{2{\color{black}L}}\gamma_\mu \gamma_{12} \chi \,,
\end{eqnarray}
where $\nabla_\mu$ is the derivative with connection, acting on a spinor as $\nabla_{\mu} \chi = (\partial_{\mu}+ \frac{1}{2} \omega_{\mu}^{12} \gamma_{12})\chi$, where the spin connection is given by
\begin{eqnarray} \label{eq: spin_conn_spindle}
    \omega_{z}^{12} = - \frac{a^2 - 3a y + 2 y^3}{6 y^{5/2}}\,.
\end{eqnarray} 
We also define\footnote{This should not be confused with the 2-form field strength, customarily also written as $F^{5d}$, which is equal to d$A^{5d}$ in our setting.}  $F^{5d} \equiv \frac{1}{2}\epsilon^{\mu\nu}F^{5d}_{\mu\nu}$  with $\epsilon_{\mu \nu}$ being the volume form on the spindle with orientation $\epsilon_{yz} = \sqrt{g_{\mathbb{\Sigma}}}$ and thus for the background gauge field given in  \eqref{eq: singular_gauge_field}, $F^{5d}=3a y^{-3/2}/2{\color{black}L}
$.  Let us set two-dimensional gamma matrix representation as $\gamma_m =(\sigma_1\,,\sigma_2)$,\footnote{In \cite{Ferrero:2020laf}, the gamma matrix along spindle direction is chosen $\gamma_{m}= (\sigma_2 \,,\sigma_1) $ . To convert it to our choice, we can use $S \sigma_1 S^{-1} = \sigma_2$ and $S \sigma_2 S^{-1} = \sigma_1$, with $S = \Big(\begin{array}{cc}
  0   & -\i \\
    1 & 0
\end{array}\Bigr)$. } which gives $\gamma_3 = \sigma_3$. Then the solution of the Killing spinor on the spindle is given by
\begin{eqnarray}\label{KSchi}
    \chi 
    = 
     \sqrt{\frac{{k_0}}{{12}}}\begin{pmatrix} 
	\sqrt{\frac{q_2(y)}{y}} \\ 	\i\, s\sqrt{\frac{q_1(y)}{y}} 
\end{pmatrix}\,,
\end{eqnarray}
where we define
\begin{equation}
    q_1(y)= - a +2 y^{3/2} + 3y \,,\quad\quad ~~q_2(y)= a + 2 y^{3/2} - 3y\,,
\end{equation}
that satisfy $q_1(y) q_2(y)= q$ and $q_1(y_1)= q_2(y_2)=0$,  and for later use we choose the normalization factor to be
\begin{equation}\label{eq:KSnormalisation}
    k_0 = {L_0}\frac{\Delta z}{2\pi}\,,
\end{equation} where $L_0$ is a reference length scale parameter.   Note that this Killing spinor does not have any dependence on the azimuthal direction $z$, so the periodicity condition is trivially satisfied.  This fact is not unrelated to the gauge choice \eqref{eq: singular_gauge_field} which is singular at $y=y_1$ and $y_2$.

We have another Killing spinor equation for the Killing spinor with opposite $U(1)$  $R$-charge, which is obtained from \eqref{5dkse} by taking charge conjugation. With the definition of conjugate spinor as  $\varepsilon^{c}\equiv B^{-1} \varepsilon^\ast$, where $B$ is defined in \eqref{5dgammaLor}, we have 
\begin{equation}
   \left[ \nabla_M -\frac{\i}{12} \left(\Gamma_M{}^{N P} - 4 \delta_M^N \Gamma^P \right)F^{5d}_{NP} +\frac{1}{2{\color{black}L}}\Gamma_M +  {\color{black}\frac{\i}{L}} A^{5d}_M \right]\varepsilon^c =0\,.
\end{equation}
This five-dimensional equation reduces using the splitting of the gamma matrix  \eqref{GammaSplit} and the spinor $\varepsilon^c = \widetilde\vartheta \otimes \widetilde{\chi}$ to 
\begin{eqnarray}\label{KSSpindle2}
  \left(  \nabla_\mu +  {\color{black}\frac{\i}{L}} A^{5d}_\mu \right)\widetilde{\chi} = s \frac{\i}{3}  F^{5d} \gamma_\mu  \widetilde{\chi}  -s  \frac{1}{2 {\color{black}L} }\gamma_\mu \gamma_{12} \widetilde{\chi} \,,
\end{eqnarray}
an equation which is solved by 
\begin{eqnarray}\label{KSchit}
    \widetilde\chi = 
    \sqrt{\frac{{k_0}}{{12}}}\begin{pmatrix}
	\i\,s\sqrt{\frac{q_1(y)}{y}} \\ 	\sqrt{\frac{q_2(y)}{y}} 
\end{pmatrix} \,.
\end{eqnarray}
 One can check that the Killing spinors \eqref{KSchi} and \eqref{KSchit} satisfy 
\begin{equation}
    \chi^\dagger = -\i \wt{\chi}^T \sigma_2\,,\qquad \wt{\chi}^\dagger = \i \chi^T \sigma_2\,.
\end{equation}

\section{Two-dimensional \texorpdfstring{$\cN =(2,2)$}{cn=(2,2)} theory on the spindle} \label{sec: D=2N=2,2_on_spindle}

\subsection{Killing spinors}
We need to compare the Killing spinor equations \eqref{KSSpindle1} and \eqref{KSSpindle2} with the general off-shell supergravity Killing spinor equation for 2d $\cN=(2,2)$ theory. As is written in \cite{GonzalezLezcano:2023cuh}, the general form of the Killing spinor equations are
\begin{subequations} \label{kse}
\begin{align} 
    D_\mu\epsilon &= - \frac{1}{2} H \gamma_\mu \epsilon - \frac{1}{2} G \gamma_\mu \gamma_{12} \epsilon\,, \label{kse1}\\
    D_\mu\widetilde{\epsilon} &= - \frac{1}{2} H \gamma_\mu \widetilde{\epsilon} + \frac{1}{2} G \gamma_\mu \gamma_{12} \widetilde{\epsilon}\,,
    \label{kse2}
\end{align}
\end{subequations}
where $D_\mu=\nabla_\mu + \i A_\mu$ for $\epsilon$ and $D_\mu=\nabla_\mu-\i A_\mu$ for $\widetilde{\epsilon}$ as the $R$-charge of $\epsilon$ and $\widetilde{\epsilon}$ are $-1$ and $+1$ respectively. Therefore, by comparison with equations \eqref{KSSpindle1} and \eqref{KSSpindle2} we identify 
\begin{equation} \label{eq: A_2d}
    A = \frac{1}{L} A^{5d} =  \frac{1}{4}\left( 1 - \frac{a}{y}\right) \textrm{d}z\, , \quad F = \frac{1}{L} F^{5d}=\frac{3a}{2L^2 y^{3/2}} 
\, ,
\end{equation} 
$\chi = \widetilde\epsilon$ and $\widetilde{\chi}= \epsilon$ as given in \eqref{eq:solutionspace},
and
\begin{equation} \label{eq: KSE_functions}
    H = - s\,\i \frac{2}{3}  {\color{black}L}F= -s\, \i \frac{a}{{\color{black}L}y^{3/2}}\,, \qquad  G = \frac{s}{\color{black}L}\,.
\end{equation} These two Killing spinors satisfy the symplectic Majorana relation as 
\begin{align}
\epsilon^\dagger = \i \widetilde{\epsilon}^T C\,, \qquad \widetilde{\epsilon}^\dagger = -\i \epsilon^T C\,,
\end{align}
with $C= \sigma_2$.
Although their numerical values are related by complex conjugation, we do not need to use the symplectic Majorana basis for spinor fields in $\cN=(2,2)$ Euclidean theory, and thus treat those Killing spinors as independent spinors.

The above solutions are expressed in a frame where the gauge field \eqref{eq: A_2d} is singular at the poles $y= y_{1,2}$. However, we can make use of constant gauge transformations such that we work in frames where the gauge field is regular in each patch, i.e. $A_\mu$ vanishes at the poles. The gauge transformations take the form 
\begin{equation} \label{eq: R_gauge_transform}
    A \rightarrow A' = A + \alpha \textrm{d}z\, ,
\end{equation}
where the explicit values of $\alpha$ in the neighbourhoods of the north ($\mathcal{U}_1$) and south poles ($\mathcal{U}_{2}$) are
\begin{equation} \label{eq: beta_1,2}
    \left. \alpha \right|_{\mathcal{U}_{1,2}} = \left(\frac{a}{4y_{1,2}} - \frac{1}{4} \right) =  \frac{2\pi}{\Delta z} \cdot \frac{1}{2n_{1,2}}\, ,
\end{equation}
and thus the regular gauge fields take the form
\begin{equation} \label{eq:gaugefield_regpatch}
     \left. A \right|_{\mathcal{U}_{1,2}} = \frac{a}{4}\left( \frac{1}{y_{1,2}} - \frac{1}{y}\right) \textrm{d}z \, .
\end{equation}
Using the fact that the Killing spinors $\widetilde{\epsilon}$, $\epsilon$ are charged (with charges $\pm1$) under the gauge transformations, we find their values in the regular gauges at each pole are
    \begin{eqnarray} \label{eq:solutionspace}
 \left. \widetilde{\epsilon} \right|_{\mathcal{U}_{1,2}} = e^{\frac{\i}{4} \left(\frac{a}{y_{1,2}}-1\right) z} \sqrt{\frac{k_0}{12}} \begin{pmatrix}
		\sqrt{\frac{q_2}{y}} \\ 	\i s \sqrt{\frac{q_1}{y}} 
	\end{pmatrix} \,,  
 \quad \left. \epsilon \right|_{\mathcal{U}_{1,2}}  = e^{-\frac{\i}{4} \left(\frac{a}{y_{1,2}}-1\right) z} \sqrt{\frac{k_0}{12}}  \begin{pmatrix}
	\i s \sqrt{\frac{q_1}{y}} \\ 	\sqrt{\frac{q_2}{y}}
\end{pmatrix}\,. \label{eq: KS_NP_gauge}
\end{eqnarray}

We note here, at the onset, that we will employ the notation ``$\ldots|_{\mathcal{U}_{1,2}}$'' frequently to denote quantities evaluated using the \textit{regular} gauge choices at the north and south poles respectively. We immediately highlight a couple of points. 
\begin{enumerate}
    \item 
As has been observed for the anti-twist case in \cite{Ferrero:2021etw}, the Killing spinors are such that opposite chirality components vanish at the north and south poles. Using $q_1(y_1) = q_2(y_2) = 0$, this fact is borne out in the above expressions and we get the following behaviour for the Killing spinors at the poles 
\begin{equation}  \label{eq: chirality_poles}
\begin{alignedat}{2}
   \widetilde{\epsilon}(y_1)|_{_{\mathcal{U}_1}} & =   e^{\frac{\i}{4} \left(\frac{a}{y_{1}}-1\right) z} \sqrt{\frac{k_0}{12}} \begin{pmatrix}
		 	 \sqrt{\frac{q_2}{y_1}}  \\ 0
	\end{pmatrix}\,,  
 \quad && \epsilon(y_1)|_{_{\mathcal{U}_1}} = e^{-\frac{\i}{4} \left(\frac{a}{y_{1}}-1\right) z} \sqrt{\frac{k_0}{12}} \begin{pmatrix}
	0 \\  \sqrt{\frac{q_2}{y_1}}
\end{pmatrix}\,, \\
 \widetilde{\epsilon}(y_2)|_{_{\mathcal{U}_2}} & =  e^{\frac{\i}{4} \left(\frac{a}{y_{2}}-1\right) z} \sqrt{\frac{k_0}{12}} \begin{pmatrix}
	0 \\ \i s \sqrt{\frac{q_1}{y_2}} 
\end{pmatrix}\,,
 \quad &&  \epsilon(y_2)|_{_{\mathcal{U}_2}}  = e^{-\frac{\i}{4} \left(\frac{a}{y_{2}}-1\right) z} \sqrt{\frac{k_0}{12}} \begin{pmatrix}
 	\i s \sqrt{\frac{q_1}{y_2}} \\ 0 
\end{pmatrix}\,. \qquad
\end{alignedat}
\end{equation}

\item Further, note that in this gauge, the Killing spinors become periodic only up to $\mathbb{Z}_{n_{1,2}}$ orbifolding around each pole. Namely, the monodromy of the Killing spinors around $y_1$ and $y_2$ respectively is
\begin{equation} \label{eq:fermionperiodicity}
    \widetilde{\epsilon} (z +  \Delta z) = e^{\i \frac{\pi}{n_{1,2}} }\widetilde{\epsilon} (z)\,,\qquad     \epsilon (z +  \Delta z) = e^{-\i \frac{\pi}{n_{1,2}} }\epsilon (z)\,.
\end{equation}
\end{enumerate}

The Killing spinors should satisfy the integrability conditions
\begin{subequations}
\begin{eqnarray}
    \Big[ D_\mu\,, D_\nu \Big]\epsilon &=& \frac{1}{4}R_{\mu\nu}{}^{ab} \gamma_{ab}\epsilon +\i F_{\mu\nu}\epsilon\,,
     \\
 \Big[ D_\mu\,, D_\nu \Big]\widetilde{\epsilon} &=& \frac{1}{4}R_{\mu\nu}{}^{ab} \gamma_{ab}\widetilde{\epsilon} -\i F_{\mu\nu}\widetilde{\epsilon} \,
    \,.
\end{eqnarray}
\end{subequations}
By using \eqref{kse1} and \eqref{kse2} and contracting the above equations with $\gamma^{\mu\nu}$, we have the following consistency relations,
\begin{subequations}
\begin{eqnarray}
    (H^2 +G^2 )\epsilon - \gamma^\mu (D_\mu H) \epsilon -\i \gamma^\mu (D_\mu G)\gamma_3 \epsilon &=& \left(- \frac{1}{2}R - 2 F \gamma_3 \right)\epsilon\,,
    \\
        (H^2 +G^2 )\widetilde{\epsilon} - \gamma^\mu (D_\mu H )\widetilde{\epsilon} + \i \gamma^\mu (D_\mu G)\gamma_3 \widetilde{\epsilon} &=& \left(- \frac{1}{2}R + 2 F \gamma_3 \right)\widetilde{\epsilon}
    \,.
\end{eqnarray}
\end{subequations}
We explicitly enter the background values $G$ and $H$ from \eqref{eq: KSE_functions}, the field strength $F=3a y^{-3/2}/2{\color{black}{L^2}}$  
and the expression for the spindle Ricci scalar \eqref{eq: Ricci_scalar_spindle} into this equation, obtaining
\begin{subequations}
\begin{eqnarray}
    \epsilon &=& \frac{1}{a-3y} \left( \i s \sqrt{q}\gamma_1 - 2 y^{3/2}\gamma_3 \right)\epsilon\,,
    \\
    \widetilde\epsilon &=& \frac{1}{a-3y} \left( \i s \sqrt{q}\gamma_1 + 2 y^{3/2}\gamma_3 \right)\widetilde\epsilon\, \label{eq: integrability_explicit_tilde},
\end{eqnarray}
\end{subequations}
and solving these equations one finds
\begin{equation}
    \epsilon = \epsilon_1(y,z) \begin{pmatrix}
	1 \\ - \i s \sqrt{\frac{q_2}{q_1}} \end{pmatrix}, \qquad \widetilde{\epsilon} = \widetilde{\epsilon}_1(y,z) \begin{pmatrix}
	1 \\  \i s \sqrt{\frac{q_1}{q_2}} \end{pmatrix},
\end{equation}
where $\epsilon_1(y,z)$ and $\widetilde{\epsilon}_1(y,z)$ are arbitrary functions. We note that the choices of 
\begin{equation}
   \epsilon_1(y,z) = \epsilon_0(z) \sqrt{\frac{k_0}{12}} \i s \sqrt{\frac{q_1}{y}}\,, \qquad \widetilde{\epsilon}_1(y,z) = \widetilde{\epsilon}_0(z) \sqrt{\frac{k_0}{12}} \sqrt{\frac{q_2}{y}} \,,
\end{equation}
allow one to obtain both the Killing spinors given in \eqref{KSchit} and \eqref{KSchi} (via choice of $\epsilon_0 = \widetilde{\epsilon}_0 =1$) as well as those of \eqref{eq: KS_NP_gauge} (via choice of $\epsilon_0 = e^{-\frac{\i}{4} \left(\frac{a}{y_{1,2}}-1\right) z}, \: \widetilde{\epsilon}_0 =  e^{\frac{\i}{4} \left(\frac{a}{y_{1,2}}-1\right) z}$). We also note that we can write the solutions as
\begin{subequations}
\begin{align}
\epsilon & = \epsilon_2(y, z) e^{is\gamma_1\Upsilon/2} \left( \Theta(3y-a) \left(
    1 \;\; 0 
\right)^T + \Theta(a-3y)(0 \;\; 1)^T\right)\,, \\
\widetilde\epsilon & = \widetilde\epsilon_2(y, z) e^{is\gamma_1\Upsilon/2}\left(\Theta(3y -a)\left(
    0 \;\; 1 \right)^T + \Theta(a - 3y)\left(
    1 \;\; 0 \right)^T
\right)\,,
\end{align}
\end{subequations}
where $\Upsilon = \tan^{-1}\left[ \sqrt{q}(a-3y)^{-1}\right]$ and again $\epsilon_2, \widetilde{\epsilon}_2$ are arbitrary functions, labelled differently to $\epsilon_1, \widetilde{\epsilon}_1$ as these functions will not be strictly the same.\footnote{Strictly speaking, we will have $\epsilon_1 = ( \Theta(3y-a) \cos\left(s\Upsilon/2\right) + \i \Theta(a-3y) \sin\left(s\Upsilon/2\right))\epsilon_2$, $\widetilde{\epsilon}_1 = ( \Theta(a-3y) \cos\left(s\Upsilon/2\right) + \i \Theta(3y-a)  \sin\left(s\Upsilon/2\right))\widetilde{\epsilon}_2$. This is just multiplication of the initial arbitrary functions so can be absorbed into the arbitrariness.} Assuming suitable properties such as no zeros or poles of $\epsilon_{2}, \widetilde{\epsilon}_2$ at $y_{1,2}$, this form of the solution makes manifest the properties of chirality/anti-chirality at the poles of the spindle as discussed in \eqref{eq: chirality_poles}.

We define multiplication of two spinors $\psi$ and $\varphi$ as $\psi\varphi \equiv \psi^T C \varphi $ with $C = \sigma_2$. Vector bispinors are defined in the same fashion, i.e. 
$\psi \gamma_A\varphi \equiv \psi^T C \gamma_A \varphi$. Then the bispinors made out of the two Killing spinors are given as follows. For $\gamma_A = (\gamma_1\,,\gamma_2\,,\gamma_3\,,1)$,
\begin{align}
\begin{split}
      \i \widetilde{\epsilon} \gamma_A \epsilon & = {\color{black} k_0}\left(0\,, -s\frac{\sqrt{q}}{6y}\,,  -\frac{a-3y}{6y}\,, \frac{\sqrt{y}}{3} \right)\,, \\
      \i \epsilon \gamma_A \epsilon & = {\color{black} k_0} \left(-\frac{\sqrt{y}}{3} \,, \i \frac{a-3y}{6y} \,, -  \i s \frac{\sqrt{q}}{6y}\,, 0 \right)\, ,\\
    \i \wt{\epsilon} \gamma_A \wt{\epsilon} & =  {\color{black} k_0}\left(\frac{\sqrt{y}}{3} \,, \i \frac{a-3y}{6y} \,, - \i s \frac{\sqrt{q}}{6y}\,, 0 \right)\,. 
\end{split}
\end{align}
In the regular gauges \eqref{eq:gaugefield_regpatch}, the bispinors constructed from the Killing spinors $\epsilon$ and $\widetilde{\epsilon}$ are given as follows
 \begin{align} \label{bispinors}
 \begin{split}
  \left. \i  \widetilde\epsilon \gamma_A \epsilon \right|_{\mathcal{U}_{1,2}}  &={\color{black} k_0}\left(0\,, -s\frac{\sqrt{q}}{6y}\,,  \frac{3y - a}{6y}\,, \frac{\sqrt{y}}{3}\right) = \i  \widetilde\epsilon \gamma_A \epsilon\,,
     \\
   \left. \i \epsilon \gamma_A \epsilon \right|_{\mathcal{U}_{1,2}} &={\color{black} k_0} \,e^{-\frac{\i}{2} \left(\frac{a}{y_{1,2}}-1\right) z} \left(-\frac{\sqrt{y}}{3}\,, \i \frac{a-3y}{6y}\,, -\i s \frac{\sqrt{q}}{6y}\,,0\right) \,, 
    \\
   \left. \i \widetilde\epsilon \gamma_A \widetilde\epsilon \right|_{\mathcal{U}_{1,2}} &={\color{black} k_0}\, e^{\frac{\i}{2} \left(\frac{a}{y_{1,2}}-1\right) z} \left(\frac{\sqrt{y}}{3}\,, \i \frac{a-3y}{6y}\,, - \i s \frac{\sqrt{q}}{6y}\,,0 \right) \,,
\end{split}
\end{align}
where we note 
\begin{equation}
    \left(\left. \i \epsilon \gamma_A \epsilon \right|_{\mathcal{U}_{1,2}} \right)^* = - \left( \left. \i \widetilde\epsilon \gamma_A \widetilde\epsilon \right|_{\mathcal{U}_{1,2}} \right)\,,
\end{equation}
and it will be of relevance that 
\begin{eqnarray} \label{eq:killingveccomp}
    \i \widetilde{\epsilon} \gamma^\mu \epsilon = {\color{black}\frac{{k_0}}{L}}(0, -s)\,, \qquad \mu = \{y,z\}\,,
\end{eqnarray} is real.
Further, from the above, we find that the Killing vector has constant components 
\begin{equation}\label{eq:KillingVector}
    {\color{black}v^{\mu}} \partial_\mu \;\equiv\; \i \widetilde\epsilon \gamma^\mu \epsilon \,\partial_\mu \= -s{\color{black} \frac{k_0}{L}} \partial_z {\color{black} = - s \frac{L_0}{L}\frac{\Delta z}{2\pi} \partial_z}\,.
\end{equation}
 {\color{black}We recall from the normalisation factor $k_0$ defined in \eqref{eq:KillingVector} that the Killing vector scales with the size of the manifold $L$ with respect to a reference scale $L_0$.}
The choice of sign $s$ gives the opposite direction of the Killing vector orbit. From now on, we choose $s= -1$.

\subsection{Multiplets}
In this subsection we give details of the supersymmetry multiplets for our theory living on the spindle. We will consider a theory consisting of one vector and one chiral multiplet, giving details of the various fields in these multiplets. 

\subsubsection{Vector multiplet} The vector multiplet consists of a vector, two real scalars, two Dirac spinors and an auxiliary real scalar 
\begin{equation} \label{eq: vector_multiplet}
    \text{Vector} : \{ \mathcal{A}_{\mu}, \sigma, \rho, \lambda, \widetilde{\lambda}, \widehat{D}\}\,,
\end{equation}
where the $R$-charge assignment is $(0,0,0,-1,1,0)$.

Let us denote the  equivariant supercharge $\qeq$ by combining the supercharge with BRST charge for the $\mathcal{G}=U(1)$ gauge symmetry as
\be\label{qeqDef}
\qeq \;\equiv \;Q + Q_{\text{brst}}\,.
\ee
where we denote the supercharge as  $Q \equiv Q_\epsilon + Q_{\widetilde{\epsilon}}$.  Under this equivariant supercharge, the transformations of  the vector multiplet fields are given by
\be\ba{lll}
\label{eq:deltaA}
\qeq \mathcal{A}_\mu &=& -\i \half (\widetilde\epsilon \gamma_\mu \lambda + \epsilon \gamma_\mu \widetilde\lambda) + \partial_\mu c\,, \\
\qeq \S  &=& -\half  (\widetilde\epsilon \lambda - \epsilon  \widetilde\lambda)\,,\\
\qeq \P  &=& -\i \half  (\widetilde\epsilon \gamma_3 \lambda + \epsilon \gamma_3 \widetilde\lambda)\,,\\
\qeq \lambda &=& \i \gamma_3 \epsilon \mathcal{F} -\widehat{D} \epsilon - \i \gamma^\mu \epsilon \,\partial_\mu \S  - \gamma_3 \gamma^\mu \epsilon \,\partial_\mu \P  + (\i H - G\gamma_3)\epsilon \,\S  +(H \gamma_3 +\i G) \epsilon\,\P 
\\
&=& \i \gamma_3 \epsilon \mathcal{F} -\widehat{D} \epsilon - \i \gamma^\mu D_\mu (\epsilon \S ) - \gamma_3 \gamma^\mu D_\mu (\epsilon \P  )\,,
\\
\qeq \widetilde\lambda &=& \i \gamma_3 \widetilde\epsilon \mathcal{F} +\widehat{D} \widetilde\epsilon +
 \i \gamma^\mu \widetilde\epsilon  \,\partial_\mu  \S  - \gamma_3 \gamma^\mu \widetilde\epsilon \,\partial_\mu   \P  -(\i H +G\gamma_3)\widetilde\epsilon  \,\S  + (H \gamma_3 -\i G) \widetilde\epsilon  \,\P  
 \\
 &=& \i \gamma_3 \widetilde\epsilon \mathcal{F} +\widehat{D} \widetilde\epsilon +
 \i \gamma^\mu D_\mu( \widetilde\epsilon  \S ) - \gamma_3 \gamma^\mu D_\mu ( \widetilde\epsilon  \P  )\,,
 \\
\qeq \widehat{D}&=& - \i\half  \widetilde\epsilon \gamma^\mu  D_\mu  \lambda + \i \half \epsilon \gamma^\mu  D_\mu  \widetilde\lambda   - \i\half  \bar\epsilon( H+\i G\gamma_3) \lambda + \i \half  \epsilon( H -\i G\gamma_3) \widetilde\lambda  

\\
&=& - \i\half  D_\mu (\widetilde\epsilon \gamma^\mu \lambda) + \i \half D_\mu (\epsilon \gamma^\mu  \widetilde\lambda  )
\,,
\ea\ee
where $\mathcal{F}= \half \epsilon^{\mu\nu}\mathcal{F}_{\mu\nu} = \epsilon^{\mu \nu} \partial_{\mu} \mathcal{A}_{\nu}$ and $c$ is an additional ghost field required for systematic treatment of gauge fixing using BRST quantisation \cite{GonzalezLezcano:2023cuh}.  

The transformation of the ghost $c$ is 
\be \label{eq: Ghost_transformation}
\qeq c \= -\Lambda^{\mathcal{G}} +\Lambda^\mathcal{G}_0\,,  \qquad \Lambda^{\mathcal{G}} \equiv - \i \widetilde{\epsilon}\gamma^\mu \epsilon \mathcal{A}_\mu - \widetilde\epsilon \epsilon \sigma - \i \widetilde{\epsilon}\gamma_3 \epsilon \rho\,,
\ee
where  $\Lambda^{\mathcal{G}}_0$ is the constant part of the field dependent parameter $\Lambda^{\mathcal{G}}$.  
Here, the covariant derivative on each field is summarised as
\begin{equation}
    D_\mu \= \nabla_\mu -i \widehat{q}_R A_{\mu}  - \i \widehat{q}_{\mathcal{G}} \mathcal{A}_{\mu}
\end{equation} 
where  $\widehat{q}_R$ is the $R$-charge and $\widehat{q}_{\mathcal{G}}$ is the gauge charge of the field upon which $D_{\mu}$ acts.

\subsubsection{Chiral multiplet} The chiral multiplet consists of two complex scalars, two Dirac fermions and two auxiliary bosonic fields
\be \label{eq: chiral_multiplet}
\text{Chiral} : \{\phi \,, \wt{\phi}\,,\psi \,,\wt{\psi}\,,\mathfrak{F}\,,\wt{\mathfrak{F}}\}\,,
\ee
where the $R$-charge assignment is $(r, -r, r-1, -r+1,r-2, -r+2 )$. We also consider abelian gauge coupling via the $U(1)_{\mathcal{G}}$ gauge field $\mathcal{A}$ in \eqref{eq: vector_multiplet}, with gauge charge assignment $ (1\,,-1\,,1\,,-1\,,1\,,-1)$.\footnote{In later sections we will take the charge of the chiral multiplet to be the more generic $q_{\mathcal{G}}$. Such a choice modifies the supersymmetry transformations \eqref{deltachiral} by rescaling the coupling between vector multiplet and chiral multiplet fields by $q_{\mathcal{G}}$.} The supersymmetry transformations of the chiral multiplet fields are given by 
\be\ba{l} \label{deltachiral}
 \qeq \phi = \widetilde\epsilon \psi+ \i c \phi \,,\\
\qeq \wt\phi = \epsilon \wt\psi - \i c\wt\phi \,,\\
\qeq \psi = \i \gamma^\mu \epsilon D_\mu \phi  
-\i \epsilon ( \S +\frac{\r}{2} H)\phi   +\gamma_3 \epsilon (\P +\frac{\r}{2}G)\phi +\widetilde\epsilon \mathfrak{F}
+ \i c \psi \,,\\
\qeq \wt\psi = \i \gamma^\mu \widetilde\epsilon D_\mu \wt\phi  
-\i \widetilde\epsilon ( \S +\frac{\r}{2} H)\wt\phi   -\gamma_3 \widetilde\epsilon ( \P +\frac{\r}{2}G)\wt\phi +\epsilon \wt{\mathfrak{F}} - \i c \wt{\psi}\,,
\\
\qeq \mathfrak{F} =    \i \epsilon\gamma^\mu D_\mu \psi  +\i  \epsilon\psi ( \S +\frac{\r}{2}H)+\epsilon\gamma_3 \psi ( \P +\frac{\r}{2}G)   -\i  \epsilon\lambda \phi + \i  c \mathfrak{F}\,,
\\
\qeq \wt{\mathfrak{F}} =    \i \widetilde\epsilon\gamma^\mu D_\mu \wt\psi  +\i  \widetilde\epsilon\wt\psi (  \S +\frac{\r}{2}H )-\wt\epsilon\gamma_3 \wt\psi ( \P +\frac{\r}{2}G)     +\i  \widetilde\epsilon \wt\lambda \,\wt\phi- \i   c \wt{\mathfrak{F}} \,.
\ea\ee 
The supersymmetry algebra is closed to 
\be \label{eq: Q^2}
Q_{eq}^2 = \cL_v + \delta_R(\Lambda^R)+ \delta_G (\Lambda^{\mathcal{G}}_0) = \cL_{v} + \i \widehat{q}_R \Lambda^R + \i \widehat{q}_{\mathcal{G}} \Lambda_0^{\mathcal{G}}\,,
\ee
where the $R$-symmetry parameter $\Lambda^R$ is given by 
\begin{eqnarray} \label{Q2algebra}
\Lambda^R &=& -v^\mu A_\mu - \frac{1}{2}(H \widetilde{\epsilon} \epsilon + \i G \widetilde{\epsilon}\gamma_3\epsilon) \,.
\end{eqnarray} 
If we use the gauge choice for the $R$-symmetry gauge field as given in \eqref{eq: A_2d}, then we find 
\begin{equation}
    \Lambda^R = 0\,,
\end{equation}
where we recall that this computation of $\Lambda^R$ uses a choice of $R$-symmetry gauge field which is singular at the poles of the spindle. If instead we perform a gauge transformation of the form \eqref{eq: R_gauge_transform} with transformation parameter $\alpha$ in \eqref{eq: beta_1,2} we obtain a gauge field which is patchwise regular at the poles as given in \eqref{eq:gaugefield_regpatch}. Using this gauge transformation, we see that in the regular gauges $\Lambda^R$ takes the form 
\begin{eqnarray} \label{Q2algebraNP}
\left. \Lambda^R \right|_{\mathcal{U}_{1,2}} =  - {\color{black}\frac{k_0}{L}} \left. \alpha \right|_{\mathcal{U}_{1,2}}  {\color{black}= - \frac{L_0}{L}  \frac{1}{2n_{1,2}} } \,,
\end{eqnarray} 
where, as highlighted before, the evaluation at $\mathcal{U}_{1,2}$ indicates that this is evaluated in regular gauges. We will later see similar relations holding for the $U(1)_{\mathcal{G}}$ symmetry parameter $\Lambda^{\mathcal{G}}$, although we postpone this discussion until we have established the form of the $U(1)_{\mathcal{G}}$ gauge field $\mathcal{A}$ which must be taken to be on the BPS locus.  
\subsection{Cohomological variables} \label{sec:Cohomological variables} It is convenient to reorganise the fields into a certain representation of supersymmetry called ``cohomological variables'', which we will do for the chiral multiplet. The cohomological variables consist of the $\qeq$-cohomology complex $(\Phi\,, \qeq \Phi\,,\Psi\,,\qeq \Psi)$, where we call $\Phi$ and $\Psi$ the elementary boson and elementary fermion, and $\qeq \Phi$ and $\qeq \Psi$ are their superpartners. For this reorganisation, we define the following twisted variables for the spinor fields
\be
\epsilon\psi\,,\quad \widetilde\epsilon \widetilde{\psi}\,,\quad \widetilde\epsilon \psi\,,\quad \epsilon\widetilde{\psi}\,,
\ee
with $R$-charge $\{ r-2\,,-r+2\,, r\,,-r \}\,$ respectively.
The inverse relation is given by 
\be\label{InverseTwist}
\psi = \frac{1}{\widetilde{\epsilon}\epsilon}\Bigl( \epsilon (\widetilde\epsilon \psi) - \widetilde{\epsilon}(\epsilon \psi)\Bigr)
\,, \qquad \widetilde{\psi} = \frac{1}{\widetilde{\epsilon}\epsilon}\Bigl(  \epsilon (\widetilde{\epsilon}\widetilde{\psi} ) - \widetilde{\epsilon}(\epsilon\widetilde{\psi}) \Bigr)\,.
\ee
This redefinition is well-defined since the Jacobian is non-zero everywhere in~$0<y_1 < y < y_2$ as
\begin{equation}
    |J| = \left|\det\frac{\partial (\psi\,,\widetilde\psi) }{\partial(\widetilde\epsilon \psi\,, \epsilon \psi\,, \widetilde\epsilon \widetilde\psi\,, \epsilon \widetilde\psi)}\right|= \frac{1}{|(\widetilde{\epsilon}\epsilon)^2|} = \frac{9}{k_0y }\,.
\end{equation}
Then the cohomological variables are defined by 
\label{eq:cohomologicalvariables}
\begin{equation}
\begin{alignedat}{2}
        \Phi & =  \{\phi\, ,\widetilde\phi\}\, ,\quad \qquad \qquad && \qeq \Phi =\{ \qeq \phi \, ,\qeq \widetilde{\phi}\}\,  \label{eq:CQarray} ,\\ 
\Psi & =  \{\epsilon \psi 
\, , \widetilde{\epsilon} \widetilde{\psi}
\}\,, \qquad\qquad\quad && \qeq \Psi = \{\qeq \left(\eps \psi\right)\, , \qeq \bigl( \widetilde{\epsilon} \widetilde{\psi}\bigr)\}\, , 
\end{alignedat}
\end{equation}
where each element in $\qeq \Phi$ and $\qeq \Psi$ is treated as an independent variable and the explicit expression in terms of the physical variables are given by 
\begin{align}
\begin{split}
    \qeq \phi &= \widetilde\epsilon \psi + \i c\phi\,,
    \\
    \qeq \widetilde{\phi}&= \epsilon \widetilde{\psi} - \i c \widetilde{\phi} \,,
    \\
    \qeq (\epsilon \psi )&= \epsilon \widetilde{\epsilon} \mathfrak{F} + \i \epsilon \gamma^\mu \epsilon D_\mu \phi + \epsilon \gamma_3 \epsilon ( \rho + \tfrac{r}{2}G)\phi + \i c (\epsilon \psi )\,,
    \\
    \qeq (\widetilde\epsilon \widetilde\psi )&= \widetilde\epsilon {\epsilon} \widetilde{\mathfrak{F}}+ \i \widetilde\epsilon \gamma^\mu \widetilde\epsilon D_\mu \widetilde\phi - \widetilde\epsilon \gamma_3 \widetilde\epsilon ( \rho + \tfrac{r}{2}G)\widetilde\phi - \i c (\widetilde\epsilon \widetilde\psi )\,.
    \end{split}
\end{align}

We note that one can also introduce cohomological variables for the vector multiplet quantities \cite{GonzalezLezcano:2023cuh}, however these will not be necessary for our localisation calculations. The explicit details will be discussed in later sections but the essential point is that the vector multiplet one-loop determinant for a theory with gauge group $\mathcal{G}$ follows from the chiral multiplet result by setting $r=2$ and looking at the adjoint representation $\mathfrak{R}_\mathcal{G} = \text{adj}_{\mathcal{G}}$ of $\mathcal{G}$ \cite{Inglese:2023tyc}.

\subsubsection*{Reality property}
In the vector multiplet, we will choose the following reality condition for the fields 
\begin{equation} \label{eq: reality_conditions_vector}
    \mathcal{A}^* = \mathcal{A}\,, \quad \sigma^* = \sigma\,, \quad \rho^* = \rho\,, \quad D^* = D\,.
\end{equation}
As discussed in \cite{GonzalezLezcano:2023cuh}, such a choice of reality condition forces us to give up compatibility of supersymmetry and reality (i.e. complex conjugation and the supersymmetry variation $Q_{eq}$ do not commute) although this choice does allow for the functional integration of the Euclidean theory to be well defined.

For the chiral multiplet, we similarly set complex conjugation of the bosonic fields as 
\begin{equation} \label{eq: chiral_bosonic_reality}
    \phi^\ast = \widetilde{\phi}\,,\qquad \mathfrak{F}^\ast = \widetilde{\mathfrak{F}}\,,
\end{equation}
which together with \eqref{eq: reality_conditions_vector} allows us to compute 
\begin{align} \label{RealityofQPsi}
\begin{split}
    ( \qeq (\epsilon \psi ))^\dagger &=  \widetilde{\epsilon}\epsilon \widetilde{\mathfrak{F}} - \i \widetilde\epsilon \gamma^\mu \widetilde\epsilon D_\mu \widetilde\phi  +\widetilde\epsilon \gamma_3 \widetilde\epsilon \left( \rho + \frac{r}{2}G\right)\widetilde\phi - \i c^\ast (\epsilon \psi)^\ast 
    \\
        &=\qeq (\widetilde\epsilon \widetilde\psi ) -2 \left[ \i \widetilde\epsilon \gamma^\mu \widetilde\epsilon D_\mu   - \widetilde{\epsilon}\gamma_3\widetilde\epsilon \left( \rho +\frac{r}{2}G \right)\right]\widetilde\phi + \i c (\widetilde\epsilon \widetilde\psi) -\i c^\ast (\epsilon \psi)^\ast\,,
        \\
 ( \qeq (\widetilde\epsilon \widetilde\psi ))^\dagger &=    
  \epsilon \widetilde{\epsilon} \mathfrak{F} - \i \epsilon \gamma^\mu \epsilon D_\mu \phi - \epsilon \gamma_3 \epsilon \left( \rho + \frac{r}{2}G \right)\phi + \i c^\ast (\widetilde\epsilon \widetilde\psi )^\ast\,
 \\
  &=\qeq (\epsilon \psi)- 2 \left[ \i \epsilon \gamma^\mu \epsilon D_\mu   + {\epsilon}\gamma_3\epsilon \left( \rho +\frac{r}{2}G \right)\right]\phi    - \i c (\epsilon \psi) + \i c^\ast (\widetilde\epsilon \widetilde\psi)^\ast\,.
\end{split}
\end{align} 
where we also used $G=G^*$, the reality properties of the bilinears in \eqref{bispinors}, and the reality of the components of the Killing vector \eqref{eq:killingveccomp}.  
We do not specify the reality condition for $c$, $\epsilon \psi$ and $\widetilde{\epsilon}\wt{\psi}$ as these are not relevant for the computation.

\subsection{Supersymmetric action on the spindle}

In order to use supersymmetric localisation to compute the partition function, we first need to specify the theory that we consider on the spindle geometry. General supersymmetric actions on curved manifolds are discussed in some detail in \cite{Closset:2014pda} and here we will consider a simple case of such an action. Following \cite{GonzalezLezcano:2023cuh}, we consider a theory of the form 
\begin{equation} \label{eq: susy_action}
    S = S_{\text{v.m.}} + S_{\text{c.m.}} + S_{\text{FI}} + S_{\text{top}}\,,
\end{equation}
where 
\begin{equation}
    S_{\text{v.m}} = \frac{1}{g^2_{\text{YM}}} \int_{\mathbb{\Sigma}} \textrm{d}^2 x  \, \sqrt{g_{\mathbb{\Sigma}}} \left[ \mathcal{L}_{\text{v.m.}}^{\text{bulk}} + D_{\mu} V_{\text{v.m.}}^{\mu}  \right] =  \frac{1}{g^2_{\text{YM}}} \int_{\mathbb{\Sigma}} \textrm{d}^2 x  \, \sqrt{g_{\mathbb{\Sigma}}}  \mathcal{L}_{\text{v.m.}}^{\text{bulk}} \,,
\end{equation}
where $g_{\text{YM}}$ is the super-renormalisable gauge coupling and the total derivative term which is required in order to make the action supersymmetric is ignored as the spindle does not have a boundary. The bulk Lagrangian is given by 
\begin{align}
\begin{split} \label{eq: vm_lagrangian}
    \mathcal{L}^{\text{bulk}}_{\text{v.m.}} & = \frac{1}{2} \left( \mathcal{F} + \i (G \sigma - H \rho ) \right)^2 + \frac{1}{2} \partial_{\mu} \sigma \partial^{\mu} \sigma + \frac{1}{2} \partial_{\mu} \rho \partial^{\mu} \rho \\
    & \qquad  + \frac{1}{2} \left( \widehat{D} - \i (H \sigma + G \rho ) \right)^2 + \frac{\i}{2} \widetilde{\lambda} \gamma^{\mu} D_{\mu} \lambda \\
    & = \frac{1}{2} \left( \mathcal{F} + \i(G \sigma - H \rho) \right)^2 + \frac{1}{2} \partial_{\mu} \sigma \partial^{\mu} \sigma + \frac{1}{2} \partial_{\mu} \rho \partial^{\mu} \rho + \frac{1}{2} D^2 + \frac{\i}{2} \widetilde{\lambda} \gamma^{\mu} D_{\mu} \lambda\,,
\end{split}
\end{align}
where as in \cite{GonzalezLezcano:2023cuh} we used the field redefinition of 
\begin{equation}
    D \equiv \widehat{D} - \i (H \sigma + G \rho)\,.
\end{equation}
The second term in \eqref{eq: susy_action} is 
\begin{equation} \label{eq: cm_action}
    S_{\text{c.m.}} = \frac{1}{g^2_{\text{YM}}} \int_{\mathbb{\Sigma}} \textrm{d}^2 x  \, \sqrt{g_{\mathbb{\Sigma}}} \left[ \mathcal{L}_{\text{c.m.}}^{\text{bulk}} + D_{\mu} V_{\text{c.m.}}^{\mu}  \right] =  \frac{1}{g^2_{\text{YM}}} \int_{\mathbb{\Sigma}} \textrm{d}^2 x  \, \sqrt{g_{\mathbb{\Sigma}}}  \mathcal{L}_{\text{c.m.}}^{\text{bulk}}\,,
\end{equation}
where 
\begin{equation} \label{eq: bulk_chiral_lagrangian}
    \mathcal{L}^{\text{bulk}}_{\text{c.m.}} = D_{\mu} \widetilde{\phi} D^{\mu} \phi + M_{\phi}^2 \widetilde{\phi} \phi + \widetilde{\mathfrak{F}} \mathfrak{F} - \i \widetilde{\psi} \gamma^{\mu} D_{\mu} \psi + \widetilde{\psi} M_{\psi} \psi - \i \widetilde{\psi} \lambda \phi - \i \widetilde{\phi} \widetilde{\lambda} \psi\,,  
\end{equation}
and mass squared of the scalar field and mass of the fermion are 
\begin{subequations}
\begin{align}
\begin{split}
    M_{\phi}^2 & = \left( \sigma + \frac{r}{2} H\right)^2  + \left( \rho + \frac{r}{2} G \right)^2 + \frac{r}{4} R_{\mathbb{\Sigma}} + \i \widehat{D}\,, \\
     M_{\psi} & = -\i \left( \sigma + \frac{r}{2} H \right) - \left(\rho + \frac{r}{2} G \right) \gamma_3\,.
     \end{split}
\end{align}
\end{subequations}
The final terms we include are the Fayet-Iliopoulos (FI) and topological terms. The combination of these two terms takes the form
\begin{equation} \label{eq: S_FI+top}
    S_{\text{FI}} + S_{\text{top}} = \i \int_{\mathbb{\Sigma}} \textrm{d}^2 x \, \sqrt{g_{\mathbb{\Sigma}}} \left( -\xi \widehat{D} + \frac{\theta}{2\pi} \mathcal{F} \right)\,,
\end{equation}
where $\xi \in \mathbb{R}$ is the FI parameter and $\theta \in (0,2\pi n_1 n_2)$ is the topological ``theta'' parameter. The action \eqref{eq: S_FI+top} will give rise to the classical contribution in our localisation calculations of later sections. We note that this term is an example of a more general class of actions generated by a \textit{twisted superpotential} \cite{Benini:2012ui, Closset:2014pda}, which is discussed in more detail in Appendix~\ref{sec: twisted_superpotential}.

\section{Supersymmetric localisation} \label{sec: localisation}
\subsection{General localisation argument}

In this paper the main object of interest is the partition function
\begin{equation} \label{eq: path_integral}
    Z_{\mathbb{\Sigma}} =  \int_{\mathcal{M}} \mathcal{D \varphi} \, e^{-S[\varphi]}  
\end{equation}
where $S$ is the action of our supersymmetric theory \eqref{eq: susy_action} and $\varphi$ denotes all \textrm{d}ynamical fields in the chiral \eqref{eq: chiral_multiplet} and vector \eqref{eq: vector_multiplet} multiplets. In order to compute this path integral exactly we will make use of the technique of \textit{supersymmetric localisation} \cite{Duistermaat:1982vw, Atiyah:1984px, Witten:1988ze, Witten:1988xj, Nekrasov:2002qd, Pestun:2007rz}, see also the review \cite{Cremonesi:2013twh}. 

We first provide a short discussion of the general procedure of localisation. One  first considers a $Q_{eq}$-exact deformation of the action in the path integral \eqref{eq: path_integral} 
\begin{equation} \label{eq: partition_function_deformation}
    Z_{\mathbb{\Sigma}}(t) =  \int_{\mathcal{M}} \mathcal{D} \varphi \, e^{-S[\varphi] - t Q_{eq} \mathcal{V}}\,,
\end{equation}
 where $t \in \mathbb{ R}_+ $ is a deformation parameter and $\mathcal{V}$ is a fermionic quantity which satisfies 
\begin{equation} \label{eq: supercharge^2_annilhilation}
    Q_{eq}^2 \mathcal{V} = (\cL_{{\color{black}v}} + \delta_R(\Lambda^R)+ \delta_G (\Lambda^{\mathcal{G}}_0)) \mathcal{V} = 0\,,
\end{equation}
as well as the bosonic part of the deformation being positive semi-definite 
\begin{equation} \label{eq: QV_positive_def}
\left. Q_{eq} \mathcal{V} \right|_{\text{bos}} \geq 0\,.
\end{equation}
Using \eqref{eq: supercharge^2_annilhilation} and \eqref{eq: QV_positive_def} we can show that the deformed path integral is in fact independent  of $t$. In order to do this, we take a derivative 
\begin{align}
\begin{split}
    Z_{\mathbb{\Sigma}}'(t) & = - \int_{\mathcal{M}} \mathcal{D} \varphi \, Q_{eq} \mathcal{V} e^{-S[\varphi] - t Q_{eq} \mathcal{V}} \\
    & = - \int_{\mathcal{M}} \mathcal{D} \varphi \, Q_{eq} \left( \mathcal{V} e^{-S[\varphi] - t Q_{eq} \mathcal{V}} \right) \\
    & =0\,,
    \end{split}
\end{align}
 where in the second line we used the fact that the theory is supersymmetric $Q_{eq} S = 0$ together with \eqref{eq: supercharge^2_annilhilation} and in the third line we used that fact that the $Q_{eq}$-exact term is a total derivative in the space of fields (the measure is invariant under both $Q$ and $Q_{\text{BRST}}$). This $t$ independence is also explained by the Ward identity as it implies expectation value of symmetry transformation of an operator is zero, i.e. $\langle \qeq \cV \rangle =0$. 

The fact that the partition function is independent of the value of $t$ means that we are free to choose any value of $t$, and in particular we may take the limit $t \rightarrow \infty$. Due to the positive semi-definite property \eqref{eq: QV_positive_def}, in this limit the stationary points of $Q_{eq} \mathcal{V}$ will dominate the path integral as all other configurations will be exponentially suppressed. These stationary points will form a moduli space known as the \textit{localisation locus}, denoted by $\mathcal{M}_{\text{loc}}$, of the theory, with the full path integral \eqref{eq: path_integral} boiling down to an integral over the localisation locus 
\begin{equation} \label{eq: Z_t_infty}
   Z_{\mathbb{\Sigma}} =  \lim_{t \rightarrow \infty} Z_{\mathbb{\Sigma}}(t) =  \int_{\mathcal{M}_{\text{loc}}} \mathcal{D} \varphi \, e^{-S[\varphi]} Z^{Q_{eq} \mathcal{V}}_{1-\text{loop}}[\varphi]\,,
\end{equation}
where $Z^{Q_{eq} \mathcal{V}}_{1-\text{loop}}$ is a one-loop super determinant arising from quadratic fluctuations in the fields around each point in $\mathcal{M}_{\text{loc}}$. We will soon see that the zeroes of $Q_{eq} \mathcal{V}$ coincide with the stationary points and thus the deformation will vanish on the localisation locus. It is important to note that this result is an \textit{exact} expression for the partition function of our theory -  no higher loop terms are present as they are suppressed by positive powers of $t$~\cite{Cremonesi:2013twh}. 

We now need to make a choice of $Q_{eq} \mathcal{V}$ which satisfies the properties \eqref{eq: supercharge^2_annilhilation} and \eqref{eq: QV_positive_def} and evaluate the form of the localisation equations that this choice gives. In what follows we will select the `canonical' choice of 
\begin{equation}
        \mathcal{V} = \mathcal{V}^{\text{v.m.}} + \mathcal{V}^{\text{c.m.}} = \frac{1}{2} \int_{\mathbb{\Sigma}} \textrm{d}^2 x  \, \sqrt{g_{\mathbb{\Sigma}}} \left[   (Q_{eq} \lambda )^{\dagger} \lambda + \widetilde{\lambda} (Q_{eq} \widetilde{\lambda} )^{\dagger} +  (Q_{eq} \psi )^{\dagger} \psi + \widetilde{\psi} (Q_{eq} \widetilde{\psi} )^{\dagger}  \right]\,,
\end{equation}
where the first two terms on the right are associated with the vector multiplet deformation and the second two terms with the chiral multiplet deformation. Acting with the equivariant supercharge and extracting the bosonic part, we find
\begin{align}
\begin{split} \label{eq: QV_bosonic}
   \left. Q_{eq} \mathcal{V} \right|_{\text{bos}}= \frac{1}{2} \int_{\mathbb{\Sigma}} \textrm{d}^2 x\, \sqrt{g_{\mathbb{\Sigma}}} \big[ &  (Q_{eq} \lambda )^{\dagger} (Q_{eq} \lambda) +( Q_{eq} \widetilde{\lambda}) (Q_{eq} \widetilde{\lambda} )^{\dagger} \\
   &+  (Q_{eq} \psi )^{\dagger} (Q_{eq} \psi) + (Q_{eq}  \widetilde{\psi}) (Q_{eq} \widetilde{\psi} )^{\dagger}  \big]\,,
   \end{split}
\end{align}
which manifestly satisfies \eqref{eq: QV_positive_def} by virtue of being a sum of squares. As this action is a sum of squares, this means that the stationary points coincide with the zeroes of the integrand, and thus the stationary points of the action \eqref{eq: QV_bosonic} are simply 
\begin{equation} \label{eq: BPS_1}
    Q_{eq} \lambda =  Q_{eq} \widetilde{\lambda} =  Q_{eq} \psi =  Q_{eq} \widetilde{\psi} = 0\,.  
\end{equation}
Note that the equations above are only the stationary points of the full deformation term $Q_{eq} \mathcal{V}$ when one also has 
\begin{align}
\begin{split}
   \left. Q_{eq} \mathcal{V} \right|_{\text{ferm}} =  \frac{1}{2} \int_{\mathbb{\Sigma}} \textrm{d}^2 x\, \sqrt{g_{\mathbb{\Sigma}}} \big[ &  (Q_{eq} (Q_{eq} \lambda )^{\dagger})  \lambda -  \widetilde{\lambda} (Q_{eq} (Q_{eq} \widetilde{\lambda} )^{\dagger}) \\
   &+  (Q_{eq}(Q_{eq} \psi )^{\dagger})  \psi -   \widetilde{\psi} (Q_{eq} (Q_{eq} \widetilde{\psi} )^{\dagger} ) \big] = 0\,,
   \end{split}
\end{align}
which is satisfied for 
\begin{equation} \label{eq: BPS_2}
    \lambda = \widetilde{\lambda} = \psi = \widetilde{\psi} = 0\,,
\end{equation}
i.e. the vanishing of all fermions. Putting \eqref{eq: BPS_1} and \eqref{eq: BPS_2} together, we see that the localisation locus is given by solutions to the BPS equations, and thus we can write 
\begin{equation} \label{eq: M_BPS}
    \mathcal{M}_{\text{loc}}  = \{ \varphi \: | \:  \text{fermions} = 0\, , \, Q_{eq} (\text{fermions}) = 0 \} \,,
\end{equation}
and from here on we will refer to the localisation locus as the BPS locus.
We also note that $ \left. Q_{eq} \mathcal{V} \right|_{\mathcal{M}_{\text{loc}}} = 0$, demonstrating explicitly why such a term does not appear in \eqref{eq: Z_t_infty}. 

This concludes the description of the general procedure of supersymmetric localisation which we will apply to our theory \eqref{eq: susy_action}. We will first solve the BPS equations \eqref{eq: BPS_1} and \eqref{eq: BPS_2}, before moving on to tackle the more challenging calculation of the one-loop determinant in Section \ref{sec: one-loop}.

\subsection{BPS locus}

Before solving \eqref{eq: BPS_1} explicitly, we begin with a discussion of the $U(1)_{\mathcal{G}}$ flux through the spindle, noting that the vector multiplet BPS locus will be characterised by such flux. The general formula for quantised $U(1)_{\mathcal{G}}$ flux is
\begin{equation} \label{eq: Gauge_flux_quantisation_spindle}
   \mathfrak{f}_{\mathcal{G}} = \frac{1}{2\pi} \int_{\mathbb{\Sigma}} \textrm{d} \mathcal{A} =   \frac{\mathfrak{m}}{n_1 n_2}\,, \qquad \mathfrak{m} \in \mathbb{Z}\,. 
\end{equation}
We can parameterise our $U(1)_{\mathcal{G}}$ gauge field on $\mathbb{\Sigma}$ as 
\begin{equation}
    \mathcal{A} = \mathcal{A}_z(y) \textrm{d}z\, ,
\end{equation}
because we can always remove the component $\mathcal{A}_y(y)$ via a gauge transformation. As discussed in \cite{Ferrero:2021etw, Inglese:2023tyc} this gauge field is the representative of an $\mathcal{O}(\mathfrak{-m})$ orbibundle on $\mathbb{\Sigma}$, which gives the values of the singular gauge field at the poles $y_{1,2}$ to be the following 
\begin{equation} \label{eq: gauge_field_poles}
    \mathcal{A}_z(y_1) = \frac{2\pi}{\Delta z} \frac{\mathfrak{m}_1}{n_1}\,, \qquad  \mathcal{A}_z(y_2) = \frac{2\pi}{\Delta z} \frac{\mathfrak{m}_2}{n_2}\,, \qquad \mathfrak{m}_{1,2} \in \mathbb{Z}\,,
\end{equation}
from which one can see that the gauge-flux quantisation condition \eqref{eq: Gauge_flux_quantisation_spindle} is satisfied and the relation 
\begin{equation}\label{mFlux}
    \mathfrak{m} = \mathfrak{m}_2 n_1 - \mathfrak{m}_1 n_2\,,
\end{equation}
must hold. As was pointed out in \cite{Inglese:2023tyc}, we note that we can further express $(\mathfrak{m}_1,\mathfrak{m}_2)$ in terms of $\mathfrak{m}$ and additional integers $(\mathfrak{a}_1, \mathfrak{a}_2)$ as
\begin{equation} \label{eq: a_i_defs}
    \mathfrak{m}_1 = \mathfrak{m} \mathfrak{a}_1\,, \qquad  \mathfrak{m}_2 = \mathfrak{m} \mathfrak{a}_2\,, \qquad 1 = \mathfrak{a}_2 n_1 - \mathfrak{a}_1 n_2\,, 
\end{equation}
where we note that the last equation implies gcd($\mathfrak{a}_1,n_1$) = gcd($\mathfrak{a}_2,n_2$) = 1. Furthermore, given a pair $(\mathfrak{a}_1, \mathfrak{a}_2)$, the pair $(\mathfrak{a}_1 + n_1 \delta \mathfrak{a}, \mathfrak{a}_2 + n_2 \delta \mathfrak{a})$ for $\delta \mathfrak{a} \in \mathbb{Z}$ also solves the constraint. We expect physical observables to be independent of $\delta \mathfrak{a}$, which we will see  in Section \ref{sec: Mirror Symmetry} to indeed be the case for our partition function. 
We can perform a gauge transformation of the form
\begin{equation} \label{eq: gauge_transform_A}
     \mathcal{A} \rightarrow \mathcal{A}' = (\mathcal{A}_z + \beta) \textrm{d}z\,,
\end{equation}
with $\beta$ a constant in each patch. We thus obtain a regular gauge field at each pole if we choose 
\begin{equation} \label{eq: betas_regular}
    \left. \beta \right|_{\mathcal{U}_{1,2}}  = -  \frac{2\pi}{\Delta z} \frac{\mathfrak{m}_{1,2}}{n_{1,2}}\,.
\end{equation}
Note that this discussion somewhat parallels the discussion around \eqref{eq: R_gauge_transform} where we considered the gauge transformations of the $U(1)_R$ field to similar effect.

We now proceed to solve the BPS equations \eqref{eq: BPS_1}, recalling that these equations arose using the `canonical' choice of the $Q_{eq}$-exact deformation 
\begin{align}
\begin{split}
    Q_{eq} \mathcal{V} & = Q_{eq} \mathcal{V}^{\text{v.m.}} + Q_{eq} \mathcal{V}^{\text{c.m.}} \\
    &= \frac{1}{2} \int _{\mathbb{\Sigma}} \textrm{d}^2 x \,\sqrt{g_{\mathbb{\Sigma}}} \, Q_{eq} \bigg[   \left( (Q_{eq} \lambda )^{\dagger} \lambda + \widetilde{\lambda} (Q_{eq} \widetilde{\lambda} )^{\dagger} \right) +  \left( (Q_{eq} \psi )^{\dagger} \psi + \widetilde{\psi} (Q_{eq} \widetilde{\psi} )^{\dagger} \right)  \bigg]\,. 
    \end{split}
\end{align}
We focus first on the vector multiplet part. Following \cite{GonzalezLezcano:2023cuh}, we can write the bosonic term as 
\begin{align}
\begin{split} \label{eq: VM_deformation_action}
  Q_{eq} \mathcal{V}^{\text{v.m.}} \bigg|_{\text{bos.}} = 2 \int _{\mathbb{\Sigma}}\textrm{d}^2 x \, \sqrt{g_{\mathbb{\Sigma}}} \,
  \epsilon^{\dagger} \epsilon \bigg[ &\left( \mathcal{F} - \i H \rho \right)^2 + \left( D + G \sigma \frac{\epsilon^{\dagger}\gamma_3 \epsilon}{\epsilon^{\dagger} \epsilon} \right)^2 \\
    & + \left( D_{\mu} \sigma + \i G \sigma \frac{\epsilon^{\dagger} \gamma_3 \gamma_{\mu} \epsilon}{\epsilon^{\dagger} \epsilon} \right)^2 + (D_{\mu} \rho)^2 \bigg]\,, 
     \end{split}
\end{align}
which is positive definite for the values of $G, H$ for the spindle, assuming real  fields as we do in \eqref{eq: reality_conditions_vector}.
Before we solve these equations, we note that unlike the case of $S^2$ \cite{Benini:2012ui, Doroud:2012xw, GonzalezLezcano:2023cuh} the deformation term is not equivalent to the vector multiplet Lagrangian \eqref{eq: vm_lagrangian}. This action is clearly not suitable to be used as the deformation term on the spindle as it is not positive definite for the values of the functions $G,H$ together with the reality conditions \eqref{eq: reality_conditions_vector}. 


We solve the equations of motion arising from the action \eqref{eq: VM_deformation_action}. Since this is a sum of squares, the equations of motion require that each quantity inside the squared brackets vanishes. Solving these equations using the Killing spinor $\epsilon=\wt{\chi}$ given in \eqref{KSchit} and recalling $G = -{\color{black} L^{-1}}, H = \i a y^{-3/2}{\color{black} L^{-1}}$ from \eqref{kse1}, we find the solution
\begin{equation} \label{eq: BPS_locus_real}
    \rho = \rho_0\,, \qquad \mathcal{F} = - \frac{\rho_0}{{\color{black} L}} \frac{a}{y^{3/2}}\,, \quad  \sigma =  \sigma_0\frac{3}{\sqrt{y}}\frac{2\pi}{\Delta z} \,, \qquad D = 3 \frac{\sigma_0}{{\color{black} L}} \frac{3y -a}{2y^2} \frac{2\pi}{\Delta z}\,,
\end{equation}
where $\rho_0$ and $\sigma_0$ are constants. We can immediately fix the value of the constant $\rho_0$ using the gauge flux quantisation condition \eqref{eq: Gauge_flux_quantisation_spindle}, obtaining 
\begin{equation} \label{eq: rho_0_quantisation}
    \rho_0 {\color{black} L} = \frac{3 \mathfrak{m}}{n_1-n_2}\,,
\end{equation}
leaving us with one undetermined constant $\sigma_0$ and the integer $\mathfrak{m}$ which together parameterise the vector multiplet contribution to the full BPS locus. When we compute the partition function \eqref{eq: Z_t_infty} we will integrate over $\sigma_0 \in \mathbb{R}$ and sum over $\mathfrak{m} \in \mathbb{Z}$.

Given the explicit form of the field strength in \eqref{eq: BPS_locus_real} we can compute the value of the gauge field $\mathcal{A}$ by integrating and using the boundary conditions \eqref{eq: gauge_field_poles}. We find two equivalent forms of the gauge field as\footnote{The singular gauge field \eqref{eq: gauge_field_explicit_real} can also be expressed as
\begin{equation} \label{eq: singular_gauge_field_democratic}
    \mathcal{A}_z = \frac{\mathfrak{m}}{n_2-n_1} \frac{1 }{2} \left( 1 - \frac{a}{y} \right) +\frac{2\pi}{\Delta z} \frac{\mathfrak{m}_2-\mathfrak{m}_1}{n_2 - n_1}\,.
\end{equation}}
\begin{eqnarray}
    \label{eq: gauge_field_explicit_real}
    \mathcal{A}_z(y) &=& \frac{\mathfrak{m}}{n_1-n_2} \frac{a }{2} \left( \frac{1}{y} - \frac{1}{y_1} \right) + \frac{2\pi}{\Delta z} \frac{\mathfrak{m}_1}{n_1}
     = \frac{\mathfrak{m}}{n_1-n_2} \frac{a }{2} \left( \frac{1}{y} - \frac{1}{y_2} \right) + \frac{2\pi}{\Delta z} \frac{\mathfrak{m}_2}{n_2}\,,
\end{eqnarray} 
where we note that this is a gauge choice which allows for a single patch description on the entire spindle, except for the poles at $y_{1,2}$ where the gauge field is singular. Due to the positions of these singularities, we refer to this as the singular gauge. In order to move into a regular gauge at one of the poles, we perform gauge transformations of the form \eqref{eq: gauge_transform_A} at each pole, with the choice of transformation parameters given in \eqref{eq: betas_regular}. We thus find that the values of the gauge field in the regular gauges
\begin{align} \label{eq: u1g_regpatch}
    \mathcal{A}_z \big|_{\mathcal{U}_{1,2}} & = \frac{\mathfrak{m}}{n_1-n_2} \frac{a }{2} \left( \frac{1}{y} - \frac{1}{y_{1,2}} \right)\,.
\end{align}

In computing the BPS value of the abelian gauge field we are also able to identify the parameter $\Lambda^{\mathcal{G}}$ through \eqref{eq: Ghost_transformation}. Explicitly on the BPS locus, we have\footnote{The gauge parameter \eqref{eq: Lambda_G_BPS} can also be expressed as
\begin{equation}
\Lambda^{\cal G} = {\color{black} \frac{L_0}{ L}} \left[ \i \sigma_0 {\color{black}L} + \frac{1}{3}\left(\frac{\mathfrak{m}_1 +\mathfrak{m}_2}{n_1+ n_2} - \frac{2\mathfrak{m}_1}{n_1}- \frac{2 \mathfrak{m}_2}{n_2}\right)\right]\,.
\end{equation}} 
\begin{align}
\begin{split} \label{eq: Lambda_G_BPS}
     \Lambda^{\mathcal{G}}  
   &= {\color{black} \frac{L_0}{ L}} \left[ \i \sigma_0 {\color{black} L}- \frac{\mathfrak{m}}{3(n_1 +n_2)}\left(\frac{1}{n_1} +\frac{2}{n_2}\right) -  \frac{\mathfrak{m}_1}{n_1}  \right]
    \\
     &={\color{black} \frac{L_0}{ L}} \left[ \i \sigma_0 {\color{black} L} + \frac{\mathfrak{m}}{3(n_1 +n_2)}\left(\frac{2}{n_1} +\frac{1}{n_2}\right) - \frac{\mathfrak{m}_2}{n_2} \right] \equiv \Lambda^{\mathcal{G}}_0\, ,
    \end{split}
 \end{align}
where the second equality uses the relation $\mathfrak{m}=\mathfrak{m}_2 n_1 - \mathfrak{m}_1 n_2$ and  the final equality arises from the fact that the non-constant part of $\Lambda^{\mathcal{G}}$ vanishes on the BPS locus. We also note that for the regular gauges we have 
\begin{align} \label{eq:lambdaG}
\begin{split}
    \Lambda^\mathcal{G} \big|_{\mathcal{U}_{1}} 
   &={\color{black} \frac{L_0}{ L}} \left[ \i \sigma_0{\color{black}L} - \frac{\mathfrak{m}}{3(n_1 +n_2)}\left(\frac{1}{n_1} +\frac{2}{n_2}\right)\right]\,,
   \end{split}
\end{align}
and
\begin{align}
\begin{split}
   \Lambda^\mathcal{G} \big|_{\mathcal{U}_2} 
     &={\color{black} \frac{L_0}{ L}} \left[ \i \sigma_0{\color{black}L} + \frac{\mathfrak{m}}{3(n_1 +n_2)}\left(\frac{2}{n_1} +\frac{1}{n_2}\right)\right]\,,
     \end{split}
\end{align}
which will be useful when computing the one-loop determinant via the orbifold index theorem. 

 Finally, we note that the bosonic part of the chiral multiplet deformation is given by
\begin{align}
\begin{split} \label{eq: CM_deformation_action}
  Q_{eq} \mathcal{V}^{\text{c.m.}} \bigg|_{\text{bos.}} = \int_{\mathbb{\Sigma}} & \textrm{d}^2 x \, \sqrt{g_{\mathbb{\Sigma}}} \, \bigg\{ \bigg| \left(\i \gamma^{\mu} D_{\mu} \phi - \i \left( \sigma + \frac{r}{2} H \right) \phi + \gamma_3 \left( \rho + \frac{r}{2} G \right) \phi\right) \epsilon \bigg|^2 \\
    & \quad + \bigg| \left( \i \gamma^{\mu} D_{\mu} \widetilde{\phi} - \i \left( \sigma + \frac{r}{2} H \right) \widetilde{\phi} - \gamma_3 \left( \rho + \frac{r}{2} G \right) \phi \right) \widetilde{\epsilon}  \bigg|^2 + \widetilde{\mathfrak{F}} \mathfrak{F}  \bigg\}\,, 
     \end{split}
\end{align}
which upon using the reality conditions for the chiral fields \eqref{eq: chiral_bosonic_reality} is again a sum of squares so each individual term must vanish. We solve these equations in Appendix~\ref{sec: CM_BPS}, finding that for generic vector multiplet on the BPS locus \eqref{eq: BPS_locus_real}, the only regular solutions are 
\begin{equation} \label{eq: Chiral_BPS_locus}
    \phi = \widetilde{\phi} = \mathfrak{F} = \widetilde{\mathfrak{F}}=  0\,,
\end{equation}
and thus the chiral multiplet BPS locus is just a single point in the space of field configurations.
In passing, we note that the chiral multiplet locus is the same as the cases of $S^2$ and $AdS_2$ \cite{GonzalezLezcano:2023cuh}, and the BPS locus evaluated above dictates that we are localising on the Coulomb branch.

\subsection{Action on the BPS locus}
In order to compute the full partition function, we need to evaluate the action \eqref{eq: susy_action} on the BPS locus. The first term in \eqref{eq: susy_action} is the vector multiplet action. Evaluation of the vector multiplet Lagrangian \eqref{eq: vm_lagrangian} on the BPS locus using \eqref{eq: BPS_2} and  \eqref{eq: BPS_locus_real} yields
\begin{equation} \label{eq: vm_bps=0}
  \left. \mathcal{L}^{\text{bulk}}_{\text{v.m.}}  \right|_{\text{BPS}} = 0\,, 
\end{equation}
and hence the vector multiplet action vanishes on the BPS locus, a result which could also be observed from the $Q_{eq}$-exactness of the vector multiplet action \cite{GonzalezLezcano:2023cuh}. 

The next (and most straightforward) term to evaluate on the BPS locus is the chiral multiplet action \eqref{eq: cm_action}. Evaluation of the chiral multiplet Lagrangian \eqref{eq: bulk_chiral_lagrangian} on the BPS locus using \eqref{eq: BPS_2} and \eqref{eq: Chiral_BPS_locus} yields
\begin{equation} \label{eq: cm_BPS=0}
    \left. \mathcal{L}^{\text{bulk}}_{\text{c.m.}} \right|_{\text{BPS}} = 0\,, 
\end{equation}
and hence the chiral multiplet action also vanishes on the BPS locus. As in the vector multiplet case, such a result can also be obtained from the $Q_{eq}$-exactness of the chiral multiplet action \cite{GonzalezLezcano:2023cuh}. 

These results demonstrate that the only classical terms which contribute to the exact partition function are those generated by the FI and topological terms. We recall that these terms take the form  \eqref{eq: S_FI+top} which we want to evaluate on the BPS locus \eqref{eq: BPS_locus_real}. We first note 
that on the BPS locus the auxillary field $\widehat{D}$ takes the form 
\begin{equation}
    \left. \widehat{D} \right|_{\text{BPS}} = \left. D + \i (H \sigma + G \rho) \right|_{\text{BPS}} = \frac{ \sigma_0}{\color{black}L} \frac{9(y-a)}{2y^2} \frac{2\pi}{\Delta z} -  {\color{black}\frac{ 1}{L^2}} \frac{ 3 \i \mathfrak{m}}{n_1 -n_2}\,,
\end{equation}
and thus
\begin{equation}
    \left. S_{\text{FI}}  \right|_{\text{BPS}}= -\i \xi \int_{\mathbb{\Sigma}} \textrm{d}^2 x\,\sqrt{g_{\mathbb{\Sigma}}}  \widehat{D} = 4 \pi \xi \gamma_{\mathcal{G}}\,, 
\end{equation}
where we define
\begin{align}
\begin{split} \label{eq: gamma_g}
    \gamma_{\mathcal{G}} \equiv 
       \i \sigma_0 {\color{black}L} +\frac{\mathfrak{m}(n_2 -n_1)}{6 n_1 n_2 (n_1 + n_2)}
       =
{\color{black}\frac{L}{L_0}}\Lambda^{\mathcal{G}} + \frac{1}{2} \left( \frac{\mathfrak{m}_1}{n_1} + \frac{\mathfrak{m}_2}{n_2} \right)  \,.
    \end{split}
\end{align}
that is related to the BPS value of the gauge parameter $\Lambda^{\cal G}$ given in \eqref{eq: Lambda_G_BPS}.  
The topological ``theta'' term can be straightforwardly analysed using \eqref{eq: Gauge_flux_quantisation_spindle}, 
\begin{equation} \label{eq: S_top_BPS}
   \left. S_{\text{top}} \right|_{\text{BPS}} = \i \theta \frac{\mathfrak{m}}{n_1 n_2}\equiv \i \theta \mathfrak{f}_{\mathcal{G}}\,,
\end{equation}
and thus we see that the combination of FI and topological terms yields 
\begin{equation}
   \left. Z_{\text{FI} + \text{top}} \right|_{\text{BPS}} = e^{\left. -S_{\text{FI}} - S_{\text{top}}\right|_{\text{BPS}}} = e^{-4 \pi \xi \gamma_{\mathcal{G}} - \i \theta \mathfrak{f}_{\mathcal{G}}}\,,
\end{equation}
which will go into the classical part of the partition function \eqref{eq: Z_t_infty}.
Recalling that our theory consists of a single chiral multiplet and a single abelian vector multiplet we have  
\begin{equation} \label{eq: partition_function_general}
    Z_{\mathbb{\Sigma}} = \sum_{\mathfrak{m} \in \mathbb{Z}} \int_{-\infty}^{\infty} \frac{\textrm{d}(\sigma_0 {\color{black}L})}{2\pi} \, Z_{\text{class}} (\sigma_0{\color{black}L},\mathfrak{m}) Z^{Q_{eq} \mathcal{V}_{\text{v.m.}}}_{\text{1-loop}}(\sigma_0{\color{black}L},\mathfrak{m}) Z^{Q_{eq} \mathcal{V}_{\text{c.m.}}}_{\text{1-loop}}(\sigma_0{\color{black}L},\mathfrak{m}) \,,
\end{equation}
where in this section we have computed the classical contribution 
\begin{equation}
    Z_{\text{class}} (\sigma_0{\color{black}L},\mathfrak{m}) =\left. Z_{\text{FI} + \text{top}} \right|_{\text{BPS}} = e^{\left. -S_{\text{FI}} - S_{\text{top}} \right|_{\text{BPS
}}} = e^{-4 \pi \xi \gamma_{\mathcal{G}} - \i \theta \mathfrak{f}_{\mathcal{G}}}\,.
\end{equation}
 Here, the scale factor in the measure $\textrm{d}(\sigma_0{\color{black}L})$ is determined to by the requirement to make the measure dimensionless. It could also be determined using Fujikawa's ultra local argument \cite{Fujikawa:1984qk} as discussed in~\cite{GonzalezLezcano:2023cuh}.
It remains to compute the chiral multiplet one-loop determinant $Z^{Q_{eq} \mathcal{V}_{\text{c.m.}}}_{\text{1-loop}}$ and vector multiplet one loop-determinant $Z^{Q_{eq} \mathcal{V}_{\text{v.m.}}}_{\text{1-loop}}$, which we will do in the next section. 

\section{One-loop determinants} \label{sec: one-loop}
\subsection{One-loop determinant from equivariant index}
We begin by reviewing how the computation of the index relates to the one-loop determinant. In the localisation formulation, the exact one-loop determinant is obtained by computing one-loop for the quadratic action of a choice of $\qeq$-exact action. We first focus on the chiral multiplet and recall the canonical choice for the $\qeq$-exact action given by 
\begin{eqnarray}\label{QVaction}
    \qeq \cV^{\text{c.m.}} &=& \int_{\mathbb{\Sigma}} \textrm{d}^2 x \, \sqrt{g_{\mathbb{\Sigma}}} \, \qeq \left( (\qeq\psi)^\dagger \psi + (\qeq \widetilde{\psi})^\dagger \widetilde{\psi} \right) \,.
\end{eqnarray}
Let us first illustrate how the one-loop determinant is related to a choice of a $\qeq \cV^{\text{c.m.}}$ action. We formally rewrite the functional $\cV^{\text{c.m.}}$ in terms of the cohomological variables given in \eqref{eq:cohomologicalvariables} as
\beqa 
 \label{Vcoho}
 \cV^{\text{c.m.}} & = & \int  \textrm{d}^2 x \, \sqrt{g_{\mathbb{\Sigma}}} \, \Biggl[\bigl(\ba{cc} \qeq \Phi & \quad  \Psi  \ea  \bigr)  \Biggl( \ba{cc} D_{00} & D_{01}  \\ D_{10} & D_{11} \ea  \Biggr) \Biggl( \ba{cc}  \Phi \\ \qeq \Psi \ea    \Biggr) \Biggr]\,,
 \eeqa
 for matrix valued differential operators $D_{ij}$. This gives the following
  \be
 \qeq \cV^{\text{c.m.}}  = \int  \textrm{d}^2 x \, \sqrt{g_{\mathbb{\Sigma}}} \,\Biggl[ \bigl(  \Phi \quad \qeq \Psi  \bigr) \mathcal{K}_b \Biggl( \ba{cc}  \Phi \\ \qeq \Psi \ea \Biggr)  +\bigl(   \qeq \Phi \quad \Psi  \bigr)\mathcal{K}_f \Biggl( \ba{cc} \qeq  \Phi \\  \Psi \ea \Biggr) \Biggr]\,, \label{eq:QVcohom}
 \ee
 where the quadratic kinetic operators are 
 \beqa \label{eq:Kb}
 \mathcal{K}_b &  = & \frac{1}{2}\Biggl( \ba{cc} - \qeq^2 & 0 \\
0 & 1 \ea \Biggr) \Biggl( \ba{cc} D_{00} & D_{01} \\
D_{10} & D_{11} \ea \Biggr) +\frac{1}{2} \Biggl( \ba{cc} D^{\text{T}}_{00} & D^{\text{T}}_{10} \\
D^{\text{T}}_{01} & D^{\text{T}}_{11} \ea \Biggr) \Biggl( \ba{cc}  \qeq^2 & 0 \\
0 & 1 \ea \Biggr)\, ,  \\ \nn
\mathcal{K}_f & = & \half \Biggl( \ba{cc} 1& 0 \\
0 & - \qeq^2 \ea \Biggr) \Biggl( \ba{cc}  D^{\text{T}}_{00} & D^{\text{T}}_{10} \\
D^{\text{T}}_{01} & D^{\text{T}}_{11} \ea \Biggr) -\half \Biggl( \ba{cc} D_{00} & D_{01} \\
D_{10} & D_{11} \ea \Biggr) \Biggl( \ba{cc} 1& 0 \\
0 &  \qeq^2 \ea \Biggr)\,.
\eeqa
The above derivation rests on the fact that $\qeq$ is a Grassmann odd operator and that the bosonic and fermionic kinetic operators are naturally symmetric and anti-symmetric respectively.  From the expression in~\eqref{eq:Kb}, we can see that the kinetic operators satisfy
\be
\biggl( \ba{cc} 1 & ~0 \\
0 & - \qeq^2 \ea \biggr) \mathcal{K}_b = \mathcal{K}_f \biggl(\ba{cc} \qeq^2 & 0\\ 
0 & ~1  \ea \biggr)  \label{eq:KbKf}\,.
\ee
Therefore, the one-loop determinant is reduced to the determinant of $\qeq^2$ over the elementary bosons $\Phi$ and fermions $\Psi$ as  
\beqa
 \label{eq:prelogNozm}
Z^{ \qeq \cV^{\text{c.m.}}}_{1\text{-loop}} &=& \sqrt{\frac{\det_{\qeq \Phi\,, \Psi} \mathcal{K}_f}{\det_{\Phi , \qeq \Psi} \mathcal{K}_b}}  =\sqrt{\frac{\text{det}_{\qeq \Psi}\qeq^2}{\text{det}_{\qeq \Phi}\qeq^2}} \= \sqrt{\frac{\text{det}_{\Psi} \qeq^2}{\text{det}_{\Phi }\qeq^2}} \,.
\eeqa
Here, we note that the zero mode sectors, $\qeq^2 =0$, do not appear in the above result
. This is because the determinant of the kinetic operators \eqref{eq:Kb} on those sectors give $(\det D_{10})^2$ both for bosons and fermions, and they are exactly cancelled by each other as the $D_{10}$ is non-degenerate \cite{GonzalezLezcano:2023cuh,Gupta:2015gga}. For the last equality of~\eqref{eq:prelogNozm} we used the fact that $\qeq^2$ and $\qeq$ commute. 
The above result can be further reduced by noting from \eqref{Vcoho} that $D_{10}$ is the operator which maps the elementary bosonic to elementary fermionic variable
\begin{eqnarray} \label{eq: D_10_map}
     D_{10} :\Phi \rightarrow \Psi\,.
 \end{eqnarray} 
Since $[D_{10}\,, \qeq^2] =0$, there exists a simultaneous eigenbasis for the operators and the eigenvalues of $\qeq^2$ in $\Phi$ and $\Psi$ are paired except the kernel and cokernel of the $D_{10}$ operator.\footnote{Note that, $D_{10}$ is then not necessarily a bijection. In cases where it is, the one-loop is trivial because of boson-fermion pairing in supersymmetry.} Therefore, after cancelling the contribution of the pairing the non-trivial contribution remains as
  \begin{eqnarray}\label{reducedone-loop}
    Z^{ \qeq \cV^{\text{c.m.}}}_{1\text{-loop}} =  \sqrt{\frac{ \textrm{det}_{\textrm{Coker($D_{10}$)}} ~ \qeq^2 }{\textrm{det}_{\textrm{Ker($D_{10}$)}} ~ \qeq^2}}\,. 
 \end{eqnarray}

 This one-loop can be computed by evaluating the equivariant index of $D_{10}$ operator with respect to the $U(1)$ rotation generated by $\qeq^2$. The index is defined as
 \begin{eqnarray}
 \label{eq:TrNozm}
  \text{ind}(D_{10}) (t)    &=&  \text{Tr}_{\text{Ker}(D_{10})}{\rm e}^{t \qeq^2} - \text{Tr}_{\text{Coker}(D_{10})}{\rm e}^{t \qeq^2}\,,    
 \end{eqnarray} 
 where $t$ is the \textit{equivariant parameter}.\footnote{This $t$ should not be confused with the deformation parameter in equation \eqref{eq: partition_function_deformation}.} 
Since this index admits a Fourier series expansion in terms of eigenvalues $\lambda_n$ of $\qeq^2$ and their degeneracy $a(n)$ in the following way
\be 
{\rm ind}( D_{10})(t) = \sum_n a(n){\rm e}^{ \lambda_n t}\,, \label{eq:indDeg}
\ee
once it is computed, we can read off the eigenvalues 
 of $\qeq^2$ with their respective degeneracies, i.e., $\{\l_n, a_n\}$ respectively. The one-loop result is then given by\footnote{Note that the one-loop determinant is independent of the equivariant parameter $t$.}
\be \label{eq:degeneracies}
 Z^{ \qeq \cV^{\text{c.m.}}}_{1\text{-loop}}   \=\prod_n \lambda_n^{-\frac{1}{2}a(n)}\,.
\ee 

The analysis above was all for the chiral multiplet one-loop determinant. In principle one should perform a similar analysis for the vector multiplet in order to compute the analagous vector multiplet one-loop determinant. However, one can use the shortcut of obtaining the vector multiplet one-loop determinant directly from the chiral multiplet result by setting $r=2$ and choosing the gauge group representation to be the adjoint. Using this knowledge we will only compute the chiral multiplet one-loop result and drop the ``c.m.'' superscript from here on.   


\subsection{Method of unpaired eigenvalues} \label{sec: Unpaired_evals}
In this subsection, we directly compute the determinant~\eqref{reducedone-loop}  or (equivalently) the index~\eqref{eq:TrNozm}   by identifying the spectrum of `unpaired eigenmodes'. The central computation will involve obtaining the kernel and cokernel of the operator $D_{10}$ and their spectrum with respect to the $\qeq^2$ on the $\Phi$ and $\Psi$ space. 

For Fredholm operators on compact spaces, the $\text{Coker}(D_{10}) =   \text{Ker}(D_{10}^\dagger)$ and therefore, we need to identify $D_{10}$ as well as its dual operator $D_{10}^\dagger$. 
To facilitate the computation, we re-express the  $\qeq$-exact action \eqref{QVaction} in terms of the cohomological variables in \eqref{eq:cohomologicalvariables} as 
\begin{align}
\begin{split}
\qeq \cV & = \frac{1}{2 } \int_{\mathbb{\Sigma}} \textrm{d}^2 x\, \sqrt{g_{\mathbb{\Sigma}}} \qeq \left[\left(\qeq (\epsilon \psi)\right)^{\dagger} \epsilon \psi + (\qeq \widetilde{\epsilon} \widetilde{\psi})^{\dagger} \widetilde{\epsilon} \widetilde{\psi}+ (\qeq (\epsilon \widetilde{\psi}))^{\dagger} \epsilon\widetilde{\psi} 
\right.    
 \\ 
&  \quad \quad \quad\quad\quad\quad\quad\quad\quad \quad
\left. 
+(\qeq \widetilde{\epsilon} \psi)^{\dagger} \widetilde{\epsilon} \psi \right]\,,\label{eq:QVcmTW}
\end{split}
\end{align}
and read off the $D_{10}$ operator from the first two terms.
Using the reality property of the bilinears as stated in ~\eqref{RealityofQPsi} and comparing with the formal expression of \eqref{Vcoho}, we find the following differential operator representation of the operator~$D_{10}$ 
\begin{subequations}
\begin{eqnarray}
      D_{10} \cdot {\phi} &=& -2 \,
      \left[  P_- 
      + \left(\rho + \frac{r}{2}G\right){\epsilon}\gamma_3 \epsilon   \right]  {\phi} \,, \label{Kerphi}
   \\
    D_{10} \cdot \widetilde{\phi} &=& - 2\, 
    \left[P_+ 
    - \left(\rho + \frac{r}{2}G\right)\widetilde{\epsilon}\gamma_3\widetilde\epsilon   \right]  \widetilde{\phi} \,.
\end{eqnarray}
\end{subequations}
By integrating by parts, we find the dual operator $D_{10}^\dagger$ as
\begin{eqnarray}
    D_{10}^\dagger (\epsilon \psi)&=& 2 \left[ P_+ 
    + \Bigl(\rho+\left(\frac{r}{2}-2\right)G \Bigr)\widetilde{\epsilon}\gamma_3
    \widetilde{\epsilon} \right](\epsilon \psi)\,,
    \\
    D_{10}^\dagger (\widetilde\epsilon \wt\psi)&=&  2 \left[ P_- 
    - \Bigl(\rho+ \left(\frac{r}{2}-2\right)G\Bigr){\epsilon}\gamma_3{\epsilon} \right](\widetilde\epsilon \wt\psi)\,.
\end{eqnarray}
 where we have defined the covariant differential operators
\begin{eqnarray}
     P_+ \wt{\phi} &=&  \i \widetilde\epsilon \gamma^\mu \widetilde\epsilon D_\mu \wt\phi \,,\qquad
    P_- \phi = \i \epsilon \gamma^\mu \epsilon D_\mu \phi \,,
\end{eqnarray}
and for the dual operator $D_{10}^\dagger$ we have used the dual relations
\begin{eqnarray}
 P_+^\dagger ({\epsilon}{\psi}) &=& - (P_+ - 2 G \widetilde\epsilon \gamma_3 \widetilde\epsilon) ({\epsilon}{\psi}) \,, \quad
     P_-^\dagger (\widetilde{\epsilon}\wt{\psi}) = - (P_- + 2 G \epsilon \gamma_3 \epsilon) (\widetilde{\epsilon}\wt{\psi})\,.
\end{eqnarray}
 Here by the subscript in $P_\pm$ we indicate that their action raises and lowers the $R$-charge by $1$ respectively. We  note that 
\begin{equation}
    \left[Q^2_{eq}, P_{\pm}\right] = 0\,.
\end{equation}

The kernel and cokernel of $D_{10}$ are obtained by solving the equations $D_{10}\Phi=0$ and $D_{10}^\dagger \Psi =0$. 
However, we will shortly see from \eqref{D10P-} that the second terms  proportional to the ${\epsilon}\gamma_3{\epsilon}$ and $\widetilde{\epsilon}\gamma_3
    \widetilde{\epsilon}$ in the operators do not affect the spectrum of the kernel with respect to  $\qeq^2$ as those terms are regular at the poles $y=y_1$ and $y=y_2$. Said in other words, they are constant shifts which keep the dimension of the kernel and cokernel the same. Thus, the relevant part of the equations will respectively be
\begin{eqnarray} \label{KerP+P-}
    \Biggl( \ba{cc} 0 & P_+  \\ P_- & 0 \ea  \Biggr) \Biggl( \ba{cc}  \phi \\ \wt{\phi} \ea    \Biggr)=0 \,,\qquad  
    \Biggl( \ba{cc} 0 & P_-  \\ P_+ & 0 \ea  \Biggr) \Biggl( \ba{cc}  \epsilon \psi \\ \widetilde{\epsilon}\wt{\psi} \ea    \Biggr)=0 \,,
\end{eqnarray}
and the computation of the index~\eqref{eq:TrNozm} is equivalent to computing the following index, 
\begin{eqnarray} \label{indexPpm}
\text{ind} &=& \Tr_{\ker(P_+)}e^{t\qeq^2}\bigr|_{\wt{\phi}}+ \Tr_{\ker(P_-)}e^{t\qeq^2}\bigr|_{{\phi}} - \Tr_{\ker(P_+)}e^{t\qeq^2}\bigr|_{{\epsilon\psi}} -\Tr_{ \ker(P_-)}e^{t\qeq^2}\bigr|_{{\widetilde\epsilon \wt\psi}}\, \nonumber \\
&=& \Tr_{\ker(P^{(\widetilde \phi)}_+)}e^{t\qeq^2}+ \Tr_{\ker(P_-^{(\phi)})}e^{t\qeq^2} - \Tr_{\ker(P^{(\widetilde \eps \widetilde \psi )}_+)}e^{t\qeq^2}-\Tr_{ \ker(P^{(\epsilon \psi)}_-)}e^{t\qeq^2}\,. \qquad
\end{eqnarray}
This is the index that was computed for the one-loop instead of \eqref{eq:TrNozm} in the literature \cite{Pittelli:2024ugf}.

\subsubsection{Spectrum of the unpaired eigenmodes}
Let us first find the kernel of $P^{(\phi)}_{-} $ in the scalar field $\phi$ by solving the first order differential equation
\be\label{KerP-phi}
P^{(\phi)}_- \phi =0\,.
\ee 
For ease of the calculation, we take the singular gauge where the $U(1)_R$ gauge field $A $ and the $U(1)_{\mathcal{G}}$ gauge field $\cal A$ are globally given as \eqref{eq: A_2d} and \eqref{eq: gauge_field_explicit_real}, respectively. These gauge choices cover the entire spindle geometry but are singular at the two poles, having non-zero values at $y=y_{1,2}$. 

In this gauge, 
the scalar field $\phi$ takes the separable form 
\be\label{ansatzphi}
\phi (y,z) = \exp\left({-2\pi \i n \frac{z}{\Delta z}}\right) \phi_n(y)\,,\qquad n\in \mathbb{Z}\,,
\ee
where we note that the separable form and exponential $z$-dependence is forced as we want $\phi$ to be an eigenfunction of $Q_{eq}^2$ \eqref{eq: Q^2}. It satisfies the periodic boundary condition
\be\label{periodicbc}
\phi (z +\Delta z, y ) = \phi (z, y)\,.
\ee 
Note that near an orbifold singularity a scalar field in general can have a twisted sector, and thus the scalar could have non-trivial monodromy $\phi(z +\Delta z , y)= \exp \left(2\pi \i \frac{k_{1,2}}{n_{1,2}}  \right)\phi(z, y)\, $ with $ k_{1,2} = 0,1,\, \ldots \,, n_{1,2}-1$ near two orbifold singular points $y_1$ and $y_2$ respectively \cite{Witten:1985xc,Hall:2001tn}. However, given that we are choosing in the singular gauge where the single chart covers the entire spindle geometry except the $y_1$ and $y_2$ points, $k_1=k_2=0$ is the only consistent choice as $n_1$ and $n_2$ are co-prime integers. 

Then the kernel equation \eqref{KerP-phi} becomes 
\begin{align}
\begin{split}
    \label{eqKer-Psi}
\partial_y \phi(y,z)  = \left[ \frac{3\i(a-3y)}{q}\left(\partial_z - \i r A_z - \i q_{\mathcal{G}} \mathcal{A}_z \right) \right]\phi(y,z)\,, \\ 
\end{split}
\end{align}
where we note that we have reinstated a generic gauge charge $q_{\mathcal{G}}$ for the chiral multiplet (in Sections \ref{sec: D=2N=2,2_on_spindle} and \ref{sec: localisation} we used $q_{\mathcal{G}}=1$). Upon use of \eqref{ansatzphi} as well as explicit values of the gauge fields via \eqref{eq: A_2d} and \eqref{eq: gauge_field_explicit_real} we obtain
\begin{align}
\frac{\phi_n'(y)}{\phi_n(y)} = 3\frac{a-3y}{q} \left( \frac{2\pi n}{\Delta z} + \frac{r}{4}\left(1- \frac{a}{y}\right) + {q_{\mathcal{G}} \left(\frac{\mathfrak{m}}{n_1-n_2} \frac{a }{2} \left( \frac{1}{y} - \frac{1}{y_1} \right) + \frac{2\pi}{\Delta z} \frac{\mathfrak{m}_1}{n_1}\right) }\right)\,,
\end{align}
which we find to be exactly soluble, with solution given by
\beqa\label{SolKerP-phi}
\phi_n(y) &=& c (y -y_1)^{\frac{3}{4} \frac{a-3y_1}{(y_2 -y_1)(y_3-y_1)}\left(\frac{2\pi n}{\Delta z}  + rA_z(y_1)+ 
{q_{\mathcal{G}}} \mathcal{A}_z(y_1) \right) }\nn
\\
&&\times (y_2 -y)^{\frac{3}{4} \frac{3y_2-a}{(y_2 -y_1)(y_3-y_2)}\left(\frac{2\pi n}{\Delta z}  +rA_z(y_2)+ q_{\mathcal{G}} \mathcal{A}_z(y_2)\right) } 
\\
&&\times (y_3 -y)^{-\frac{3}{4} \frac{3y_3-a}{(y_3 -y_1)(y_3-y_2)}\left(\frac{2\pi n}{\Delta z}  + r A_z(y_3)+{q_{\mathcal{G}} \mathcal{A}_z(y_3)} \right)} y^{\frac{1}{4} (3 r-2 \rho_0 q_{\mathcal{G}})} \,,
\nn
\eeqa
 where $c$ is a constant of integration. 
 As $y_1  \leq y \leq y_2 <y_3$, the third line is regular, but from the first and second line we have to restrict the range of the quantum number $n$ by imposing regularity.
 We note that near the north pole $y=y_1$, the solution to this equation takes the form 
\begin{equation} \label{eq: phi_N}
    \phi_n(y) \sim (y -y_1)^{\frac{3}{4}  \frac{2\pi}{\Delta z}  \frac{a-3y_1}{(y_2 -y_1)(y_3-y_1)}\left(n  -\frac{r}{2n_1}+ 
{ \frac{\mathfrak{m}_1 q_{\mathcal{G}}}{n_1}} \right)}\,,
\end{equation}
and near the south pole $y=y_2$, 
\begin{equation} \label{eq: phi_S}
     \phi_n(y) \sim (y -y_2)^{\frac{3}{4} \frac{2\pi}{\Delta z} \frac{3y_2-a}{(y_2 -y_1)(y_3-y_2)}\left( n  -\frac{r}{2n_2} +{ \frac{\mathfrak{m}_2 q_{\mathcal{G}}}{n_2}}\right) }\,.
\end{equation}In order to avoid the singular configurations near $y=y_1$ and $y=y_2$, the exponents in both~\eqref{eq: phi_N} and \eqref{eq: phi_S} must be greater than or equal to zero. We first introduce the notation
\begin{equation} \label{eq: p12}
    \mathfrak{p}_{1,2} = \frac{r}{2} - q_{\mathcal{G}} \mathfrak{m}_{1,2}\, ,
\end{equation}By noting that $3y_2 > a > 3y_1$, we obtain the following condition for the quantum number 
\be\label{spectrumphi}
{n \geq \max \left( \frac{\mathfrak{p}_1}{n_1}\,, \frac{\mathfrak{p}_2}{n_2} \right)}\,.
\ee

Before solving the spectrum of the kernel for the other fields, let us comment about the equivalence between the ind$(D_{10})$ in \eqref{eq:TrNozm} and the ind in~\eqref{indexPpm}. For this, we demonstrate that the regularity conditions imposed by the equation $D_{10}\phi =0$ are the same as those of  $P_- \phi =0$. All of the kernels for $\wt{\phi}\,, \epsilon \psi\,,\widetilde{\epsilon}\wt{\psi}$ follow via identical arguments. From the relation of $D_{10}$ and $P_-$ as in \eqref{Kerphi},  we start by writing 
\begin{equation}\label{D10P-}
    D_{10} \phi = \left( P_- + \frac{r}{2} G \epsilon \gamma_3 \epsilon \right) \phi = \left( P_- - \frac{r}{12} \frac{k_0}{L} \frac{\sqrt{q}}{y}  \right) \phi = 0\,,  
\end{equation}
where we have set $G= -{\color{black}L^{-1}}$ and neglect the coupling with $\rho$.
Denoting $\phi^{(D_{10})} \in \text{Ker}(D_{10})$ and recalling that the $P_- \phi=0 $ is solved by $\phi$, we immediately reach the result  
\begin{equation}
\phi^{(D_{10})} =  y^{-\frac{r}{4}} \phi\,,
\end{equation}
where the power of $y$ is regular at both $y_{1,2}$ and thus the analysis of regularity at the poles of $\mathbb{\Sigma}$ is unchanged from the $P_-$ case. As the regularity is the only possible reason for discrepancy between $\text{ind} D_{10}(t)$ and ind, we conclude that they are equivalent. In fact such a result is expected on general grounds as $[D_{10}\,,\qeq^2]= [P_\pm\,,\qeq^2]=0$ and both $D_{10}$ and $P_\pm$ map between elementary boson and fermion (similarly $D_{10}^\dagger$ and $P_{\pm}^\dagger$ map elementary fermion to boson).

Now we turn to find the kernel of $\wt{\phi}$ by solving
\be \label{eq: kernel_phi_tilde}
 P_+ \wt{\phi} = 0\,,
\ee
among ~\eqref{KerP+P-}. Again, the eigenfunction requirement together with the periodicity condition for singular gauge fields \eqref{periodicbc} implies
\be
\wt\phi (y,z) = \exp \left(  2\pi i m \frac{z}{\Delta z}\right)\wt{\phi}_m(y)\,,
\ee 
and one finds that the equation \eqref{eq: kernel_phi_tilde} is the same equation as the $P_- \phi =0$ due to the fact that $\wt{\phi}$ has opposite $R$-charge $-r$ {and opposite gauge charge $-q_{\mathcal{G}}$}. Therefore, we obtain the same conclusion as \eqref{spectrumphi}, namely 
\be
{m \geq \max \left( \frac{\mathfrak{p}_1}{n_1}\,, \frac{\mathfrak{p}_2}{n_2} \right)}\,.
\ee

We also need to perform the same analysis for $\Psi = (\epsilon\psi\,, \widetilde{\epsilon}\wt{\psi})$. From \eqref{KerP+P-}, we solve the following kernel equations 
\be\label{KerP+P-Psi}
P_- (\widetilde{\epsilon}\wt{\psi})=0\,,\qquad P_+ (\epsilon\psi)=0\,.
\ee
With
\be
\widetilde{\epsilon}\wt{\psi}(y,z) = \exp \left({-} 2\pi \i l \frac{z}{\Delta z} \right)\widetilde{\epsilon}\wt{\psi}_l(y)\,,\quad {\epsilon}{\psi}(y,z) = \exp \left( 2\pi \i p \frac{z}{\Delta z} \right){\epsilon}{\psi}_p(y)\,,\quad l,p \in \mathbb{Z}\,,
\ee
one can find that both equations in \eqref{KerP+P-Psi} are the same as the $P_-\phi =0$ that is given in~\eqref{eqKer-Psi}, replacing the $R$-charge parameter $r$ by $-r+2$ {and the gauge charge parameter $q_{\mathcal{G}}$ with $-q_{\mathcal{G}}$}. Therefore we obtain the condition for the quantum numbers as 
\be
{l,p \geq \max \left( \frac{1- \mathfrak{p}_1 }{n_1}\,, \frac{ 1-\mathfrak{p}_2}{n_2}\right)}\,.
\ee
In summary, we have for the kernels in the singular gauge, the following constraints: 
\begin{equation}  \label{eq: range_P-}
\begin{alignedat}{2}
    \text{Ker}(P_-) & : \phi \sim e^{ {-} \frac{2\pi \i n}{\Delta z} z} \phi_n (y)\,, && \quad {n \geq \max \left( \frac{\mathfrak{p}_1}{n_1}\,, \frac{\mathfrak{p}_2}{n_2} \right)}\, , \\
    \text{Ker}(P_+) & : \widetilde{\phi} \sim e^{\frac{2\pi \i m}{\Delta z} z} \widetilde{\phi}_m (y)\,, && \quad {m \geq \max \left( \frac{\mathfrak{p}_1}{n_1}\,, \frac{\mathfrak{p}_2}{n_2} \right)} \,, \\
     \text{Ker}(P^\dagger_+) & : \epsilon \psi  \sim e^{\frac{2\pi \i p}{\Delta z} z} \epsilon \psi_p (y)\,,&&  \quad {p \geq \max \left( \frac{1- \mathfrak{p}_1 }{n_1}\,, \frac{ 1-\mathfrak{p}_2}{n_2}\right)}\, , \\
      \text{Ker}(P^\dagger_-) & : \widetilde{\epsilon} \widetilde{\psi}  \sim e^{ {-}\frac{2\pi \i l}{\Delta z} z} \widetilde{\epsilon} \widetilde{\psi}_l (y)\,, &&  \quad {l \geq \max \left( \frac{1- \mathfrak{p}_1 }{n_1}\,, \frac{ 1-\mathfrak{p}_2}{n_2}\right)}\, .
\end{alignedat}
\end{equation}
We also note the eigenvalues under the action of the $Q^2_{eq}$ operator: 
\begin{align} \label{eq: Q^2_evals}
\begin{split}
Q^2_{eq} \phi &  =-\i\frac{L_0}{L}\left( n - q_{\cal G} \frac{L}{L_0}\Lambda^{\cG}\right) \phi \,,  \\
     Q^2_{eq} \widetilde{\phi} & =\i \frac{L_0}{L}\left( m - q_{\cal G} \frac{L}{L_0}\Lambda^{\cG}\right) \widetilde{\phi}\,, \\
     Q^2_{eq} (\epsilon \psi) & = \i \frac{L_0}{L}\left( p + q_{\cal G} \frac{L}{L_0}\Lambda^{\cG}\right) (\epsilon \psi)\,, \\
     Q^2_{eq}(\widetilde{\epsilon} \widetilde{\psi}) & = -\i \frac{L_0}{L}\left( l + q_{\cal G} \frac{L}{L_0}\Lambda^{\cG}\right) (\widetilde{\epsilon} \widetilde{\psi})\,, 
     \end{split}
\end{align}
 which will be our main ingredients in computing the one-loop determinant. Before we do this, we comment on a quantisation condition which will greatly simplify this calculation.

\subsubsection{Charge quantisation condition}
 
Up to this point we have been working entirely generically with $r \in \mathbb{R}, \: q_{\mathcal{G}} \in \mathbb{R}$. However, \cite{Inglese:2023wky} proposed charge quantisation conditions given by\footnote{Note also that a weaker quantisation condition of $r(n_+ + n_-) \in 2\mathbb{Z}$ appeared in \cite{Pittelli:2024ugf} for the twist case only. } 
\begin{equation} \label{eq: charge_quantisation}
    r \in 2\mathbb{Z}\,, \qquad q_{\mathcal{G}} \in \mathbb{Z}\,,
\end{equation} 
which are required for the scalar $\phi$ to transform as a section of a well-defined $U(1)_R \times U(1)_{\mathcal{G}}$ valued line orbibundle $L^{r,q_{\mathcal{G}}}$ on the spindle. We now provide two brief arguments for this charge quantisation condition. 

First, we note that $\phi$ is charged under $U(1)_R \times U(1)_{\mathcal{G}}$, and thus transforms as 
\begin{equation} \label{eq: phi_transformation}
     \phi \rightarrow e^{\i (r \alpha +{q_{\mathcal{G}} \beta}) z}  \phi \, ,
\end{equation}
under gauge transformations of the form $A \rightarrow A + \alpha \textrm{d}z$ and ${\mathcal{A} \rightarrow \mathcal{A}+ \beta \textrm{d}z}$ \cite{Inglese:2023tyc}. As discussed in \cite{Ferrero:2021etw}, one can obtain the order of the line orbibundle directly from the formulae for flux through the spindle. In particular, if the flux takes the form $\lambda/n_{1} n_{2}$, then the order of the orbibundle is $\mathcal{O}(-\lambda)$. 
Using the flux formulae \eqref{eq: R-flux} and \eqref{eq: Gauge_flux_quantisation_spindle} for the $R$ and gauge fluxes respectively, we compute 
\begin{equation}
    \lambda = \frac{1}{2} ( n_2 - n_1 ) +{\mathfrak{m}}\,,
\end{equation}
where we utilised the direct product structure of $U(1)_R \times U(1)_{\mathcal{G}}$ in adding the degrees of the respective orbibundles. Including the values for the charges, the scalar $\phi$ is thus a section of the orbibundle 
\begin{equation}
    \phi: L^{r, q_{\mathcal{G}}} = \mathcal{O} \left(-\frac{r}{2}(n_2 - n_1)- {q_{\mathcal{G}} \mathfrak{m}}\right)\,,
\end{equation}
where for well-definedness, the order must be integer
\begin{equation} \label{eq: quantisation_r_q}
    \frac{r}{2}(n_2 - n_1) + {q_{\mathcal{G}} \mathfrak{m}} \in \mathbb{Z}\,, 
\end{equation}
which if we demand this holds $\forall$ $n_{1,2} \in \mathbb{Z}$ s.t. $\text{gcd}(n_1,n_2)=1$ then we must have 
\begin{equation} \label{eq: r_quantisation}
    r \in 2 \mathbb{Z} \quad \textrm{and} \quad {q_{\mathcal{G}}} \in \mathbb{Z}\,, 
\end{equation}
 which reproduces \eqref{eq: charge_quantisation} precisely. 

Another, more geometrical, way to understand this quantisation condition is to look at the monodromy of the scalar field around the north and south poles of the spindle. In order to do this, we move into the regular gauges (for both $R$ and gauge fields) at the poles and have \begin{align}
    \left.A \right|_{\mathcal{U}_{1,2}} = \frac{a}{4}\left( \frac{1}{y_{1,2}} - \frac{1}{y}\right) \textrm{d}z\,, \qquad {\left. \mathcal{A} \right|_{\mathcal{U}_{1,2}}= \frac{\mathfrak{m}}{n_2-n_1} \frac{a }{2} \left( \frac{1}{y_{1,2}} - \frac{1}{y} \right) \textrm{d}z} \,,
\end{align}
now if we use \eqref{eq: phi_transformation} together with the values of $\alpha, \beta$ given in \eqref{eq: beta_1,2} and \eqref{eq: betas_regular} respectively  we have
\begin{equation}
    \left. \phi(y,z)\right|_{\mathcal{U}_{1,2}} = \exp \left({\frac{2\pi \i}{\Delta z} \left( \frac{\frac{r}{2} - q_{\mathcal{G}} \mathfrak{m}_{1,2}}{n_{1,2}} \right) z } \right) \phi(y,z) = \exp\left({\frac{2\pi\i}{\Delta z} \frac{\mathfrak{p}_{1,2}}{n_{1,2}}z}\right) \phi(y,z) \,, 
\end{equation}
for the values of the charged scalar in the regular gauge fields at each pole. We note that regular gauge fields for the $U(1)_R$ are related by 
\begin{equation}
 \left. A \right|_{\mathcal{U}_1} =  \left. A \right|_{\mathcal{U}_2} + \frac{a}{4} \left (\frac{1}{y_1} - \frac{1}{y_2} \right) \textrm{d}z = \left. A \right|_{\mathcal{U}_2} + \frac{2\pi}{\Delta z} \left(\frac{1}{2n_{1}} - \frac{1}{2n_{2}}\right) \textrm{d}z\,, 
 \end{equation}
 and for $U(1)_{\mathcal{G}}$ gauge fields
 \begin{equation}
 \left. \mathcal{A} \right|_{\mathcal{U}_1}  =  \left. \mathcal{A} \right|_{\mathcal{U}_2} + \frac{\mathfrak{m}}{n_1-n_2} \frac{a }{2} \left( \frac{1}{y_2} - \frac{1}{y_1} \right) \textrm{d}z = \left. \mathcal{A} \right|_{\mathcal{U}_2} + \frac{2\pi}{\Delta z} \frac{\mathfrak{m}}{n_1 n_2} \textrm{d}z\,, 
\end{equation}
thus 
\begin{equation}
     \left. \phi(y,z)\right|_{\mathcal{U}_{1}} = e^{\frac{2\pi\i}{\Delta z} \left(\frac{\mathfrak{p}_{1}}{n_{1}} - \frac{\mathfrak{p}_{2}}{n_{2}} \right)  z} \left. \phi(y,z)\right|_{\mathcal{U}_{2}}\,.
\end{equation}
Thus if we perform a gauge transformation from the regular gauge at the north pole of the spindle to the regular gauge at the south pole of the spindle, the scalar transforms as 
\begin{equation}
     \left. \phi(y,z)\right|_{\mathcal{U}_{1}} \rightarrow \left. \phi(y,z)\right|_{\mathcal{U}_{2}} = e^{-\frac{2\pi\i}{\Delta z} \left(\frac{\mathfrak{p}_{1}}{n_{1}} - \frac{\mathfrak{p}_{2}}{n_{2}} \right)  z} \left. \phi(y,z)\right|_{\mathcal{U}_{1}}\,,
\end{equation}
taking the value of this across one period of $z$ we find
\begin{equation}
   \left. \phi(z+\Delta z, y) \right|_{\mathcal{U}_{1}} \rightarrow e^{-2\pi\i\left(\frac{\mathfrak{p}_{1}}{n_{1}} - \frac{\mathfrak{p}_{2}}{n_{2}} \right) }  \left.\phi(z+\Delta z, y) \right|_{\mathcal{U}_{1}}\,,
\end{equation}
which will lead to a constraint on the value of the argument of the exponential in order to give suitable periodicity. Recalling that the spindle is an orbifold with conical singularities of deficit angle $2\pi(1-1/n_{1,2})$ and that we have performed a gauge transformation from one pole to the other, we deduce that the correct quantisation condition is 
\begin{equation}
  \frac{\mathfrak{p}_{1}}{n_{1}} - \frac{\mathfrak{p}_{2}}{n_{2}} \in \frac{\mathbb{Z}}{n_1 n_2} \iff  n_2 \mathfrak{p}_1 - n_1 \mathfrak{p}_2 \in \mathbb{Z}\,,
\end{equation}
rewriting this constraint using the definitions \eqref{eq: p12} leads to
    \begin{equation}
    \frac{r}{2}(n_2 - n_1) + {q_{\mathcal{G}} \mathfrak{m}} \in \mathbb{Z}\,, 
\end{equation}
which is equivalent to \eqref{eq: quantisation_r_q} and thus solved by the quantisation condition \eqref{eq: r_quantisation}. We note that as a consequence of the quantisation condition we have
\begin{equation}
    \mathfrak{p}_{1,2} \in \mathbb{Z} \, ,
\end{equation}
which will be important in simplifying our result for the one-loop determinant.

\subsubsection{One-loop determinant via unpaired eigenvalues}
We can now use the regularity conditions in order to establish the formulae for the one-loop determinant 
\begin{equation}
    Z^{ \qeq \cV}_{1\text{-loop}} =     \sqrt{\frac{ \textrm{det} ~ (\qeq^2)_{_{\Psi}} }{\textrm{det} ~ (\qeq^2)_{_{\Phi}}}} = \sqrt{\frac{\textrm{det}_{\text{Ker}(P^\dagger_+)} ~ \qeq^2 \cdot \textrm{det}_{\text{Ker}(P^\dagger_-)} ~ \qeq^2}{\textrm{det}_{\text{Ker}(P_+)} ~ \qeq^2 \cdot \textrm{det}_{\text{Ker}(P_-)} ~ \qeq^2}} \,,
\end{equation}
using the product ranges \eqref{eq: range_P-} and eigenvalues \eqref{eq: Q^2_evals}, we see that the one-loop determinant can be written as 
\begin{align} \label{eq: one-loop}
\begin{split}
    Z^{ \qeq \cV}_{1\text{-loop}} & = \sqrt{ \frac{\prod\limits_{p= 0 }^{\infty } \i \frac{L_0}{L}\left(p+p_{\min} +{q_{\mathcal{G}}\frac{L}{L_0} \Lambda^{\mathcal{G}}} \right) \prod\limits_{l= 0 }^{\infty }-\i\frac{L_0}{L} \left( l + l_{\min} +{q_{\mathcal{G}}\frac{L}{L_0} \Lambda^{\mathcal{G}}} \right)}{\prod\limits_{{n}=0}^{\infty} -\i\frac{L_0}{L} \left( n+n_{\min}  -{q_{\mathcal{G}}\frac{L}{L_0} \Lambda^{\mathcal{G}}} \right) \prod\limits_{{m}=0}^{\infty}\i\frac{L_0}{L}\left(m+m_{\min} - {q_{\mathcal{G}}\frac{L}{L_0} \Lambda^{\mathcal{G}}} \right) } } \, ,
    \end{split}
\end{align}
where we define
\begin{align}
    n_{\min}= m_{\min} = \max \left(\ceil*{\frac{\mathfrak{p}_1}{n_{1}}}\,,\ceil*{\frac{\mathfrak{p}_2}{n_{2}}} \right)\,,\:  p_{\min}= l_{\min} = \max \left(\ceil*{\frac{1-\mathfrak{p}_1}{n_{1}}}\,,\ceil*{\frac{1-\mathfrak{p}_2}{n_{2}}} \right)\,,
\end{align}
and the ceiling function $\ceil*{x}$ is introduced since $n_{\min}$ and $p_{\min}$ are integers. We note that the product over $n$ and $p$  and the product over $l$ and $m$ are from the chiral and  anti-chiral parts of the multiplet respectively. We use Zeta function regularisation of infinite products~\cite{e684b263-94c7-31f3-8829-20ce2a1af791} and in particular we make use of the regularisation 
\begin{equation}
 \prod_{n=0}^{\infty} \left( \frac{n+x}{Y} \right) = \frac{Y^{-\frac{1}{2} + x}}{\Gamma(x)}\,,
\end{equation} and see that the phase factor from the product over $\i$ is exactly cancelled. Up to an overall numerical factor we find 
\begin{equation}
    Z^{ \qeq \cV}_{1\text{-loop}}  = \left(\frac{L}{L_0}\right)^{p_{\min}-n_{\min}+ 2 q_{\cal G} \frac{L}{L_0}\Lambda^{\cal G}   } \sqrt{\frac{\Gamma \left( n_{\min} - q_{\mathcal{G}}\frac{L}{L_0} \Lambda^{\mathcal{G}}\right)^2}{\Gamma \left( p_{\min} +q_{\mathcal{G}}\frac{L}{L_0} \Lambda^{\mathcal{G}}\right)^2}}\,,
\end{equation}
where the unusual appearance of a square root of a square is dealt with next.
To  proceed further, 
we use
\begin{align}
\begin{split}
       &\max (x,y) = \frac{1}{2}(x + y + |x-y|)\,,\\
       \end{split}
\end{align}
and the following properties of the ceiling and floor function 
\begin{align}
\begin{split}
       &\ceil*{\frac{x}{y}}= \floor*{\frac{x-1}{y}}+1 \quad \mbox{for }x,y \in \mathbb{Z} \mbox{ and } x\neq 0\,, \qquad
       \ceil*{a}= - \floor*{-a}\,, 
       \end{split}
\end{align}
and introduce the notation $\llbracket \bullet \rrbracket_{\diamond}$ to denote the remainder after dividing $\bullet$ by $\diamond$ such that 
\begin{equation}
\floor*{\frac{\bullet}{n_{1,2}}}  = \frac{\bullet}{n_{1,2}}- \frac{\llbracket \bullet \rrbracket_{n_{1,2}}}{n_{1,2}}\,.  
\end{equation}
Then the result of the one-loop 
\begin{equation} 
    Z^{ \qeq \cV}_{1\text{-loop}} =  \textcolor{black}{\left( \frac{L}{L_0} \right)^{1-\frac{r}{2} \chi + 2 q_{\mathcal{G}} \gamma_{\mathcal{G}} - \mathfrak{c}}} \sqrt{ \frac{\Gamma \left(\frac{1}{2} (\mathfrak{c} + |\mathfrak{b}-1| ) + \frac{r}{4} \chi -{q_{\mathcal{G}} \gamma_{\mathcal{G}}} \right)^2 }{\Gamma \left(1-\frac{1}{2} (\mathfrak{c} - |\mathfrak{b} -1| ) - \frac{r}{4} \chi +{q_{\mathcal{G}} \gamma_{\mathcal{G}}}\right)^2}}\,,
\end{equation}
where we recall $\chi$ is the Euler number of the spindle \eqref{eq: Euler_number}, $\gamma_{\mathcal{G}}$ is defined in \eqref{eq: gamma_g} and we also introduced the following notation, in order to facilitate comparison with \cite{Inglese:2023wky, Inglese:2023tyc}, 
\begin{equation}\label{eq: compact_notation}
\begin{alignedat}{2}
        \mathfrak{b} = 1 + \floor*{ - \frac{\mathfrak{p}_2}{n_2}} - \floor*{ - \frac{\mathfrak{p}_1}{n_1} }\,,  \,  \qquad  \mathfrak{c} = \frac{\llbracket - \mathfrak{p}_2 \rrbracket_{n_2}}{n_2} + \frac{\llbracket - \mathfrak{p}_1 \rrbracket_{n_1}}{n_1}\,.
\end{alignedat}
\end{equation}
Finally, we note that due to Euler's reflection formula: $\Gamma(z)\Gamma(1-z)= \pi/ \sin(\pi z)\,,$ for $z \notin \mathbb{Z}$, we can substitute
\begin{equation} \label{eq: b-1-replacement}
    |\mathfrak{b} -1| \rightarrow \mathfrak{b}-1\,,
\end{equation}
and obtain the result of 1-loop partition function as 
\begin{eqnarray} \label{eq: one-loop-eigenmodes}
     Z^{ \qeq \cV}_{1\text{-loop}} = \textcolor{black}{\left( \frac{L}{L_0} \right)^{1-\frac{r}{2} \chi + 2 q_{\mathcal{G}} \gamma_{\mathcal{G}} - \mathfrak{c}}} \frac{\Gamma\left( \frac{\mathfrak{b}+\mathfrak{c}-1}{2} + \frac{r}{4} \chi  - {q_{\mathcal{G}} \gamma_{\mathcal{G}}} \right) }{\Gamma\left( \frac{\mathfrak{b}-\mathfrak{c}+1}{2} - \frac{r}{4} \chi + {q_{\mathcal{G}} \gamma_{\mathcal{G}}} \right) }\,.
\end{eqnarray}

We now provide a careful comparison with the notation of Inglese, Martelli, and Pittelli (IMP) \cite{Inglese:2023tyc} in order to demonstrate that our results are consistent. The dictionary between our notation and theirs is\footnote{Note that the $k_0$ that appears in \cite{Pittelli:2024ugf} (denoted here as $k_0^{\text{(IMP)}}$) is different from the normalisation $k_0$ that we introduce in \eqref{KSchi}. As we use axial coordinates on the spindle with different periodicities, the map is $k_0 = (\Delta z/2\pi) k_0^{(\text{IMP})}$.} 
\begin{equation} \label{eq:dictionaryIMP}
    n_1 = n_+\,, \quad n_2 = n_-\,, \quad \mathfrak{m}_1 = - \mathfrak{m}_+\,, \quad \mathfrak{m}_2 = -\mathfrak{m}_-\,, \quad \Lambda^{\mathcal{G}} = k_0^{\text{(IMP)}}\varphi_G=   {\color{black}{L_0}}\varphi_G\,,
\end{equation} 
where the $\mathfrak{m}_{1,2}$ pick up additional signs in order to account for the conventions of the signs of the R-symmetry and gauge fields is opposite of ours and hence the fluxes through $\mathbb{\Sigma}$ in \cite{Inglese:2023tyc} take the opposite sign.\footnote{See equations (2.38) and (3.1) of \cite{Inglese:2023tyc}, note that $\mathfrak{m} = - \mathfrak{m}^{(\text{IMP})}$.} Upon application of this dictionary, we find 
\begin{equation} \label{eq: dictionaryIMP2}
\mathfrak{m} = - \mathfrak{m}^{\text{(IMP)}}\,, \quad    \mathfrak{b} = \mathfrak{b}^{\text{(IMP)}}\,, \quad \mathfrak{c} = \mathfrak{c}^{\text{(IMP)}}\,, \quad \gamma_{\mathcal{G}} = -  {\color{black}L}\gamma_{G}\,, 
\end{equation}
where the minus sign in the final equation above shows that the ratio of Gamma functions in \eqref{eq: one-loop-eigenmodes} reproduces (3.85) of \cite{Inglese:2023tyc}, where the one-loop determinant was computed via the index theorem instead. However, we note that our result differs from that of \cite{Inglese:2023tyc} via the inclusion of the moduli-dependent power of $L/L_0$ which multiplies the ratio of Gamma functions in \eqref{eq: one-loop-eigenmodes}. Although this factor can be removed via a different choice of Killing spinor normalisation (which was seemingly done in the prior work), we will treat it carefully as the overall power of this ratio of scales is an important object related to the central charge and conformal anomaly of the field theory \cite{Benini:2016qnm,Vassilevich:2003xt}. The dependence of the power of $L/L_0$ on the vector multiplet moduli $(\sigma_0, \mathfrak{m})$ will renormalise the FI parameter $\xi$ as we will see in Section~\ref{sec: Mirror Symmetry}. We also note that the one-loop determinant~\eqref{eq: one-loop-eigenmodes} matches the one-loop determinant for $S^2$ \cite{GonzalezLezcano:2023cuh} in the limit $n_{1,2} \rightarrow 1$. Finally, we note that our calculation makes the natural identification of 
\begin{equation}
   k_0^{(\text{IMP})} \omega =  \frac{L_0}{L}\,,
\end{equation}
where $\omega$ is defined in equation (2.17) of \cite{Inglese:2023tyc} as the fugacity for the angular momentum relative to the additional $S^1$ factor present in that work. From a 3d perspective, $L_0$ may be related to the radius of $S^1$ in the analogous localisation calculation on $\mathbb{\Sigma}\times S^1$.

One advantage of our calculation relative to \cite{Inglese:2023wky, Inglese:2023tyc, Pittelli:2024ugf} is that by working with the cohomological complex and the square root form of the one-loop determinant \eqref{eq:prelogNozm}, we are able to entirely fix the various factors which appear in the computation of the one-loop determinant (notice that as well as the powers of $L/L_0$, a factor of $\sigma_- =-1$ is ignored in equation (3.83) of \cite{Inglese:2023tyc}). We will see that a similar accounting occurs when computing the one-loop determinant using the index method on the fields in the cohomological complex. 

The one-loop determinant \eqref{eq: one-loop-eigenmodes} is for the chiral multiplet with gauge/flavour symmetry group $\mathcal{G} = U(1)$ but can also be leveraged in a straightforward manner in order to compute the analogous expressions for an arbitrary gauge/flavour group, as well as the vector multiplet one-loop result. For a gauge/flavour symmetry group $\mathcal{G}$ where the fields transform in a representation $\mathfrak{R}_\mathcal{G}$ with weight $\varrho$, the chiral multiplet one-loop determinant can be obtained from the abelian case via the substitutions 
\begin{equation}
\mathfrak{m}_{1,2} \rightarrow \varrho(\mathfrak{m}_{1,2})\,, \qquad \gamma_{\mathcal{G}} \rightarrow \varrho( \gamma_{\mathcal{G}})\,,
\end{equation}
allowing us to write the chiral multiplet one loop determinant as
\begin{equation} \label{eq: chiral_one_loop_G}
     Z^{ \qeq \cV}_{1\text{-loop}} = \left( \frac{L}{L_0} \right)^{1-\frac{r}{2} \chi + 2 \varrho( \gamma_{\mathcal{G}} ) - \mathfrak{c}(\varrho(\mathfrak{m}))} \frac{ \Gamma \left(\frac{1}{2} (\mathfrak{c}(\varrho(\mathfrak{m})) + \mathfrak{b}(\varrho(\mathfrak{m}))-1 ) + \frac{r}{4} \chi - {\varrho( \gamma_{\mathcal{G}})} \right)}{\Gamma \left(\frac{1}{2} ( \mathfrak{b}(\varrho(\mathfrak{m})) - \mathfrak{c}(\varrho(\mathfrak{m})) +1  ) - \frac{r}{4} \chi +  {\varrho( \gamma_{\mathcal{G}})}\right)}\,.
\end{equation}

Finally, we can obtain the vector multiplet one-loop determinant directly from \eqref{eq: chiral_one_loop_G} by setting $r=2$ and $\mathfrak{R}_\mathcal{G} = \text{adj}_\mathcal{G}$. Upon doing this, one obtains
\begin{align}
\begin{split}
 \left. Z^{ \qeq \cV}_{1\text{-loop}} \right|_{r=2, \mathfrak{R}_\mathcal{G} =  \text{adj}_\mathcal{G}} & =  \left( \frac{L}{L_0} \right)^{1-\chi + 2 \varrho_{\text{ad}}( \gamma_{\mathcal{G}} ) - \mathfrak{c}(\varrho_{\text{ad}}(\mathfrak{m}))} \\
 & \quad \quad  \times \frac{ \Gamma \left( \frac{1}{2} (\mathfrak{c}(\varrho_{\text{ad}}(\mathfrak{m})) + \mathfrak{b}(\varrho_{\text{ad}}(\mathfrak{m}))-1 ) + \frac{\chi}{2} - {\varrho_{\text{ad}}( \gamma_{\mathcal{G}})} \right)}{\Gamma \left(\frac{1}{2} ( \mathfrak{b}(\varrho_{\text{ad}}(\mathfrak{m})) - \mathfrak{c}(\varrho_{\text{ad}}(\mathfrak{m})) +1  ) - \frac{\chi}{2} +  {\varrho_{\text{ad}}( \gamma_{\mathcal{G}})}\right)}\,,
        \end{split}
\end{align}
 where $\varrho_{\text{ad}}$ represents the weight of the fields transforming in the adjoint representation.
 
\subsection{Fixed point formula}\label{IndexMethod}
 
An alternative approach to derive the one-loop determinant formula \eqref{eq: one-loop-eigenmodes} is to use the fixed point formula \cite{atiyah1966lefschetz,10.2307/1970694,Atiyah:1974obx}, applied to orbifolds in \cite{10.1215/S0012-7094-96-08226-5}, in order to derive the equivariant index of the operator which pairs bosons and fermions in the cohomological complex \cite{Pestun:2007rz}. In this section we closely follow \cite{Inglese:2023wky, Inglese:2023tyc,  Pittelli:2024ugf} in our analysis of this index (see also \cite{MEINRENKEN1998240} for a mathematical perspective on this topic). However, our computation is fundamentally different in a crucial way. We will directly derive the index on $\mathbb{\Sigma}$ using our conventions, rather than obtain it as a dimensional reduction of a higher dimensional index. To reiterate, nowhere did we envisage $\mathbb{\Sigma}$ as a sub-manifold of any higher dimensional manifold.

First we set the notation. Our first object of interest is the index of the operator $D_{10}$ with respect to the action of the group $g$ on a \textit{manifold} $\mathbb{M}$\footnote{To distinguish it from the index defined in \eqref{eq: index_orbifold}, we shall refer to this index as the \textit{manifold~index}.}  
\begin{equation} \label{eq: index_manifold}
    \text{ind}(D_{10}; g) = \sum_{p \in \mathbb{M}^g} \frac{\text{tr}_{\Gamma_0} g - \text{tr}_{\Gamma_1} g}{\det (1 - J_p)}\,,
\end{equation} where all notation will be defined shortly. To begin with, we note that for our purposes,
\begin{align}
\begin{split} \label{eq: g_def}
    g = \exp \left( t Q^2_{eq} \right)\,,
    \end{split}
\end{align}
where the parameter $t$ is the equivariant parameter. Recalling \eqref{eq: Q^2} we have
\begin{equation} \label{eq: Q^2_2}
    Q^2_{eq} = \mathcal{L}_{v} + \i \widehat{q}_R \Lambda^R + \i {\widehat{q}_{\mathcal{G}} \Lambda^{\mathcal{G}}}\,,\qquad {v= \frac{k_0}{L}\partial_z \,,}
\end{equation}
where $\Lambda^R, {\Lambda^{\mathcal{G}}}$ are gauge-dependent terms which we will fix by choice of gauge shortly. We note that from the structure of \eqref{eq: g_def} and \eqref{eq: Q^2_2}, $g$ is a map from the total space to the total space and it induces both a map on the manifold $\mathbb{M}$: a $U(1)$ rotation via the Killing vector $\partial_z$, as well as a map on the sections of the $U(1)_R \times U(1)_{\mathcal{G}}$ valued line orbibundle $L^{\widehat{q}_{R},\widehat{  q}_{\mathcal{G}}}$: a free action via the $R$ and gauge symmetry parameters. 

The structure of this map aides us in defining the other quantities which appear in the index \eqref{eq: index_manifold}. The summation in \eqref{eq: index_manifold} is taken over the fixed points $p$ of the group action on the manifold, denoted $\mathbb{M}^g$ (these will turn out to be the poles of the spindle $y_{1,2}$). The action of $g$  on the manifold is given by the following passive transformations: $g \circ z_p = z'_p$, and $J_p = \partial z'_p/\partial z_p$ is the Jacobian of this transformation.  Finally, $\Gamma_0$ is the space of sections of the $U(1)_R \times U(1)_{\mathcal{G}}$ valued line bundle $L$ over $\mathbb{M}$ (the elementary boson $\Phi$ in the cohomological complex), and $\Gamma_1 = D_{10} (\Gamma_0)$, the image of sections $\Gamma_0$ under the operator $D_{10}$ (the elementary fermion $\Psi$ in the cohomological complex).

In the case of a spindle, the neighbourhoods of the fixed points are locally $\mathcal{U}_{1,2} \cong \mathbb{C}/\mathbb{Z}_{n_{1,2}}$ and thus we need to use the index of an \textit{orbifold}, first given in \cite{10.1215/S0012-7094-96-08226-5}. For the spindle, the orbifold index \cite{Inglese:2023wky, Inglese:2023tyc, Pittelli:2024ugf} is given by 
\begin{equation} \label{eq: index_orbifold}
    \text{ind}_{\text{orb}}(D_{10}; g) = \sum_{p=1}^2 \frac{1}{n_p} \sum_{w \in \mathbb{Z}_{n_p}} \frac{\text{tr}_{\Gamma_0} (w g) - \text{tr}_{\Gamma_1}( w g)}{\det (1 - W_p J_p)}\,,
\end{equation}
where
 $w$ is the additional map required to take into account the fact that the orbifold is locally a quotient of $\mathbb{C}$ by $\mathbb{Z}_{n_p}$, and the action of which we note is analogous to the map $g$ in~\eqref{eq: g_def} except the transformation is now naturally weighted by elements of $\mathbb{Z}_{n_p}$ rather than the equivariant parameter $t$ (an explicit formula for this map is given in \eqref{eq: w_map} below). $W_p$ is the Jacobian of $w$ acting on the coordinates. We will now work through the computation of the index theorem step-by-step, computing the index \eqref{eq: index_orbifold} before using this to derive the one-loop determinant via the arguments of equations \eqref{eq:indDeg} and \eqref{eq:degeneracies}. We will see that the orbifold index approach leads to an identical result for the one-loop determinant as reported in \eqref{eq: one-loop-eigenmodes}, serving as a stringent consistency check of our results.

\subsubsection{Regular gauges at the poles}

In order to compute \eqref{eq: index_orbifold} we utilise suitable gauge choices for our $U(1)_R$ and $U(1)_{\mathcal{G}}$ gauge fields in the neighbourhoods $\mathcal{U}_{1,2}$. In contrast to the choices \eqref{eq: A_2d}, \eqref{eq: gauge_field_explicit_real} which we employed in the unpaired eigenvalue approach, we will use gauge fields which are regular at the poles of the spindle. This leads us to employ the regular gauge choices \eqref{eq:gaugefield_regpatch} and \eqref{eq: u1g_regpatch}, repeated below for convenience
\begin{equation}
     \left. A \right|_{\mathcal{U}_{1,2}} = \frac{a}{4}\left( \frac{1}{y_{1,2}} - \frac{1}{y}\right) \textrm{d}z\,, \qquad  {\left. \mathcal{A} \right|_{\mathcal{U}_{1,2}} = \frac{\mathfrak{m}}{n_1-n_2} \frac{a }{2} \left( \frac{1}{y} - \frac{1}{y_{1,2}} \right) \textrm{d}z}\,,
\end{equation}
from which we can compute 
\begin{equation}
  \left. \Lambda^{\mathcal{G}} \right|_{\mathcal{U}_{1,2}} = \Lambda^{\mathcal{G}} +\frac{k_0}{L}\frac{2\pi}{\Delta z} \frac{\mathfrak{m}_{1,2}}{n_{1,2}} = \Lambda^{\mathcal{G}} + \frac{L_0}{L}\frac{\mathfrak{m}_{1,2}}{n_{1,2}}\,.
\end{equation}
Using the above equation together with \eqref{Q2algebraNP} we have for the equivariant supersymmetry squared operator \eqref{eq: Q^2}
{\color{black}
\begin{align}\label{eq: Qeq^2_regular}
    \begin{split}
        \left. Q_{eq}^2\right|_{\mathcal{U}_{1,2}}  &= \frac{k_0}{L}\mathcal{L}_{\partial_z}-  \i \widehat{q}_R  \frac{k_0}{L}\frac{2\pi }{\Delta z} \cdot  \frac{1}{2n_{1,2}} + \i \widehat{q}_{\mathcal{G}} \left( \Lambda^{\mathcal{G}} +\frac{k_0}{L}\frac{2\pi}{\Delta z} \frac{\mathfrak{m}_{1,2}}{n_{1,2}}\right)
        \\
        &=\frac{L_0}{L}\left[ \frac{\Delta z}{2\pi} \cL_{\partial_z} - \i\widehat{q}_R   \frac{1}{2n_{1,2}} + \i \widehat{q}_{\mathcal{G}} \left( \Lambda^{\mathcal{G}}\frac{L}{L_0} + \frac{\mathfrak{m}_{1,2}}{n_{1,2}}\right) \right]\,,
    \end{split}
\end{align} 
 }
and the map $g$ appearing in the orbifold index \eqref{eq: index_orbifold} is 
{\color{black}\begin{align}
   \left. g  \right|_{\mathcal{U}_{1,2}} & = \exp \left(  t \frac{L_0}{L} \left[\frac{\Delta z}{2\pi} \mathcal{L}_{\partial_z} -  \i \widehat{q}_R  \frac{1}{2n_{1,2}} + \i \widehat{q}_{\mathcal{G}} \left( \Lambda^{\mathcal{G}} \frac{L}{L_0} + \frac{\mathfrak{m}_{1,2}}{n_{1,2}}\right) \right] \right)\,.
\end{align}}
We will now examine suitable choices of coordinates near the poles $y_{1,2}$.
\subsubsection*{Coordinates near the poles}

In order to compute \eqref{eq: index_orbifold}, it will be useful to examine how the coordinates in \eqref{eq: spindle_metric} relate to the coordinates of the $\mathbb{C}/\mathbb{Z}_{n_{1,2}}$ to which the local neighbourhoods of the poles $\mathcal{U}_{1,2}$ are homeomorphic. A further comparison of the coordinates \eqref{eq: spindle_metric} and those of \cite{Inglese:2023tyc} is given in Appendix~\ref{app: coordinates}.  We start by performing the mappings 
\begin{equation}
    y = y_{1,2} + x_{1,2}\,,
\end{equation}
near $y=y_{1,2}$ respectively. Note that for our chosen ordering of $y_1$ and $y_2$, we have  $x_1 >0 >x_2$. This coordinate transformation brings the metric \eqref{eq: spindle_metric} into the form: 
\begin{equation} \label{eq: spindle_line_element_epsilon}
   {\color{black}\frac{1}{L^{2}}} \left. \textrm{d}s^2_{\mathbb{\Sigma}} \right|_{\mathcal{U}_{1,2}} = \frac{1}{x_{1,2}} \frac{y_{1,2}}{q'(y_{1,2})} \textrm{d}x_{1,2}^2 + \frac{x_{1,2} q'(y_{1,2})}{36 y_{1,2}^2} \textrm{d}z^2 + \mathcal{O}(x_{1,2}^2)\,.
\end{equation}
Now defining the coordinates $r_{1,2}$ by 
\begin{equation}
    x_{1,2} = \frac{r_{1,2}^2}{4} \frac{q'(y_{1,2})}{y_{1,2}}\,,
\end{equation}
we recast the metric \eqref{eq: spindle_line_element_epsilon} into the form (ignoring higher order corrections in $x_{1,2}$) 
\begin{equation}
  {\color{black}\frac{1}{L^{2}}} \left. \textrm{d}s^2_{\mathbb{\Sigma}} \right|_{\mathcal{U}_{1,2}} = \textrm{d}r_{1,2}^2 + \frac{r_{1,2}^2 q'(y_{1,2})^2}{144 y_{1,2}^3} \textrm{d}z^2\,,
\end{equation}
which takes the suggestive form akin to that of plane polar coordinates on $\mathbb{R}^2$. We can now perform the simple re-scalings of the coordinates via 
\begin{equation} \label{eq: zeta_z_map}
    \rho_{1,2} = \frac{r_{1,2}}{n_{1,2}}\,, \qquad \varphi = z n_{1,2} \frac{|q'(y_{1,2})|}{12 y_{1,2}^{3/2}} = z \frac{2\pi}{\Delta z}\,, 
\end{equation} 
giving us the final form of the metric
\begin{equation} \label{eq: rho_zeta_metric}
    {\color{black}\frac{1}{L^{2}}} \left. \textrm{d}s^2_{\mathbb{\Sigma}} \right|_{\mathcal{U}_{1,2}} =  n_{1,2}^2 \textrm{d}\rho_{1,2}^2 + \rho_{1,2}^2 \textrm{d}\varphi^2\,,
\end{equation}
where $\rho_{1,2}>0$ and $ \varphi \in [0,2\pi)$, matching the coordinates used near the poles in \cite{Pittelli:2024ugf}. The complex coordinates in the near the north pole of the spindle are 
\begin{equation}
    w_1 = \rho_1 \exp(\i\varphi/n_1)\,, \qquad  \bar{w}_1 = \rho_1 \exp(-\i\varphi/n_1)\,, \qquad w_1 \sim e^{2\pi \i /n_1} w_1\,,
\end{equation}
and those near the south pole $\mathbb{C}/\mathbb{Z}_{n_2}$ are 
\begin{equation}
    w_2 = \rho_2 \exp(-\i\varphi/n_2)\,, \qquad  \bar{w}_2 = \rho_2 \exp(\i\varphi/n_2)\,, \qquad w_2 \sim e^{-2\pi \i /n_2} w_2\,,
\end{equation}
where the minus sign in the exponent arises from the fact that the neighbourhoods around the poles should have the opposite orientation.
\subsubsection*{Equivariant supersymmetry squared operator}
 We now want to examine the form of the operator $Q^2_{eq}$ in the $\rho, \varphi$ coordinates \eqref{eq: rho_zeta_metric}. We start with the equivariant supersymmetry squared operator \eqref{eq: Qeq^2_regular} in the regular gauges and apply the coordinate transformation \eqref{eq: zeta_z_map}, finding
{\color{black}
\begin{align}
\begin{split}
    \left. Q^2_{eq} \right|_{\mathcal{U}_{1,2}} = \frac{L_0}{L} \left[ \partial_{\varphi} - \i \widehat{q}_R \frac{1}{2n_{1,2}} + {\i \widehat{q}_{\mathcal{G}} \frac{\mathfrak{m}_{1,2}}{n_{1,2}} + \i \widehat{q}_{\mathcal{G}} \frac{L}{L_0} \Lambda^{\mathcal{G}}} \right]\,,
    \end{split}
\end{align}}
upon which we can use the notation defined in \eqref{eq: p12} to write
{\color{black}\begin{equation} \label{eq: Q^2_zeta}
\left. Q^2_{eq} \right|_{\mathcal{U}_{1,2}} =   \frac{L_0}{L} \left[ \partial_{\varphi} - \i \frac{\widehat{\mathfrak{p}}_{1,2}}{n_{1,2}}+ \i \widehat{q}_{\mathcal{G}} \frac{L}{L_0} \Lambda^{\mathcal{G}}\right]\,,
\end{equation}}where $\widehat{\mathfrak{p}}_{1,2}$ can be obtained from $\mathfrak{p}_{1,2}$ given in \eqref{eq: p12} by performing the replacements $r \rightarrow \widehat{q}_R,\,  q_{\mathcal{G}} \rightarrow \widehat{q}_{\mathcal{G}}$. We note that this operator takes a similar form to that found in the literature~\cite{Inglese:2023tyc, Pittelli:2024ugf}. For direct comparison, one can compare our equations \eqref{eq: g_def} and \eqref{eq: Q^2_zeta} with equations (3.44) and (3.45) of \cite{Pittelli:2024ugf}, where the map between their notation and ours is 
\begin{equation}
    \epsilon = t\,, \quad k^{(\text{IMP})}_0 = {\color{black}{L_0}}\,, \quad \omega = {\color{black}\frac{1}{L}}\,,
\end{equation}
 where our choice of $k_0$ arises from our chosen normalisation of the Killing spinors \eqref{KSchi}, which leads directly to the normalisation of the Killing vector \eqref{eq:killingveccomp}.

\subsubsection{One-loop determinant via fixed point formula}

Using the form of the operator \eqref{eq: Q^2_zeta} above we now perform the computations of the index~\eqref{eq: index_orbifold} on $\mathbb{\Sigma}$. We first divide the map $g$ as
\begin{equation}
   \left. g  \right|_{\mathcal{U}_{1,2}} = g_{(0)} \left. g_{\mathbb{\Sigma}} \right|_{\mathcal{U}_{1,2}}  \, ,
\end{equation}
where  
\begin{equation}
    \left. g_{\mathbb{\Sigma}} \right|_{\mathcal{U}_{1,2}} = e^ { {\color{black}{\frac{L_0}{L}}} t  \left( \partial_{\varphi} - \i \frac{\widehat{\mathfrak{p}}_{1,2}}{n_{1,2}} \right) }\,, \qquad {g_{(0)} = e^{t \i \widehat{q}_{\mathcal{G}}  \Lambda^{\mathcal{G}}}}\,.
\end{equation}
 We have followed \cite{Pittelli:2024ugf} in separating the action of $g$ into a spindle piece with fixed points, $g_{\mathbb{\Sigma}}$, and a term that acts freely, $g_{(0)}$. The $g_{(0)}$ piece is a gauge fugacity, and will factor out through the calculation. We will now proceed with the index calculation, focusing primarily on $g_{\mathbb{\Sigma}}$.

The first step is to determine how the operator $g_{\mathbb{\Sigma}}$ acts on the coordinates $w_{1,2}$ and sections of the line orbibundle $L^{\widehat{q}_R,\widehat{q}_{\mathcal{G}}}$. We recall that neighbourhoods around the northern and southern poles are homeomorphic respectively to $\mathbb{C}/\mathbb{Z}_{n_{1,2}}$ and therefore, in the northern patch, we have
\begin{equation}
    \left. g_{\mathbb{\Sigma}} \right|_{\mathcal{U}_{1}} \circ w_1 = q_1 w_1\,, \qquad q_1 = e^{-\i {\color{black}\frac{L_0}{L}} \frac{t}{n_1}}\,, 
\end{equation} 
and the equivariant action on the bosonic fields $\Phi= \{ \phi, \widetilde{\phi}\}  \in \Gamma_0$ is
\begin{equation}
    \left. g_{\mathbb{\Sigma}} \right|_{\mathcal{U}_{1}} \circ \phi = q_1^{\mathfrak{p}_1} \phi\,, \qquad  \left. g_{\mathbb{\Sigma}} \right|_{\mathcal{U}_{1}} \circ \widetilde{\phi} = q_1^{-\mathfrak{p}_1} \widetilde{\phi}\,,
\end{equation}
where the difference in the signs in the exponents arises from the opposite $R$ and gauge charge signs between $\phi$ and $\widetilde{\phi}$. The relevant definitions are found in \eqref{eq: p12}. 
In order to compute the equivariant action on fields in $\Gamma_1 = D_{10}(\Gamma_0)$ we can use the result \eqref{eq: D_10_map} that $D_{10}$ maps from bosonic to fermionic quantities in the cohomological complex. Thus the action on fermions $\Psi = \{\epsilon \psi, \widetilde{\epsilon} \widetilde{\psi} \}$ is
\begin{equation}
     \left. g_{\mathbb{\Sigma}} \right|_{\mathcal{U}_{1}} \circ (\epsilon \psi) = q_1^{\mathfrak{p}_1-1}(\epsilon \psi)\,, \qquad \left. g_{\mathbb{\Sigma}} \right|_{\mathcal{U}_{1}} \circ (\widetilde{\epsilon}\widetilde{\psi}) = q_1^{-(\mathfrak{p}_1-1)}(\widetilde{\epsilon}\widetilde{\psi})\,. \qquad 
\end{equation}
Finally, we note that the action of the gauge element $g_{(0)}$ is 
\begin{equation}
    \begin{alignedat}{2}
& \left. g_{(0)} \right|_{\mathcal{U}_{1}} \circ \phi  = \mathfrak{q}^{-q_{\mathcal{G}}{\color{black}\frac{L}{L_0}} \Lambda^{\mathcal{G}}}\phi\,, \qquad && \left. g_{(0)} \right|_{\mathcal{U}_{1}} \circ \widetilde{\phi}  = \mathfrak{q}^{q_{\mathcal{G}} {\color{black}\frac{L}{L_0}} \Lambda^{\mathcal{G}}}\widetilde{\phi}\,, \\
  &  \left. g_{(0)} \right|_{\mathcal{U}_{1}} \circ (\epsilon \psi)  = \mathfrak{q}^{-q_{\mathcal{G}} {\color{black}\frac{L}{L_0}} \Lambda^{\mathcal{G}}}(\epsilon \psi)\,, \qquad && \left. g_{(0)} \right|_{\mathcal{U}_{1}} \circ (\widetilde{\epsilon}\widetilde{\psi}) = \mathfrak{q}^{q_{\mathcal{G}} {\color{black}\frac{L}{L_0}} \Lambda^{\mathcal{G}}}(\widetilde{\epsilon}\widetilde{\psi})\,,
    \end{alignedat}
\end{equation} where we have defined $\mathfrak{q} = q_1^{n_1} = e^{-\i {\color{black}\frac{L_0}{L}}t}\,.$

 Continuing in this manner, we find the contribution to the manifold index \eqref{eq: index_manifold} from $y_1$ is 
\begin{align}
\begin{split}
    I_1 & = \frac{\mathfrak{q}^{-q_{\mathcal{G}} {\color{black}\frac{L}{L_0}} \Lambda^{\mathcal{G}}}(q_1^{\mathfrak{p}_1}- q_1^{\mathfrak{p}_1-1})+\mathfrak{q}^{q_{\mathcal{G}} {\color{black}\frac{L}{L_0}} \Lambda^{\mathcal{G}}}(q_1^{-\mathfrak{p}_1} -q_1^{1-\mathfrak{p}_1})}{(1-q_1)(1-q_1^{-1})} \\
    &= \mathfrak{q}^{-q_{\mathcal{G}}{\color{black}\frac{L}{L_0}} \Lambda^{\mathcal{G}}}\frac{q_1^{\mathfrak{p}_1}}{1-q_1} - \mathfrak{q}^{q_{\mathcal{G}} {\color{black}\frac{L}{L_0}} \Lambda^{\mathcal{G}}}\frac{q_1^{-\mathfrak{p}_1+1}}{1-q_1}\,, 
    \end{split}
\end{align}
where the first term in the final equality is the chiral contribution and the second the anti-chiral contribution. This will be important to keep track of as the gauge fugacity piece $g_{(0)}$ will act differently on each. The index above needs to be modified accordingly to take into account the action of the orbifold map $w$ 
on coordinates and sections. The map $w$ is given by \cite{Inglese:2023tyc, Pittelli:2024ugf} 
\begin{equation} \label{eq: w_map}
    \left. w \right|_{\mathcal{U}_{1,2}} = \left. w_{\mathbb{\Sigma}} \right|_{\mathcal{U}_{1,2}}  =  e^ { -2\pi l_{1,2}  \left( \partial_{\varphi} - \i \frac{\mathfrak{p}_{1,2}}{n_{1,2}} \right) }\,, \qquad l_{1,2} \in \mathbb{Z}_{n_{1,2}}\,,
\end{equation}and thus this acts on coordinates as
\begin{equation}
 \left.  w \right|_{\mathcal{U}_{1}} \circ w_1 = \exp \left( 2\pi \i l/n_{1} \right) w_1 = u_1^l w_1\,, \qquad u_1 = e^{2\pi i/n_1}\,,
\end{equation}
on sections of $\Gamma_0$ as
\begin{equation}
     \left.  w \right|_{\mathcal{U}_{1}} \circ \phi = u_1^{l\mathfrak{p}_1} \phi\,, \qquad \left.  w \right|_{\mathcal{U}_{1}} \circ \widetilde{\phi} = u_1^{-l\mathfrak{p}_1} \widetilde{\phi}\,,
\end{equation}
and on sections of $\Gamma_1$ as 
\begin{equation}
     \left.  w \right|_{\mathcal{U}_{1}} \circ (\epsilon \psi) = u_1^{l(\mathfrak{p}_1-1)} (\epsilon \psi)\,, \qquad \left.  w \right|_{\mathcal{U}_{1}} \circ (\widetilde{\epsilon} \widetilde{\psi}) = u_1^{-l(\mathfrak{p}_1-1)} (\widetilde{\epsilon} \widetilde{\psi})\,.
\end{equation}
Putting these results together, one sees that the substitution rule in moving from the manifold index to the \textit{orbifold} index should be $q_1 \rightarrow q_1 u_1^j$ followed by averaging over $j = 0\,, \ldots\,, n_1-1$. In performing this procedure, the action of $g_{(0)}$ factors out, resulting in
\begin{align}
\begin{split} \label{eq: orbifold_index_north}
    I_{\mathcal{U}_1} &  = \mathfrak{q}^{-q_{\mathcal{G}}  {\color{black}\frac{L}{L_0}} \Lambda^{\mathcal{G}}} \frac{1}{n_1} \sum_{j=0}^{n_1-1} \frac{ u_1^{j \mathfrak{p}_1} q_1^{\mathfrak{p}_1}}{1-u_1^{j} q_1} - \mathfrak{q}^{q_{\mathcal{G}} {\color{black}\frac{L}{L_0}} \Lambda^{\mathcal{G}}} \frac{1}{n_1} \sum_{j=0}^{n_1-1} \frac{ u_1^{j( 1-\mathfrak{p}_1)} q_1^{1-\mathfrak{p}_1}}{1-u_1^{j} q_1} \\
    &= \textcolor{black}{-} \mathfrak{q}^{-q_{\mathcal{G}}  {\color{black}\frac{L}{L_0}} \Lambda^{\mathcal{G}}} \frac{\mathfrak{q}^{\textcolor{black}{-1}-\floor{-\mathfrak{p}_1/n_1}}}{1-\mathfrak{q}^{\textcolor{black}{-1}}} \textcolor{black}{+}  \mathfrak{q}^{q_{\mathcal{G}} {\color{black}\frac{L}{L_0}}\Lambda^{\mathcal{G}}} \frac{\mathfrak{q}^{\floor{-\mathfrak{p}_1/n_1}}}{1-\mathfrak{q}^{\textcolor{black}{-1}}}\,,
    \end{split}
\end{align}
where the manipulations in moving from the first to the second line are proved in Appendix~\ref{sec: sums}. Note that up to the gauge fugacity term the second term matches equation (3.70) of \cite{Pittelli:2024ugf}, but we also have the additional first term arising from the chiral fields in the cohomological complex. Using the fact that the denominator of the equation above is the sum of a geometric progression series in $\mathfrak{q}^{\textcolor{black}{-1}}$, we have 
\begin{equation}
    I_{\mathcal{U}_1} = \sum_{\ell = 0}^{\infty} e^{ \textcolor{black}{-} \i {\color{black}\frac{L_0}{L}} t  \left(-\ell + \floor{-\mathfrak{p}_1/n_1} + {q_{\mathcal{G}} {\color{black}\frac{L}{L_0}} \Lambda^{\mathcal{G}}}\right)} - \sum_{\widetilde{\ell} = 0}^{\infty} e^{ \textcolor{black}{-}\i  {\color{black}\frac{L_0}{L}} t  \left(-\widetilde{\ell} - \floor{-\mathfrak{p}_1/n_1} - 1-{q_{\mathcal{G}} {\color{black}\frac{L}{L_0}}\Lambda^{\mathcal{G}}}\right)}\,,
\end{equation}
which translates into the infinite product:
\begin{equation} \label{eq: one-loop_North}
    Z_1 = \left(\frac{\prod_{\widetilde{\ell} = 0}^{\infty} \i  {\color{black}\frac{L_0}{L}} \left( \widetilde{\ell} + 1 + \floor*{-\frac{\mathfrak{p}_1}{n_1}} + {q_{\mathcal{G}}{\color{black}\frac{L}{L_0}} \Lambda^{\mathcal{G}}}\right) }{\prod_{\ell = 0}^{\infty} \i {\color{black}\frac{L_0}{L}} \left(\ell - \floor*{-\frac{\mathfrak{p}_1}{n_1}} - {q_{\mathcal{G}} {\color{black}\frac{L}{L_0}} \Lambda^{\mathcal{G}}} \right)}\right)^{1/2}\,.
\end{equation}
This gives half of the overall contribution to the one-loop determinant. 

We now follow a similar calculation in order to compute the contribution to the one-loop determinant from the neighbourhood $\mathcal{U}_{2}$ of the south pole. As before, we note that the group element $g$ acts on the coordinates as   
\begin{equation}
    \left. g \right|_{\mathcal{U}_{2}} \circ w_2 = q_2^{-1} w_2\,, \qquad q_2 = e^{-\i {\color{black}\frac{L_0}{L}} \frac{t}{n_2}}\,, \qquad \mathfrak{q} = q_1^{n_1} = q_2^{n_2} = e^{-\i {\color{black}\frac{L_0}{L}} t}\,.
\end{equation}
The operator $g_{\mathbb{\Sigma}}$ acts on the elementary boson $\Phi = \{ \phi, \widetilde{\phi}\}$ as 
\begin{equation}
    \left. g_{\mathbb{\Sigma}} \right|_{\mathcal{U}_{2}} \circ \phi = q_2^{\mathfrak{p}_2} \phi\,, \qquad  \left. g_{\mathbb{\Sigma}} \right|_{\mathcal{U}_{2}} \circ \widetilde{\phi} = q_2^{-\mathfrak{p}_2} \widetilde{\phi}\,,
\end{equation}
and the elementary fermion $\Psi = \{\epsilon \psi, \widetilde{\epsilon} \widetilde{\psi} \}$ as  
\begin{equation}
     \left. g_{\mathbb{\Sigma}} \right|_{\mathcal{U}_{2}} \circ (\epsilon \psi) = q_2^{\mathfrak{p}_2-1} (\epsilon \psi)\,, \qquad \left. g_{\mathbb{\Sigma}} \right|_{\mathcal{U}_{2}} \circ (\widetilde{\epsilon} \widetilde{\psi}) = q_2^{1-\mathfrak{p}_2} (\widetilde{\epsilon} \widetilde{\psi})\,.
\end{equation}
Finally, we note that the gauge fugacity element $g_{(0)}$ acts on bosons and fermions as
\begin{equation}
    \begin{alignedat}{2}
& \left. g_{(0)} \right|_{\mathcal{U}_{2}} \circ \phi  = \mathfrak{q}^{-q_{\mathcal{G}} {\color{black}\frac{L}{L_0}} \Lambda^{\mathcal{G}}}\phi\,, \qquad && \left. g_{(0)} \right|_{\mathcal{U}_{2}} \circ \widetilde{\phi}  = \mathfrak{q}^{q_{\mathcal{G}}{\color{black}\frac{L}{L_0}} \Lambda^{\mathcal{G}}}\widetilde{\phi}\,, \\
  &  \left. g_{(0)} \right|_{\mathcal{U}_{2}} \circ (\epsilon \psi)  = \mathfrak{q}^{-q_{\mathcal{G}}  {\color{black}\frac{L}{L_0}} \Lambda^{\mathcal{G}}}(\epsilon \psi)\,, \qquad && \left. g_{(0)} \right|_{\mathcal{U}_{2}} \circ (\widetilde{\epsilon}\widetilde{\psi}) = \mathfrak{q}^{q_{\mathcal{G}}{\color{black}\frac{L}{L_0}} \Lambda^{\mathcal{G}}}(\widetilde{\epsilon}\widetilde{\psi})\,.
    \end{alignedat}
\end{equation} 
As in the case of the north pole, we can put these results together in order to evaluate the manifold index 
\begin{align}
\begin{split}
    I_2 & = \frac{\mathfrak{q}^{-q_{\mathcal{G}} {\color{black}\frac{L}{L_0}}\Lambda^{\mathcal{G}}}(q_2^{\mathfrak{p}_2}- q_2^{\mathfrak{p}_2-1})+\mathfrak{q}^{q_{\mathcal{G}}{\color{black}\frac{L}{L_0}} \Lambda^{\mathcal{G}}}(q_2^{-\mathfrak{p}_2} -q_2^{1-\mathfrak{p}_2})}{(1-q_2^{-1})(1-q_2)} \\
    &= \mathfrak{q}^{-q_{\mathcal{G}}{\color{black}\frac{L}{L_0}}\Lambda^{\mathcal{G}}}\frac{q_2^{\mathfrak{p}_2}}{1-q_2} - \mathfrak{q}^{q_{\mathcal{G}} {\color{black}\frac{L}{L_0}}\Lambda^{\mathcal{G}}}\frac{q_2^{-\mathfrak{p}_2+1}}{1-q_2}\,. 
    \end{split}
\end{align}
The action of $w$ on coordinates and sections will again lead to the substitution rule $q_2 \rightarrow q_2 u_2^j$ ($u_2 = \exp(2\pi \i/n_2)$) followed by averaging over $j = 0\,, \ldots\,, n_2-1$.
\begin{align}
\begin{split}
    I_{\mathcal{U}_2} &  =  \mathfrak{q}^{-q_{\mathcal{G}} {\color{black}\frac{L}{L_0}} \Lambda^{\mathcal{G}}} \frac{1}{n_2} \sum_{j=0}^{n_2-1} \frac{ u_2^{j \mathfrak{p}_2} q_2^{\mathfrak{p}_2}}{1-u_2^{j} q_2} - \mathfrak{q}^{q_{\mathcal{G}} {\color{black}\frac{L}{L_0}} \Lambda^{\mathcal{G}}} \frac{1}{n_2} \sum_{j=0}^{n_2-1} \frac{ u_2^{j( 1-\mathfrak{p}_2)} q_2^{1-\mathfrak{p}_2}}{1-u_2^{j} q_2} \\
    &=   \textcolor{black}{-} \mathfrak{q}^{q_{\mathcal{G}} {\color{black}\frac{L}{L_0}} \Lambda^{\mathcal{G}}} \frac{\mathfrak{q}^{\textcolor{black}{1+}\floor{-\mathfrak{p}_2/n_2}}}{1-\mathfrak{q}}\textcolor{black}{+}\mathfrak{q}^{-q_{\mathcal{G}} {\color{black}\frac{L}{L_0}} \Lambda^{\mathcal{G}}} \frac{\mathfrak{q}^{-\floor{-\mathfrak{p}_2/n_2}}}{1-\mathfrak{q}}\,,
    \end{split}
\end{align}
we now expand this index in powers of $\mathfrak{q}$ (as opposed to $\mathfrak{q}^{\textcolor{black}{-1}}$ at the north pole) following~\cite{Closset:2013sxa,Hama:2011ea, Alday:2013lba}. This expansion gives
\begin{equation}
    I_{\mathcal{U}_2} = \sum_{k = 0}^{\infty} e^{\textcolor{black}{-} \i {\color{black}\frac{L_0}{L}} t \left(k- \floor{-\mathfrak{p}_2/n_2} - {q_{\mathcal{G}} {\color{black}\frac{L}{L_0}} \Lambda^{\mathcal{G}}}\right)} - \sum_{\widetilde{k} = 0}^{\infty} e^{\textcolor{black}{-} \i {\color{black}\frac{L_0}{L}} t \left(\widetilde{k}+1+ \floor{-\mathfrak{p}_2/n_2} + {q_{\mathcal{G}} {\color{black}\frac{L}{L_0}} \Lambda^{\mathcal{G}}}\right)}\,,
\end{equation}
which translates into the infinite product:
\begin{equation}
        Z_2 =  \left( \frac{\prod_{\widetilde{k} = 0}^{\infty} \textcolor{black}{-}\i {\color{black}\frac{L_0}{L}} \left( \widetilde{k} + 1 + \floor*{-\frac{\mathfrak{p}_2}{n_2}} + {q_{\mathcal{G}} {\color{black}\frac{L}{L_0}} \Lambda^{\mathcal{G}}}\right) }{\prod_{k = 0}^{\infty} \textcolor{black}{-} \i {\color{black}\frac{L_0}{L}} \left(k - \floor*{-\frac{\mathfrak{p}_2}{n_2}} - {q_{\mathcal{G}}  {\color{black}\frac{L}{L_0}} \Lambda^{\mathcal{G}}} \right)} \right)^{1/2}\,,  
\end{equation}
which gives the other half of the one-loop determinant. Putting this together with \eqref{eq: one-loop_North} we have
\begin{align}
\begin{split}
     Z^{ \qeq \cV}_{1\text{-loop}} & = Z_1 Z_2 = \left( \frac{\displaystyle \prod_{\widetilde{\ell} = 0}^{\infty} \left( \i {\color{black}\frac{L_0}{L}} \left( \widetilde{\ell} + 1 + \floor*{-\frac{\mathfrak{p}_1}{n_1}} + {q_{\mathcal{G}} {\color{black}\frac{L}{L_0}} \Lambda^{\mathcal{G}}}\right) \right)}{\displaystyle \prod_{\ell = 0}^{\infty} \left(\i {\color{black}\frac{L_0}{L}} \left(\ell - \floor*{-\frac{\mathfrak{p}_1}{n_1}} - {q_{\mathcal{G}} {\color{black}\frac{L}{L_0}}\Lambda^{\mathcal{G}}} \right)\right)} \right. \\
    &\quad \quad \quad \quad \quad \quad \times \left. \frac{\displaystyle \prod_{\widetilde{k} = 0}^{\infty} \left(\textcolor{black}{-}\i {\color{black}\frac{L_0}{L}} \left( \widetilde{k} + 1 + \floor*{-\frac{\mathfrak{p}_2}{n_2}} + {q_{\mathcal{G}} {\color{black}\frac{L}{L_0}} \Lambda^{\mathcal{G}}}\right) \right)}{\displaystyle \prod_{k = 0}^{\infty} \left( \textcolor{black}{-}\i {\color{black}\frac{L_0}{L}} \left(k - \floor*{-\frac{\mathfrak{p}_2}{n_2}} - {q_{\mathcal{G}}  {\color{black}\frac{L}{L_0}}\Lambda^{\mathcal{G}}} \right)\right)} \right)^{1/2}\,,
    \end{split}
\end{align}
which needs to be suitably regularised in order to give the correct one-loop determinant. We use Zeta function regularisation of infinite products as in \cite{e684b263-94c7-31f3-8829-20ce2a1af791} and in particular we make use of the regularisation 
\begin{equation}
 \prod_{n=0}^{\infty} \left( \frac{n+x}{Y} \right) = \frac{Y^{-\frac{1}{2} + x}}{\Gamma(x)}\,,
\end{equation}
which gives 
\begin{align}
\begin{split}
    Z^{ \qeq \cV}_{1\text{-loop}}  & = \textcolor{black}{ \left(\frac{L}{L_0}\right)^{1-\frac{r}{2} \chi + 2 q_{\mathcal{G}} \gamma_{\mathcal{G}} - \mathfrak{c}}} (-1)^{\frac{1}{2}\floor*{- \frac{\mathfrak{p}_2}{n_2}}\textcolor{black}{-}\frac{1}{2}\floor*{- \frac{\mathfrak{p}_1}{n_1}}} \\ 
    & \quad \quad \times\left( \frac{\Gamma\left( - \floor*{- \frac{\mathfrak{p}_1}{n_1}}-{q_{\mathcal{G}}  {\color{black}\frac{L}{L_0}}  \Lambda^{\mathcal{G}}}\right)  \Gamma\left( - \floor*{- \frac{\mathfrak{p}_2}{n_2}}-{q_{\mathcal{G}}  {\color{black}\frac{L}{L_0}}\Lambda^{\mathcal{G}}}\right)} {\Gamma\left(1 + \floor*{- \frac{\mathfrak{p}_1}{n_1}}+{q_{\mathcal{G}}{\color{black}\frac{L}{L_0}} \Lambda^{\mathcal{G}}}\right)\Gamma\left(1 + \floor*{- \frac{\mathfrak{p}_2}{n_2}}+{q_{\mathcal{G}}{\color{black}\frac{L}{L_0}}\Lambda^{\mathcal{G}}}\right)} \right)^{1/2}\,, 
    \end{split}
\end{align}
upon which we can apply Euler's reflection formula for Gamma functions
\begin{equation}
    \Gamma(z)\Gamma(1-z) = \frac{\pi}{\sin(\pi z)}\, ,\quad z \notin \mathbb{Z}\,,
\end{equation}
and write
\begin{align}
\begin{split}
    Z^{ \qeq \cV}_{1\text{-loop}} &  = \textcolor{black}{}  \textcolor{black}{ \left(\frac{L}{L_0}\right)^{1-\frac{r}{2} \chi + 2 q_{\mathcal{G}} \gamma_{\mathcal{G}} - \mathfrak{c}}}\frac{\Gamma\left( - \floor*{- \frac{\mathfrak{p}_1}{n_1}}-{q_{\mathcal{G}} {\color{black}\frac{L}{L_0}}\Lambda^{\mathcal{G}}}\right) } {\Gamma\left(1 + \floor*{- \frac{\mathfrak{p}_2}{n_2}}+{q_{\mathcal{G}} {\color{black}\frac{L}{L_0}} \Lambda^{\mathcal{G}}}\right)} \\
    & = \textcolor{black}{  \textcolor{black}{ \left(\frac{L}{L_0}\right)^{1-\frac{r}{2} \chi + 2 q_{\mathcal{G}} \gamma_{\mathcal{G}} - \mathfrak{c}}}}\frac{\Gamma\left( \frac{\mathfrak{b}+\mathfrak{c}-1}{2} + \frac{r}{4} \chi  - {q_{\mathcal{G}} \gamma_{\mathcal{G}}} \right) }{\Gamma\left( \frac{\mathfrak{b}-\mathfrak{c}+1}{2} - \frac{r}{4} \chi + {q_{\mathcal{G}} \gamma_{\mathcal{G}}} \right) }\,,
    \end{split}
\end{align}
in precise agreement with the one-loop result arising from the method of unpaired eigenvalues \eqref{eq: one-loop-eigenmodes}. As mentioned before, this agreement provides us with a non-trivial test for the validity and consistency of our computations.
\section{Partition function} \label{sec: Mirror Symmetry}
\subsection{\texorpdfstring{$U(1)$}{u(1)} gauge theory with a charged chiral multiplet} \label{sec:evaloffullpartfunc}

Recalling \eqref{eq: partition_function_general}, the partition function for a $U(1)$ gauge theory on $\mathbb{\Sigma}$ takes the form 
\begin{align} \label{eq: partition_function_FI}
\begin{split}
      Z_{\mathbb{\Sigma}} & = \textcolor{black}{ \left( \frac{L}{L_0} \right)^{1-\frac{r}{2} \chi}} \sum_{\mathfrak{m} \in \mathbb{Z}} \left( e^{-\mathfrak{m}\left( \frac{2(n_2-n_1)\pi \xi}{3n_1n_2(n_1+n_2)} + \frac{\i \theta}{n_1 n_2}  \right)} \textcolor{black}{ \left( \frac{L}{L_0} \right)^{q_{\mathcal{G}}\mathfrak{m} \frac{n_2-n_1}{3n_1n_2 (n_1+n_2)}-\mathfrak{c}}} \right. \\
      & \quad \quad \int_{-\infty}^{\infty}  \frac{\textrm{d} \color{black}{(L \sigma_0)}}{2\pi} ~  \left. \textcolor{black}{ \left( \frac{L}{L_0} \right)^{2\i q_{\mathcal{G}} L \sigma_0}} e^{-4 \pi \i \xi \textcolor{black}{L} \sigma_0} 
      \frac{\Gamma\left(\frac{\mathfrak{b}-1}{2}+\frac{\mathfrak{c}}{2} + \frac{r}{4} \chi  -  q_{\mathcal{G}} \gamma_{\mathcal{G}} \right) }{\Gamma\left( 1+ \frac{\mathfrak{b}-1}{2} - \left(\frac{\mathfrak{c}}{2} + \frac{r}{4} \chi - q_{\mathcal{G}} \gamma_{\mathcal{G}} \right) \right) } \right)\,, 
      \end{split}
\end{align}where the vector multiplet one-loop determinant is trivial for abelian theory and hence the partition function is determined by the classical contributions and the chiral multiplet one-loop determinant. We note again that our field configuration on the BPS locus is entirely real and thus we are not integrating over the real part of a complex moduli space, in contrast with \cite{Inglese:2023tyc}. We also note that the FI parameter $\xi$ gets renormalised as
\begin{equation} \label{eq: fi_ren}
    {\color{black}\xi_{\text{ren}} = \xi - \frac{q_{\mathcal{G}}}{2\pi} \ln \left( \frac{L}{L_0} \right)}
\end{equation}
which allows us to absorb several of the $L/L_0$ factors which appear in the integral above, namely
\begin{align}
\begin{split} \label{eq: Z_xi_ren}
    Z_{\mathbb{\Sigma}} & = \textcolor{black}{ \left( \frac{L}{L_0} \right)^{1-\frac{r}{2} \chi}}  \sum_{\mathfrak{m} \in \mathbb{Z}} \left( e^{-\mathfrak{m}\left( \frac{2(n_2-n_1)\pi \xi_{\textcolor{black}{\text{ren}}}}{3n_1n_2(n_1+n_2)} + \frac{\i \theta}{n_1 n_2}  \right)} \textcolor{black}{ \left( \frac{L}{L_0} \right)^{-\mathfrak{c}}} \right. \\
    & \quad \quad \quad \quad \quad \quad \quad \quad\quad \left. \int_{-\infty}^{\infty} \frac{\textrm{d}\bar{\sigma}_0}{2\pi} ~ e^{-4\pi \i \xi_{\textcolor{black}{\text{ren}}} \bar{\sigma}_0} \frac{\Gamma\left(\kappa + \upsilon \bar{\sigma}_0 + \iota\right)}{\Gamma\left(1+\kappa - (\upsilon \bar{\sigma}_0 + \iota)\right)} \right)\,, 
    \end{split}
\end{align}
with
\begin{align}
   \label{eq: alpha} \bar{\sigma}_0 = L \sigma_0\,,\qquad \upsilon = - \i  q_{\mathcal{G}}\,, \qquad  \iota  = \frac{\mathfrak{c}}{2}+\frac{r}{4}\chi - \frac{\mathfrak{m}(n_2-n_1)}{6(n_1+n_2)n_1n_2}q_{\mathcal{G}}\,, \qquad   \kappa  = \frac{1}{2}( \mathfrak{b}-1)\,. 
\end{align}    
Note that $\upsilon$ is imaginary whereas $\iota$ and $\kappa$ are real. The coefficients $\{\upsilon, \iota, \kappa \}$ in the above equation can be calculated by using the expression for $\gamma_{\mathcal{G}}$ given in \eqref{eq: gamma_g}.

We will evaluate the above line integral by a contour integral in the complex $\bar{\sigma}_0$ plane. Therefore, the integral we are interested in is 
\begin{align} \label{eq: Z_contour}
\begin{split}
        Z_{\mathbb{\Sigma}} & = \textcolor{black}{ \left( \frac{L}{L_0} \right)^{1-\frac{r}{2} \chi}} \sum_{\mathfrak{m} \in \mathbb{Z}} \left( e^{-\mathfrak{m}\left( \frac{2(n_2-n_1)\pi \xi_{\textcolor{black}{\text{ren}}}}{3n_1n_2(n_1+n_2)} + \frac{\i \theta}{n_1 n_2}  \right)} \textcolor{black}{ \left( \frac{L}{L_0} \right)^{-\mathfrak{c}}} \right. \\
         &\qquad \qquad \qquad \qquad \quad \left. \oint_\mathcal{C} \frac{\textrm{d}\bar{\sigma}_0}{2\pi} ~ e^{-4\pi \i \xi_{\textcolor{black}{\text{ren}}} \bar{\sigma}_0} \frac{\Gamma\left(\kappa + \upsilon \bar{\sigma}_0 + \iota\right)}{\Gamma\left(1+\kappa - (\upsilon \bar{\sigma}_0 + \iota)\right)} \right)\,,
        \end{split}
\end{align}
 with the contour $\mathcal{{C}}$ being closed in the upper or lower half plane. We note the integral has the symmetry property $Z_{\mathbb{\Sigma}}(q_{\mathcal{G}}, \xi_{\text{ren}}, \theta) = Z_{\mathbb{\Sigma}}(-q_{\mathcal{G}},-\xi_{\text{ren}}, -\theta)$, similar to the partition function for $S^2$ given in \cite{Benini:2012ui}. To see this, we use $\iota(q_{\mathcal{G}},\mathfrak{m}) = \iota(-q_{\mathcal{G}}, -\mathfrak{m})$ (same for $\kappa$)  and flip the signs of the dummy variables in both the summation and integration.
To evaluate the above integral, let us recall that 

\begin{enumerate}
    \item Inverse Gamma functions are entire functions, which means that the Gamma functions have no zeroes. They have poles whenever the argument is either 0 or a negative integer. 
    \item Asymptotic series for Gamma function \cite{Gradshteyn:1702455} for large $|z|$:
    \begin{eqnarray} \label{eq: Gamma_asymptotic}
        \log \Gamma(z) = z \log z - z - \frac{1}{2} \log \left( \frac{z}{2 \pi}\right)  + \mathcal{O}\left(\frac{1}{z}\right) \,.
    \end{eqnarray}
    \item Res$_{z=-n} \Gamma(z) = \dfrac{(-1)^n}{n!}\quad \forall ~ n\in \mathbb{Z}_{>0}$\,. 
    \item Res$_{z=z_*} \Gamma(a z + b) = \frac{1}{a}$ Res$_{z=a z_* + b} \Gamma(z)$\,. 
\end{enumerate}
We note that the coefficient $\upsilon$ \eqref{eq: alpha} has sign selected by the sign of $q_{\mathcal{G}}$. This is an important feature in determining how one closes the countour of integration in \eqref{eq: Z_contour}. In order to determine how to close the contour, we stu\textrm{d}y the asymptotic behaviour of the integrand
\begin{equation}
    I_0 = e^{-4\pi \i \xi_{\textcolor{black}{\text{ren}}} \bar{\sigma}_0} \frac{\Gamma\left(\kappa + \upsilon \bar{\sigma}_0 + \iota\right)}{\Gamma\left(1+\kappa - (\upsilon \bar{\sigma}_0 + \iota)\right)}\,, \qquad |\bar{\sigma}_0| \rightarrow \infty\,.
\end{equation}
In order to make use of \eqref{eq: Gamma_asymptotic}, we consider $|\bar{\sigma}_0| \rightarrow \infty$ of
\begin{align}
\begin{split} \label{eq: log_I_0_asymptotic}
    \log I_0 & = - 4\pi \i \xi_{\textcolor{black}{\text{ren}}} \bar{\sigma}_0 + \log \left[ \Gamma(\kappa + \upsilon \bar{\sigma}_0 + \iota) \right] - \log\left[ \Gamma(1+c-(\upsilon \bar{\sigma}_0 + b)) \right]\\
 & = \upsilon \bar{\sigma}_0 \left( \log (\upsilon \bar{\sigma}_0) +  \log (-\upsilon \bar{\sigma}_0 )\right) + \mathcal{O}(\bar{\sigma}_0)\,,
 \end{split}
\end{align}
where we used \eqref{eq: Gamma_asymptotic} in going to the second line. We now write $\bar{\sigma}_0$ in exponential form
\begin{equation}
    \bar{\sigma}_0 = \mathcal{R} e^{\i \varsigma}\,, \qquad \mathcal{R}> 0\,, \qquad \varsigma \in (-\pi, \pi)\,
\end{equation}
and see that the leading order behaviour in \eqref{eq: log_I_0_asymptotic} can be written as 
\begin{equation}
    \log I_0 =2 \upsilon \mathcal{R} e^{\i \varsigma}  \log| \upsilon \mathcal{R} | + \mathcal{O}(\mathcal{R}) = -2 \i q_{\mathcal{G}} \mathcal{R} (\cos \varsigma + \i \sin \varsigma) \log|q_{\mathcal{G}} \mathcal{R}| + \mathcal{O}(\mathcal{R})\,.
\end{equation}
 Recall we are interested in whether or not $I_0 = e^{\log I_0}$ converges as $\mathcal{R} \rightarrow \infty$ and thus we will not need to analyse the imaginary part of the term above as it will simply contribute an oscillating factor. The real part is 
\begin{equation}
    \text{Re}(\log I_0) = 2 q_{\mathcal{G}} \mathcal{R} \sin \varsigma \log | q_{\mathcal{G}} \mathcal{R}| + \mathcal{O}(\mathcal{R})\,,
\end{equation}
so in order to ensure convergence we require  
\begin{equation}
    q_{\mathcal{G}} \sin \varsigma < 0\,,
\end{equation}\begin{subequations}which splits the closure of the contour into two separate cases depending on the sign of $q_{\mathcal{G}}$:
\begin{align}
    q_{\mathcal{G}} & > 0 \implies \sin \varsigma < 0 \iff \varsigma \in (-\pi,0) \implies \text{close in LHP}\,, \\
    q_{\mathcal{G}} & < 0 \implies \sin \varsigma > 0 \iff \varsigma \in (0,\pi) \implies \text{close in UHP}\,.
\end{align}
\end{subequations}
It is particularly important to note that when the convergence of $I_0$ is ensured, it falls off as 
\begin{equation}
    I_0 \sim e^{{2q_{\mathcal{G}} \sin \varsigma}|z|\log|z|} = |z|^{{ 2 q_{\mathcal{G}} \sin \varsigma}|z|}\,,
\end{equation}
 which in particular is faster than $\mathcal{O}(|z|^{-1})$ as required for the part of the contour along the semicircle to vanish \cite{garfken67:math}. We also note that this convergent factor comes entirely from the ratio of Gamma functions which arose from the one-loop determinant factor in the partition function, and thus we do not need to include the non-zero FI parameter in order to ensure convergence.
In summary, via the asymptotic behaviour \eqref{eq: Gamma_asymptotic} we see that we should close the contour of integration in the upper-half-plane when $q_{\mathcal{G}} < 0$ and the lower-half-plane when $q_{\mathcal{G}}>0$. 

Following point 1 above, the Gamma function in the numerator of the integrand of \eqref{eq: Z_contour} has poles at 
\begin{equation}
   \bar{\sigma}_0 = \sigma^{(Num)}_* = -\frac{1}{\upsilon}\left(k + \iota + \kappa\right)\,,\quad \forall ~ k \geq 0, \quad k \in \mathbb{Z}\,, 
    \end{equation}
and the Gamma function in the denominator of the integrand has poles at 
    \begin{equation}
    \bar{\sigma}_0 = \sigma^{(Den)}_* = \frac{1}{\upsilon}\left(h - \iota + \kappa +1\right)\,,\quad \forall ~ h\geq 0, \quad h \in \mathbb{Z}\,. 
\end{equation} Now, as argued, there are no zeroes coming from the Gamma function of the denominator. However, the poles coming from the two Gamma functions could effectively cancel and it can be shown that they do. To check when this happens, we set
\begin{eqnarray}
    \sigma^{(Num)}_* = \sigma^{(Den)}_*\,, 
\end{eqnarray} and check if the equation can be satisfied for non-negative integers $k$ and $h$. This implies
\begin{eqnarray}
    h =  - k - 2\kappa -1 \geq 0\,, \qquad h \geq 0\,,
\end{eqnarray} which will only give a non-trivial cancellation when the equation above has solutions for $h$, i.e. when $2\kappa\leq -1$ ($\mathfrak{b}\leq -1$). When $2\kappa \leq -1$ all potential poles from $k=0,1,\ldots, -(2\kappa +1)$ cancel and higher values of $k$ will survive as genuine poles. We can combine the two cases into a single formula for the location of the poles
\begin{eqnarray}
       \bar{\sigma}_0 = \sigma^{(\text{p})}_k &=&  -\frac{1}{\upsilon}\left(k + \iota + |\kappa|\right)\,,\quad \forall ~ k \geq 0\,.
\end{eqnarray} Since $\upsilon$ is imaginary and $b$ and $c$ are real, the poles are all on the  imaginary axis. 

Using points 3. and 4. given above we compute the residue from the numerator in the contour \eqref{eq: Z_contour} as
\begin{equation}
    \text{Res}_{\sigma^{(\text{p})}_k} \Gamma(\kappa+\upsilon \bar{\sigma}_0 + \iota) = \frac{1}{\upsilon} \frac{(-1)^{k+|\kappa|-\kappa}}{(k+|\kappa|-\kappa)!}\,,
\end{equation}
and from the denominator we obtain the term
\begin{equation} \label{eq: Gamma_den_value}
    \Gamma(1+\kappa - (\upsilon \sigma^{(\text{p})}_k + \iota)) = (\kappa + |\kappa| +k )!\,.
\end{equation}
Defining
\begin{equation}
    \zeta \equiv  \frac{2\pi \xi_{\textcolor{black}{\text{ren}}}}{q_{\mathcal{G}}}\,,
\end{equation}
 we can directly evaluate the contour integral, noting that we have to deform the contour in order to include all of the poles for $\iota \leq 0$. We obtain 
\begin{align} \label{eq: Z_contour_computed} 
\begin{split}
Z_{\mathbb{\Sigma}} & = \left( \frac{L}{L_0} \right)^{1-\frac{r}{2} \chi} \frac{\textcolor{black}{1}}{|q_{\mathcal{G}}|} \sum_{\mathfrak{m} \in \mathbb{Z}} \left( e^{-\mathfrak{m}\left( \frac{2(n_2-n_1)\pi \xi_{\text{ren}}}{3n_1n_2(n_1+n_2)} + \frac{\i \theta}{n_1 n_2}  \right)} (-1)^{|\kappa|-\kappa} e^{-2\zeta \iota} \textcolor{black}{ \left( \frac{L}{L_0} \right)^{-\mathfrak{c}}} \right. \\
& \qquad \qquad \qquad \qquad \qquad \quad \left. \sum_{k=0}^{\infty} \frac{(-1)^k (e^{-\zeta})^{2k+2|\kappa|} }{k!(k+2|\kappa|)!} \right)\,.
\end{split}
\end{align}
In order to proceed, we follow \cite{Benini:2012ui} in noting that the Bessel function of the first kind $J_\mu$ may be expressed as 
\begin{equation}
    J_\mu(x) = \sum_{k=0}^{\infty} \frac{(-1)^k}{k!\Gamma(k+\mu+1)} \left( \frac{x}{2} \right)^{2k+\mu} \stackrel{\mu \in \mathbb{Z}}{=} \frac{1}{2\pi} \int_{-\pi}^{\pi} e^{-\i \mu y} e^{\i x \sin y} \, \textrm{d}y\,,
\end{equation}
which is produced via the sum over $k$ in \eqref{eq: Z_contour}. We readily identify $\mu = 2|\kappa|$, $x=2e^{-\zeta}$ and upon use of $J_{-n}(x) = (-1)^n J_n (x)$ for $n \in \mathbb{Z}$ we obtain 
\begin{equation}
    Z_{\mathbb{\Sigma}} =  \left( \frac{L}{L_0} \right)^{1-\frac{r}{2} \chi} \frac{\textcolor{black}{1}}{|q_{\mathcal{G}}|} \sum_{\mathfrak{m} \in \mathbb{Z}} e^{-\mathfrak{m}\left( \frac{2(n_2-n_1)\pi \xi_{\text{ren}}}{3n_1n_2(n_1+n_2)} + \frac{\i \theta}{n_1 n_2}  \right)} e^{-2\zeta \iota} \textcolor{black}{ \left( \frac{L}{L_0} \right)^{-\mathfrak{c}}} J_{2\kappa} (2e^{-\zeta}) \,,
\end{equation}
which can be further simplified using the definition \eqref{eq: alpha} to obtain 
\begin{align} \label{eq:startingpoint}
\begin{split}
    Z_{\mathbb{\Sigma}} & = \textcolor{black}{ \left( \frac{L}{L_0} \right)^{1-\frac{r}{2} \chi}} \frac{e^{-\zeta \frac{r}{2} \chi}}{2\pi |q_{\mathcal{G}}|}  \int_{-\pi}^{\pi} \, \textrm{d}y \sum_{\mathfrak{m} \in \mathbb{Z}} e^{-\mathfrak{m}\frac{\i \theta}{n_1n_2} - \zeta \mathfrak{c} - 2\i y \kappa  + 2\i e^{-\zeta} \sin y } \textcolor{black}{ \left( \frac{L}{L_0} \right)^{-\mathfrak{c}}} \\
& =  \frac{L}{L_0} \frac{e^{-\frac{2\pi\xi}{q_{\mathcal{G}}} \frac{r}{2} \chi}}{2\pi |q_{\mathcal{G}}|}  \int_{-\pi}^{\pi} \, \textrm{d}y \sum_{\mathfrak{m} \in \mathbb{Z}} e^{-\mathfrak{m}\frac{\i \theta}{n_1n_2} - \frac{2\pi\xi}{q_{\mathcal{G}}} \mathfrak{c} - 2\i y \kappa  + 2\i e^{-\zeta} \sin y } \,,
\end{split}
\end{align}
where we note that both $\mathfrak{c}$ and $\kappa$ depend explicitly on $\mathfrak{m}$ and so will be important in performing the summation over $\mathfrak{m}$. Focusing on this sum first, we have 
{\color{black}\begin{align}
\begin{split} \label{eq: Z_m_sum_part}
    &\sum_{\mathfrak{m} \in \mathbb{Z}} e^{-\mathfrak{m}\frac{\i \theta}{n_1n_2} - \frac{2\pi \xi}{q_{\mathcal{G}}} \mathfrak{c} - 2\i y \kappa} = e^{-\i y r \frac{\chi_{-}}{2}} \sum_{\mathfrak{m} \in \mathbb{Z}} e^{-\i \mathfrak{m} \frac{y q_{\mathcal{G}}  +\theta}{n_1n_2} +  \frac{\llbracket - \mathfrak{p}_2 \rrbracket_{n_2}}{n_2} \left(\i y -\frac{2\pi\xi}{q_{\mathcal{G}}}\right) + \frac{\llbracket - \mathfrak{p}_1 \rrbracket_{n_1}}{n_1} \left(-\i y -\frac{2\pi\xi}{q_{\mathcal{G}}}\right)}\,,
    \end{split}
\end{align}}
where we recall $\chi_- = 1/n_1 - 1/n_2$. Note that the first term in the sum on the RHS is all that contributes to the $S^2$ partition function, and the other terms are novel contributions arising for $\mathbb{\Sigma}$. We also note that the partition function is manifestly invariant under the shift $\mathfrak{a}_i \rightarrow 
\mathfrak{a}_i + n_i \delta \mathfrak{a}$ for $i\in \{1,2\}$, $\delta \mathfrak{a} \in \mathbb{Z}$. Physical observables should be invariant under such a shift, a property which was originally discussed below \eqref{eq: a_i_defs}. We need to perform the sum over $\mathfrak{m}$ and a useful technique in doing this is Poisson resummation \cite{Benini:2012ui}:
\begin{equation} \label{eq: Poisson_resummation}
    \sum_{\mathfrak{m} \in \mathbb{Z}} f(\mathfrak{m}) = \sum_{n \in \mathbb{Z}} \widehat{f}(n)\,,
\end{equation}
where 
\begin{equation}
\widehat{f}(n) = \int_{-\infty}^{\infty} f(\mathfrak{m}) e^{-2\i \pi n\mathfrak{ m}} \, \textrm{d}\mathfrak{m}\,,
\end{equation}
essentially the Fourier transform of $f(\mathfrak{m})$. 

Using this approach we will try and evaluate the partition function in the most general case, finding that some convenient parameter choices will be helpful in evaluating the final form explicitly. The general formula for $n_1,n_2,r, q_{\mathcal{G}}$ all included is 
{\color{black}
\begin{align}
\begin{split}
\hspace{-1.5em} Z_{\mathbb{\Sigma}}& =  \frac{L}{L_0} \frac{e^{-\frac{2\pi \xi}{q_{\mathcal{G}}} \frac{r}{2} \chi}}{2\pi |q_{\mathcal{G}}|} \int_{-\pi}^{\pi}  \textrm{d}y \, \left( e^{- \i y r \frac{\chi_-}{2} + 2 \i  e^{-\zeta} \sin y} \right. \\
&\qquad \qquad \qquad \quad \quad \quad \quad\left.\sum_{\mathfrak{m} \in \mathbb{Z}}  e^{- \i \mathfrak{m}\frac{q_{\mathcal{G}} y+\theta}{n_1 n_2}-\frac{\left(\frac{2\pi \xi}{q_{\mathcal{G}}} +\i y\right) \llbracket q_{\mathcal{G}} \mathfrak{m}_1 - \frac{r}{2} \rrbracket_{n_1}}{n_1} - \frac{\left(\frac{2\pi \xi}{q_{\mathcal{G}}} -\i y\right) \llbracket q_{\mathcal{G}} \mathfrak{m}_2 - \frac{r}{2} \rrbracket_{n_2}}{ n_2}} \right)\,. \quad 
\end{split}
\end{align}}The next step is to remove $\mathfrak{m}_{1,2}$ using equation \eqref{eq: a_i_defs} and then we can further split $\mathfrak{m}$ up using 
\begin{equation}
    \mathfrak{m} = \mathfrak{m}' n_1 n_2 + \mathfrak{l}\,, \qquad \mathfrak{m}' \in \mathbb{Z} \, , \qquad \mathfrak{l} = 0\,, \ldots ,\, n_1 n_2 -1\,,
\end{equation}
which gives
{\color{black}\begin{align}
\begin{split}
   Z_{\mathbb{\Sigma}} & =  \frac{L}{L_0}  \frac{e^{-\frac{2\pi \xi}{q_{\mathcal{G}}} \frac{r}{2}}}{2\pi|q_{\mathcal{G}}|} \int_{-\pi}^{\pi}  \textrm{d}y \, \bigg( e^{- \i y r \frac{\chi_-}{2} + 2 \i  e^{-\zeta } \sin y }  \sum_{\mathfrak{m}' \in \mathbb{Z}} e^{- \i \mathfrak{m}'(q_{\mathcal{G}} y+\theta)} \\
   & \qquad \qquad \qquad \qquad \times \sum_{\mathfrak{l}=0}^{n_1n_2-1} e^{  - \i \mathfrak{l}  \frac{q_{\mathcal{G}} y+\theta}{n_1 n_2} -\frac{\left(\frac{2\pi \xi}{q_{\mathcal{G}}} +\i y\right) \llbracket q_{\mathcal{G}} \mathfrak{l} \mathfrak{a}_1 - \frac{r}{2} \rrbracket_{n_1}}{n_1} - \frac{\left(\frac{2\pi \xi}{q_{\mathcal{G}}} -\i y\right) \llbracket q_{\mathcal{G}} \mathfrak{l} \mathfrak{a}_2 - \frac{r}{2} \rrbracket_{n_2}}{ n_2}}  \bigg)\,. 
   \end{split}
\end{align}}
We now apply Poisson resummation \eqref{eq: Poisson_resummation} in order to perform the summation over $\mathfrak{m}'$:
{\color{black}\begin{align}
\begin{split}
   Z_{\mathbb{\Sigma}} & =  \frac{L}{L_0}  \frac{e^{-\frac{2\pi \xi}{q_{\mathcal{G}}} \frac{r}{2} \chi}}{q_{\mathcal{G}}^2}  \int_{-\pi}^{\pi}  \textrm{d}y \, \left( e^{- \i y r \frac{\chi_-}{2} + 2 \i  e^{-\zeta} \sin y}  \sum_{n \in \mathbb{Z}} \delta \left(y + \frac{\theta + 2\pi n}{q_{\mathcal{G}}} \right) \right. \\
   & \qquad \qquad \qquad  \qquad \quad \left. \sum_{\mathfrak{l}=0}^{n_1n_2-1} e^{ - \i \mathfrak{l}  \frac{q_{\mathcal{G}} y+\theta}{n_1 n_2} -\frac{\left(\frac{2\pi\xi}{q_{\mathcal{G}}} +\i y\right) \llbracket q_{\mathcal{G}} \mathfrak{l} \mathfrak{a}_1 - \frac{r}{2} \rrbracket_{n_1}}{n_1} - \frac{\left(\frac{2\pi \xi}{q_{\mathcal{G}}} -\i y\right) \llbracket q_{\mathcal{G}} \mathfrak{l} \mathfrak{a}_2 - \frac{r}{2} \rrbracket_{n_2}}{ n_2} }  \right)\,, 
   \end{split}
\end{align}}
and then perform the integral over $y$
{\color{black}\begin{align}
\begin{split}
   Z_{\mathbb{\Sigma}} & =   \frac{L}{L_0}  \frac{e^{-\frac{r}{2q_{\mathcal{G}}} \left(2\pi \xi \chi - \i  \theta \chi_-\right)}}{q_{\mathcal{G}}^2}   \sum_{n =n^*}^{n^* +|q_{\mathcal{G}}| -1} \bigg( e^{- 2 \i  e^{-\zeta} \sin \left( \frac{\theta + 2\pi n}{q_{\mathcal{G}}} \right)} \\
   &\qquad \times \sum_{\mathfrak{l}=0}^{n_1n_2-1} e^{ \frac{2\pi \i n}{q_{\mathcal{G}}} \left( \floor*{\frac{q_{\mathcal{G}} \mathfrak{l} \mathfrak{a}_2 - \frac{r}{2}}{n_2}} - \floor*{\frac{q_{\mathcal{G}} \mathfrak{l} \mathfrak{a}_1-\frac{r}{2}}{n_1} } \right) -\frac{\left(2\pi \xi -\i \theta\right) \llbracket q_{\mathcal{G}} \mathfrak{l} \mathfrak{a}_1 - \frac{r}{2} \rrbracket_{n_1}}{q_{\mathcal{G}}n_1} - \frac{\left(2\pi \xi +\i \theta \right) \llbracket q_{\mathcal{G}} \mathfrak{l} \mathfrak{a}_2 - \frac{r}{2} \rrbracket_{n_2}}{q_{\mathcal{G}} n_2}}  \bigg)\,,
   \end{split}
\end{align}}
where $n^*$ is an integer determined by the constraint 
\begin{equation}
     \frac{\theta + 2\pi n}{q_{\mathcal{G}}} \in (-\pi, \pi) \,, \,\, \textrm{for} \,\, n \in \{n^*, n^* +1, \ldots, n^*+|q_{\mathcal{G}}| -1\}\,. 
    \end{equation} 
In order to make some progress in evaluating this term, \textit{henceforth} we set $q_{\mathcal{G}} =1$. This brings the partition function into the form
{\color{black}\begin{eqnarray}
    Z_{\mathbb{\Sigma}}  = \frac{L}{L_0} e^{-\frac{r}{2} \left(2\pi \xi \chi - \i  \theta \chi_-\right)}  e^{-2 \i e^{-\zeta} \sin \theta} \sum_{\mathfrak{l}=0}^{n_1n_2-1} e^{- \left(2\pi \xi - \i \theta\right) \frac{\llbracket \mathfrak{l} \mathfrak{a}_1 - \frac{r}{2} \rrbracket_{n_1}}{n_1}- \left(2\pi \xi + \i \theta\right) \frac{\llbracket \mathfrak{l} \mathfrak{a}_2 - \frac{r}{2} \rrbracket_{n_2}}{n_2}} \,,
\end{eqnarray}}
and then we use the following decomposition 
\begin{equation}
    \mathfrak{l} = \mathfrak{l}' n_1 + \mathfrak{o}\,, \qquad \mathfrak{l}' \in \{0, 1, \ldots, n_2-1 \}\,, \qquad \mathfrak{o} \in \{ 0,1, \ldots, n_1-1 \}\,,
\end{equation}
under which we see 
{\color{black}\begin{eqnarray}
    Z_{\mathbb{\Sigma}}  =  \frac{L}{L_0} e^{\frac{r}{2} \left(\i  \theta \chi_-- 2\pi \xi \chi \right)} e^{-2 \i e^{-\zeta} \sin \theta} \sum_{\mathfrak{l}'=0}^{n_2-1} \sum_{\mathfrak{o} = 0}^{n_1-1}e^{ - \left(2\pi \xi - \i \theta\right) \frac{\llbracket \mathfrak{o} \mathfrak{a}_1 - \frac{r}{2} \rrbracket_{n_1}}{n_1}- \left(2\pi \xi + \i \theta\right) \frac{\llbracket \mathfrak{l}' + \mathfrak{o} \mathfrak{a}_2 - \frac{r}{2} \rrbracket_{n_2}}{n_2} } \,,
\end{eqnarray}}
where we again made use of \eqref{eq: a_i_defs}. We notice that the double sum above factorises as 
{\color{black}\begin{eqnarray} \label{eq: Spindle_partition_function_final_sum}
    Z_{\mathbb{\Sigma}}  = \frac{L}{L_0} e^{-\frac{r}{2} \left( \frac{2\pi \xi -\i \theta}{n_1} + \frac{2\pi \xi + \i \theta}{n_2} \right)}  e^{-2 \i e^{-\zeta} \sin \theta} \sum_{k_1=0}^{n_1-1}  e^{- \left(2\pi \xi - \i \theta\right)\frac{k_1}{n_1}}  \sum_{k_2 = 0}^{n_2-1}  e^{-\left(2\pi \xi + \i \theta\right){\frac{k_2}{n_2}}} \,,
\end{eqnarray}}
where in evaluating the summation over $\mathfrak{o}$ we made use of 
\begin{equation}
    \{ \llbracket \mathfrak{o} \mathfrak{a}_1 \rrbracket_{n_1} | \mathfrak{o} = 0,\ldots n_1 -1 \} = \{ \llbracket \mathfrak{o} \rrbracket_{n_1} | \mathfrak{o} = 0,\ldots n_1 -1 \}\,,
\end{equation}
which is true due to the fact that gcd$(\mathfrak{a}_1, n_1) =1$ which can be seen immediately from \eqref{eq: a_i_defs} together with B\'ezout's identity. Using the summation of geometric progressions we arrive at the final result
{\color{black}\begin{equation} \label{eq: partition_function_xi_unren}
     Z_{\mathbb{\Sigma}} =  \frac{L}{L_0}  e^{-\frac{r}{2} \left( \frac{2\pi \xi -\i \theta}{n_1} + \frac{2\pi \xi + \i \theta}{n_2} \right)}  e^{-2 \i  \frac{L}{L_0}  e^{-2\pi \xi} \sin \theta} \frac{(1-e^{-(2\pi \xi-\i\theta)})(1-e^{-(2\pi \xi+\i\theta)})}{(1-e^{-(2\pi \xi-\i\theta)/n_1})(1-e^{-(2\pi \xi+\i\theta)/n_2})}\,,
\end{equation}}
which we can also write entirely in terms of the renormalised FI parameter as
\begin{align} \label{eq:partitionfunc_qg=1}
\begin{split}
   Z_{\mathbb{\Sigma}} & = \textcolor{black}{ \left( \frac{L}{L_0} \right)^{1-\frac{r}{2} \chi}} e^{-\frac{r}{2} \left( \frac{\zeta -\i \theta}{n_1} + \frac{\zeta + \i \theta}{n_2} \right)}  e^{-2 \i e^{-\zeta} \sin \theta} \frac{\left(1-\textcolor{black}{ \frac{L_0}{L}}e^{-\zeta+\i\theta}\right)\left(1-\textcolor{black}{ \frac{L_0}{L}}e^{-\zeta-\i\theta}\right)}{\left(1-\left(\textcolor{black}{ \frac{L_0}{L}}e^{-\zeta+\i\theta}\right)^{1/n_1}\right)\left(1-\left(\textcolor{black}{ \frac{L_0}{L}}e^{-\zeta-\i\theta}\right)^{1/n_2}\right)}\,.
   \end{split}
\end{align}
We note that the dependence on the ratio of scales $L/L_0$ does not appear as an overall factor in \eqref{eq:partitionfunc_qg=1} as it also appears non-trivially in the term on the second line.
As a consistency check, we note that setting $n_1 = n_2 =1$ in  reduces \eqref{eq:partitionfunc_qg=1} to
\begin{equation} \label{eq: S^2_partition_limit}
   Z_{S^2} = Z_{\mathbb{\Sigma}} \big|_{n_1 = n_2 =1} = \textcolor{black}{\left(\frac{L}{L_0}\right)^{1-r}}  e^{-\zeta r}  e^{-2 \i e^{-\zeta} \sin \theta}\,,
\end{equation}
 which, up to an unimportant numerical factor, matches \cite{Benini:2012ui} up to $\theta \rightarrow -\theta$, a sign change which arises from our convention of $\mathfrak{m} = -\mathfrak{m}^{(\text{IMP})}$ \eqref{eq: dictionaryIMP2}. Note that the $L/L_0$ does appear as an overall factor for this case, providing a contrast between the partition functions for $S^2$ and $\mathbb{\Sigma}$.

\subsection{Vanishing classical contribution and vacuum degeneracy}

Our spindle partition function \eqref{eq:partitionfunc_qg=1} allows us to compute some simple subcases, one of which is the $S^2$ limit of $n_{1,2} \rightarrow 1$ as discussed in \eqref{eq: S^2_partition_limit}. Another simple case is that of vanishing action on the BPS locus 
{\color{black}\begin{align}
\begin{split}
  \left. Z_{\mathbb{\Sigma}} \right|_{\xi_{\text{ren}}=\theta=0} & = \textcolor{black}{ \left( \frac{L}{L_0} \right)^{1-\frac{r}{2} \chi}}  \sum_{\mathfrak{m} \in \mathbb{Z}} \left( \textcolor{black}{ \left( \frac{L}{L_0} \right)^{-\mathfrak{c}}} \int_{-\infty}^{\infty} \frac{\textrm{d}\bar{\sigma}_0}{2\pi} \frac{\Gamma\left(\kappa + \upsilon \bar{\sigma}_0 + \iota\right)}{\Gamma\left(1+\kappa - (\upsilon \bar{\sigma}_0 + \iota)\right)} \right)\,,
    \end{split}
\end{align}}{\color{black}which can be obtained directly from \eqref{eq: Z_xi_ren} by setting $\xi_{\text{ren}} = \theta = 0$. Note that we switch the \textit{renormalised} FI parameter off, rather than the bare parameter $\xi$. Even though we introduced the classical FI Lagrangian \eqref{eq: S_FI+top} with parameter $\xi$, the renormalisation of $\xi$ which results from factors in the one-loop determinant means that vanishing classical contribution is taken to be $\xi_{\text{ren}} =0$, see related discussion in \cite{Benini:2016qnm}.}
{\color{black}We can obtain the result of this partition function directly from \eqref{eq:partitionfunc_qg=1}, obtaining  
\begin{equation}
     \left. Z_{\mathbb{\Sigma}} \right|_{\xi_{\text{ren}}=\theta=0} = \left(\frac{L}{L_0}\right)^{1-\frac{r}{2}\chi}\frac{\left(1-\textcolor{black}{ \frac{L_0}{L}} \right)^2}{\left(1-\left(\textcolor{black}{ \frac{L_0}{L}}\right)^{1/n_1}\right)\left(1-\left(\textcolor{black}{ \frac{L_0}{L}}\right)^{1/n_2}\right)}\,,
\end{equation}
where an interesting limit of the above formula is that of $L \rightarrow L_0$, where we find
\begin{equation}
    \lim_{L \rightarrow L_0}  \left. Z_{\mathbb{\Sigma}}  \right|_{\xi_{\text{ren}}=\theta=0} = n_1 n_2\,,
\end{equation}
from which we can observe an increase in vacuum degeneracy relative to the unit round $S^2$ case of $n_1=n_2=1$ \cite{Benini:2012ui, GonzalezLezcano:2023cuh},
and is reminiscent of similar features found for the three-dimensional $\mathbb{\Sigma} \times S^1$ partition functions computed in \cite{Inglese:2023tyc}.}

\subsection{Comment on the structure of the partition function} 
A brief comment is in order about the nature of localisation we have performed here. Clearly, the solutions of the BPS equations, especially as seen from that of the chiral matter in \eqref{eq: Chiral_BPS_locus}, suggest that we are doing localisation in the Coulomb branch. Yet, the full partition function in \eqref{eq:partitionfunc_qg=1} separates nicely into the product of two contributions coming from the north and the south pole respectively. This is reminiscent of a similar phenomenon that occurs in $\mathcal{N}=(2,2)$ theory on the round 2-sphere $S^2$ \cite{Benini:2012ui, Doroud:2012xw} as well as for the $\mathcal{N}=2$ Maxwell-Chern Simons theory on the squashed sphere $S_b^{3}$ as shown in \cite{Pasquetti:2011fj}, whereby the partition function admits a block-wise factorisation receiving contributions only from the fixed points. In the 2d case, this was interpreted as Higgs branch localisation, where the contributions were associated with vortices and anti-vortices at the fixed points.  Although we have not performed localisation on the Higgs branch - specifically, as seen from the vanishing BPS values of the scalars $(\phi, \widetilde \phi)$  in \eqref{eq: Chiral_BPS_locus} - the factorisation of our partition function is suggestive of a similar  mechanism. 

It would also be interesting to see if one can reproduce \eqref{eq:partitionfunc_qg=1} by performing analogous localisation calculations on a singular disc $\mathbb{D}^2$ \cite{Bah:2021mzw, Bah:2021hei}. The factorised form suggests that one may be able to reproduce the same result by gluing the partition functions of two discs (with interior conical singularities of deficit angle $2\pi(1-1/n_{1,2})$ respectively) along their respective boundaries, {\color{black}see \cite{Kim:2025ziz} for a related calculation at the level of the off-shell central charges}. While the general structure of the partition function is suggestive of such a construction, there may be additional subtleties involved due to the additional degrees of freedom that live at the boundary of $\mathbb{D}^2$ \cite{Beem:2012mb, Longhi:2019hdh}. Such an undertaking may involve specifying certain boundary conditions for the fields and altering the form of the allowed terms in the action in order to preserve supersymmetry. We leave these explicit checks as future directions to be pursued.

\section{Conclusions and outlook} \label{sec: conclusions}

In this paper we have performed the first entirely two-dimensional computation of supersymmetric localisation on an orbifold background, allowing us to compute the full partition function for abelian gauge theories defined on the spindle $\mathbb{\Sigma}$. We obtained our 2d theory, in particular the defining functions $G$ and $H$ of the Killing spinor equation \eqref{eq: KSE_functions}, by considering the spindle part of the solution to $D=5$ minimal gauged supergravity \cite{Ferrero:2020laf}. This choice means that our analysis is entirely restricted to theories which preserve supersymmetry on the spindle via the \textit{anti-twist} condition \eqref{eq: R-flux}. We considered an abelian theory of vector and chiral multiplets together with an FI term \eqref{eq: susy_action} and used localisation in order to compute the partition function of such a theory, with our main result being that of a charged chiral multiplet coupled to a \textrm{d}ynamical vector multiplet, given in \eqref{eq:partitionfunc_qg=1}. 

Although our approach in focusing directly on two dimensional $\mathcal{N} =(2,2)$ theories is original, we are able to confirm several of our results with those obtained via the dimensional reduction of the three dimensional $\mathcal{N}=2$ theories considered in \cite{Inglese:2023tyc}. In particular, the one loop determinants for 2d $\mathcal{N}=(2,2)$ theories on $\mathbb{\Sigma}$ were anticipated in equation (3.83) of that paper, a result which precisely matches our equation \eqref{eq: one-loop-eigenmodes}. We also note that the 2d results presented there were entirely computed using the spindle index, and in this work we have provided an additional check via computation of the one-loop determinants using both the index and the unpaired eigenvalue method. 

While our work closely follows that of \cite{Inglese:2023wky, Inglese:2023tyc, Pittelli:2024ugf}, there are several interesting differences in our calculations which may be of note for future work. Firstly, by utilizing the cohomological complex \eqref{eq:cohomologicalvariables} consisting of both chiral and anti-chiral fields together with the formula for the one-loop determinant \eqref{eq:prelogNozm} we are able to entirely fix the factors appearing in the one-loop determinants for the anti-twisted spindle. Factors of $\sigma_{-} = -1$ were ignored in their one-loop results, and have been treated more carefully in this work. {\color{black}In addition to this, we took our spindle to have size $L$, and saw that moduli dependent powers of $\textcolor{black}{L/L_0}$ enter the one-loop determinants.} We observed that a careful accounting of such factors is necessary in computing the correct form of the partition function. The FI parameter gets renormalised \eqref{eq: fi_ren} and $\textcolor{black}{L/L_0}$ enters the final answer \eqref{eq: partition_function_xi_unren} in non-trivial fashion.

Secondly, our fields in both vector and chiral multiplets are taken to be entirely real. This is in contrast to \cite{Inglese:2023tyc}, where the fields (including the metric) are allowed to be complex and the BPS locus is also complex. From the perspective of localisation, the complex BPS locus there appears to correspond to a different choice of localisation scheme, as it still satisfies the 3d $\mathcal{N} =2$ BPS equations of the form \eqref{eq: M_BPS}. One of the main advantages of our real BPS locus is that it is immediately clear how to perform the integration over the moduli space $(\sigma_0 \in \mathbb{R})$ of the vector multiplet BPS locus  in the partition function \eqref{eq: partition_function_FI}. If the moduli space was taken to be complex then it is less clear how to construct the integration contour, with one possibility being to integrate over the real fluctuations as suggested in that work. Despite these difficulties, it would be interesting to see whether a relaxation of the reality conditions on our fields allows us to reproduce the same results as this work via localisation around a complex saddle, or whether such conditions are not necessary in studying 2d $\mathcal{N} = (2,2)$ theories on the spindle. 

There are several generalisations of our work which would be of interest to stu\textrm{d}y in the future. Possibly the most natural direction to take is to perform localisation for the \textit{twisted} spindle, as this work focused solely on the anti-twisted case. One possible way to do this would be to start from the $D=5$ spindle solution to the STU model \cite{Ferrero:2021etw}, which includes both the twist and the anti-twist solutions instead of the minimal gauged model. Alternatively, one could attempt to work even more generically as in \cite{Inglese:2023wky, Inglese:2023tyc, Pittelli:2024ugf} and \textcolor{black}{consider a generic metric on $\mathbb{\Sigma}$ i.e. not inherited from any particular supergravity solution. One could then attempt to solve the Killing spinor equations \eqref{kse} \textit{algebraically} before proceeding with the localisation procedure. In general it should be possible to compute the full partition function for both twisted and anti-twisted $\mathbb{\Sigma}$ without specifying the explicit metric, since the result should only depend on topology and not the local geometric structure. However, we emphasise that our approach of choosing an explicit metric not only makes the computational steps more tractable, but also  allows us to gain an immediate understanding of how our anti-twisted spindle fits into the ``target space'' picture by being a solution to minimal gauged $D=5$ SUGRA.}  

Additional extensions include those of non-abelian theories with a generic gauge group $\mathcal{G}$ (a conceptually straightforward undertaking, as the one-loop determinants are alrea\textrm{d}y given in this work), or localisation in the presence of additional multiplets, following along the lines of analogous calculations considered on $S^2$ \cite{Benini:2012ui, Doroud:2012xw,   Gomis:2012wy}. This would pave the way for applications to Calabi-Yau sigma models (where $\sum q_{\mathcal{G}} =0$) with the spindle (somewhat exotically) playing the role of the string worldsheet.  

On a related note, it would be interesting to investigate whether the spindle partition function can be used to understand the physics of (orbifold) worldsheet instantons. This idea follows from an analogous construction on $S^2$ \cite{Jockers:2012dk, Gomis:2012wy}, where the exact partition function computes the contributions of worldsheet instantons in Calabi-Yau sigma models via Gromov-Witten invariants \cite{Gromov, Witten:1988xj}. The stu\textrm{d}y of Gromov-Witten invariants on orbifolds is a somewhat active area \cite{huang2025gromovwitteninvariantsmathbbp1orbifoldstopological}, and it would be interesting to see if similar results can be obtained using suitable extensions of the partition function computed in this work.

The additional extensions mentioned above will be important in the stu\textrm{d}y of two-dimensional supersymmetric dualities \cite{Hori:2000kt, Hori:2006dk} (see also \cite{Hori:2002fa} and Section 5 of \cite{Benini:2016qnm} for reviews) involving the spindle partition function. A first duality to explore is the abelian Mirror symmetry of our partition function \eqref{eq:partitionfunc_qg=1}, as worked out in \cite{Hori:2000kt}. This duality states that our theory should admit an equivalent ``mirror pair'' description in terms of a twisted chiral multiplet $\Sigma$ as introduced in \eqref{eq: Sigma_multiplet} as well as another twisted chiral multiplet $Y$. It was shown that for supersymmetric theories in flat space the mirror theory is generated by a twisted superpotential of the form
\begin{equation}
    \widehat{W} = \frac{\i}{4\pi} \left[ \Sigma \left(t - \i \theta - q_{\mathcal{G}} Y \right) - \i \mu e^{-Y} \right]\,,  
\end{equation}
which can also be extended to supersymmetric theories on curved manifolds e.g. the $\mathcal{N}=(2,2)$ theories on $S^2$ as considered in \cite{Benini:2012ui}. This duality is checked by evaluating the action for this superpotential \eqref{eq: superpotential_lagrangian} on the BPS locus and finding equivalence with the partition function of an abelian gauge theory with charged chiral multiplet. It would be natural to attempt to arrive at a similar result for $\mathbb{\Sigma}$, although we state that our first attempt at such a calculation leads us to the conclusion that the \textit{twisted superpotential for the mirror dual to the spindle requires modification}. This is perhaps unsurprising due to the fact that $\mathbb{\Sigma}$ is a bad orbifold without a universal cover and thus one would expect the periodicity of the twisted chiral multiplet $Y$ to be suitably modified in order to account for such structure. In addition to this, the non-trivial values of the functions which appear in the Killing spinor equation \eqref{eq: KSE_functions} make the process of conjecturing the form of the twisted superpotential somewhat subtle. It is possible that the works \cite{Hosomichi:2015pia, Okuda:2015yra, Hosomichi:2017dbc} may be helpful in developing the notion of mirror symmetry for the spindle. We leave the details of this duality to future work but feel this represents an interesting avenue for further stu\textrm{d}y, potentially opening a window to a wider variety of ``orbifold'' dualities for supersymmetric field theories.

Finally, it would be interesting to see if there are holographic applications of our spindle partition function, akin to those found in higher dimensions. It is well known that the theories living at the conformal boundary of accelerating asymptotically locally AdS$_4$ black hole solutions \cite{Ferrero:2020twa, Cassani:2021dwa, Kim:2023ncn} are examples of  $\mathcal{N}=2$ field theories on $\mathbb{\Sigma} \times S^1$ and it has recently been shown \cite{Colombo:2024mts} that one can reproduce the accelerating black hole entropy via the large-$N$ limit of the spindle index. In the case of AdS$_3$/CFT$_2$, the situation is far less clear. The dual gravitational description to our $\mathcal{N}=(2,2)$ theory on $\mathbb{\Sigma}$ will consist of asymptotically locally AdS$_3$ solutions with boundary $\mathscr{I} \cong \mathbb{\Sigma}$ but examples of such solutions are not explicitly known. They will certainly be different from the known accelerating black holes in three spacetime dimensions \cite{Astorino:2011mw, Arenas-Henriquez:2022www} which have $\mathscr{I} \cong S^1 \times S^1_{\text{pinched}}$. It may be of interest to perform a  large-$N$ calculation analogous to that of \cite{Colombo:2024mts} for the 2d $\mathcal{N}=(2,2)$ theory on $\mathbb{\Sigma}$ as a stepping stone in constructing the holographic duals, potentially resulting in additional precision checks of entropy counting and a possible quantum origin of accelerating AdS$_3$ black hole entropy.

\acknowledgments

We would like to thank Heng-Yu Chen, Christopher Couzens,  Junho Hong, Seyed Morteza Hosseini, Seok Kim, Johanna Knapp, Dario Martelli, Joseph McGovern, Antonio Pittelli, Matthew Roberts, James Sparks, and Piljin Yi for discussions. This work is supported by the National Research Foundation of Korea under the grants: NRF-2021R1F1A1048531 (I.J.), NRF-2022R1A2B5B02002247 (H.K, N.K, and A.P.), 2021R1A2C2012350 (A.R.), RS-2023-00243491(H.K.), and JRG Program at the Asia Pacific Center for Theoretical Physics (APCTP) through the Science and Technology Promotion Fund and Lottery Fund of the Korean Government (I.J.). A.P. is supported in part by the National Science and Technology Council, R.O.C. (NSTC 112-2112-M-002-024-MY3 and NSTC 113-2112-M-002-040-MY2), and by National Taiwan University. A.P. acknowledges the hospitality of The University of Oxford and The University of Melbourne, as well as the conferences ``Eurostrings 2024'' at The University of Southampton, ``2024 East Asia Joint Workshop on Fields and Strings'' at National Sun Yat-sen University and the ``\nth{12} Bangkok workshop on High-Energy Theory'' at Chulalongkorn University, where preliminary results from this work were presented.  
I.J. acknowledges the hospitality of Junggi Yoon, Matthew Roberts and Kanghoon Lee at APCTP, as well as the conferences ``Ensemble Average Theories in HET'' at Shanghai University and ``Some formal aspects of Field theories and Holography'' at University of Science and Technology of China. 
H.K. thanks the string theory group at National Taiwan University for the hospitality. A.R. acknowledges the hospitality of KIAS as well as the conference on ``New Frontiers in Strings and Field theories 2024'', Republic of Korea where parts of the results were presented.



\appendix

\section{Definitions and conventions} \label{app: Defs}
\subsection{Indices}
We set five-dimensional curved spacetime indices as $M,N,\ldots$ and local indices as $A,B,\ldots$. We denote the indices corresponding to $AdS_3$ direction and $\mathbb{\Sigma}$ direction as  
\begin{equation}
    M = (\alpha\,, \mu)\,, \qquad A = (a\,,m )\,.
\end{equation}

\subsection{Gamma matrix conventions}
In five dimension with Lorentzian signature, a consistent choice of gamma matrix satisfies the following relations:
\begin{equation}\label{5dgammaLor}
\begin{array}{llll}
\Gamma_{A}^{\dagger}\=-A\Gamma_{A}A^{-1}\,,~&~A\=\Gamma_{0}\,,~&~A^{\dagger}\=A^{-1}\=-\Gamma_{0}\,,~&\\
\Gamma_{A}^{T}\={\cal{C}}\Gamma_{A}{\cal{C}}^{-1}\,,~&~{\cal{C}}^{T}\=-{\cal{C}}\,,~&~{\cal{C}}^{\dagger}={\cal{C}}^{-1}\,,~&\\
\Gamma_{A}^{*}\=-B\Gamma_{A}B^{-1}\,,~&~B^{T}\={\cal{C}}A^{-1}\,,~&~B^{\dagger}\=B^{-1}\,,~~~~B^{*}B\=-1\,.
\end{array}\end{equation}
This is followed by the property, regarding the charge conjugation matrix $\cC$, 
\begin{equation}
({\cal{C}}\Gamma_{A_1 A_2\cdots A_p})^{T}= - (-)^{p(p-1)/2}{\cal{C}}\Gamma_{A_1 A_2\cdots A_p}\,.
\end{equation} 
Due to the property of the charge conjugation matrix, we can use the spinor representation satisfying the symplectic-Majorana condition
\be\label{SMLor}
(\psi^i)^\dagger \gamma_0 \=  \varepsilon_{ij}(\psi^j)^T \mathcal{C}\,,\quad \Leftrightarrow \quad (\psi^i)^\ast = \varepsilon_{ij} B \psi^j\,, 
\ee
where~$i$ is an SU(2)$_R$ index
with $\varepsilon_{ij}$ being the SU(2) symplectic metric $\varepsilon_{12} = -\varepsilon_{21} = 1$.\\

\section{Coordinates on the spindle} \label{app: coordinates}

In this Appendix we demonstrate that the spindle has conical singularities of deficit angle $2\pi(1-1/n_{1,2})$ at the north and south poles $y_{1,2}$ respectively. To do this, we give the mapping between the coordinates \eqref{eq: spindle_metric} and those used in the works \cite{Inglese:2023wky, Inglese:2023tyc, Pittelli:2024ugf}, where the metric on ($L=1$) $\mathbb{\Sigma}$ takes the form 
\begin{equation} \label{eq: spindle_theta_phi}
    \textrm{d}s^2_{\mathbb{\Sigma}} = f^2(\theta) \textrm{d}\theta^2 + \sin^2 \theta \textrm{d}\varphi^2\,, \qquad f(0) = n_1\,, \quad f(\pi) = n_2\,, 
\end{equation}
where $\theta \in [0, \pi]$ and $\varphi \in [0, 2\pi)$.
In order to map between the coordinates, we note immediately the straightforward relation between the axial coordinates 
\begin{equation}
    \varphi = \frac{2\pi}{\Delta z} z\,, 
\end{equation}
which leaves us with a relation between the azimuthal coordinates of 
\begin{equation} \label{eq: y_theta_map}
    \frac{q}{36 y^2} \left( \frac{\Delta z}{2\pi} \right)^2 = \sin^2 \theta\,. 
\end{equation}
Comparing the line elements we can also read off $f$ via 
\begin{equation} \label{eq: y_f_relation}
    \frac{y}{q} \textrm{d}y^2 = f^2(\theta) \textrm{d}\theta^2\,, 
\end{equation}
which when combined with \eqref{eq: y_theta_map} allows us to obtain 
\begin{equation}
    f^2(\theta(y)) = \frac{144}{y} \left( \frac{2\pi}{\Delta z} \right)^2 \left[ \partial_y \left( \frac{q}{y^2} \right)\right]^{-2}  \left( 1 - \frac{q}{36y^2} \left( \frac{\Delta z}{2\pi} \right)^2 \right)\,.  
\end{equation}

One can check that this $f$ satisfies the desired properties at the poles of the spindle given in \eqref{eq: spindle_theta_phi}, namely
\begin{align}
    f^2(0)  = f^2 (\theta(y_1)) = 9 \left( \frac{2\pi}{\Delta z} \right)^2 \frac{y_1^3}{(y_1-y_2)^2\left( y_1 - \frac{a^2}{4 y_1 y_2} \right)^2} = n_1^2\,, 
\end{align}
    and
\begin{align}
    f^2(\pi)  = f^2 (\theta(y_2)) = 9 \left( \frac{2\pi}{\Delta z} \right)^2 \frac{y_2^3}{(y_2-y_1)^2\left( y_2 - \frac{a^2}{4 y_1 y_2} \right)^2} = n_2^2\,. 
\end{align}

\section{\texorpdfstring{$AdS_3$}{ads3} Killing spinor equation}
\label{KSEAdS3}
In the main text we only made use of the components in the $\mathbb{\Sigma}$ directions, resulting in the Killing spinor equation for the spindle \eqref{KSSpindle1}. In this appendix we stu\textrm{d}y all other components of the Killing spinor equation \eqref{5dkse}, namely the components along the $AdS_3$ directions. We will demonstrate
\begin{equation} \label{eq: KSE_AdS_3}
     \left[ \nabla_{\alpha} -\frac{\i}{12}\left(\Gamma_{\alpha}{}^{N P} - 4 \delta_{\alpha}^N \Gamma^P \right)F^{5d}_{NP} -\frac{1}{2 {\color{black}{L}}}\Gamma_{\alpha} - \frac{\i}{L}  A^{5d}_{\alpha} \right]\varepsilon =0\,,
\end{equation}
and thus the full Killing spinor equation \eqref{5dkse} is solved, when the Killing spinor takes the form 
\begin{equation} \label{eq: spinor_decomposition}
    \varepsilon = \vartheta \otimes \chi\,, 
\end{equation}
where $\vartheta$ is a Killing spinor for $AdS_3$ i.e.\footnote{There is a difference in sign in the $AdS_3$ part of the Killing spinor equation relative to \cite{Ferrero:2020laf} which originates from the differences in our choices of $\Gamma$-matrices. We can recover the typical sign in the $AdS_3$ Killing spinor equation by taking $\rho^a \rightarrow -\rho^a$, an alternative choice of 3-dimensional gamma matrices.} 
\begin{equation} \label{eq: ads_3_kse}
     \left( \bar{\nabla}_{\alpha} + \frac{1}{2  {\color{black}{L}}} \bar{\rho}_{\alpha} \right) \vartheta  = 0\,,
\end{equation}
where bars denote $AdS_3$ quantities i.e. $\bar{\nabla}$ is the $AdS_3$ covariant derivative and $\bar{\rho}_{\alpha} =  \bar{e}_{\alpha}^a \rho_a$ where $\bar{e}^a_{\alpha}$ form a basis of connection 1-forms for a {\color{black}radius $L$} $AdS_3$. 

We start with the left hand side of \eqref{eq: KSE_AdS_3}  
\begin{equation} \label{eq: KSE_AdS_3_LHS}
     \left[ \nabla_{\alpha} -\frac{\i}{12}\left(\Gamma_{\alpha}{}^{N P} - 4 \delta_{\alpha}^N \Gamma^P \right)F^{5d}_{NP} -\frac{1}{2 {\color{black}{L}}}\Gamma_{\alpha} - \frac{\i}{L} A^{5d}_{\alpha} \right]\varepsilon \,,
\end{equation}
as our aim is to manipulate this into showing that it vanishes. Using \eqref{eq: singular_gauge_field}, we immediately see that 
\begin{equation}
    A^{5d}_{\alpha} = F^{5d}_{\alpha P} =0\,,
\end{equation}
and thus \eqref{eq: KSE_AdS_3_LHS} immediately simplifies to 
\begin{equation} \label{eq: KSE_AdS_3_2}
     \left[ \nabla_{\alpha} -\frac{\i}{12}\Gamma_{\alpha}{}^{N P} F^{5d}_{NP} -\frac{1}{2 {\color{black}{L}}}\Gamma_{\alpha}  \right]\varepsilon \,.
\end{equation}
We now note that 
\begin{equation} \label{eq: Gamma_F_product_intermediate}
    \Gamma_{\alpha}{}^{N P} F^{5d}_{NP} = 2 g_{\alpha \beta} \Gamma^{\beta y z} F^{5d}_{y z} = 2 g_{\alpha \beta} \Gamma^{\beta 3 4} F^{5d}_{3 4} = 2 g_{\alpha \beta} \Gamma^{\beta} \Gamma^3 \Gamma^4 F^{5d} = 2\Gamma_{\alpha} \Gamma^3 \Gamma^4 F^{5d}\,,
\end{equation} 
and using \eqref{GammaSplit} one can write $\Gamma_{\alpha}{}^{N P} F^{5d}_{NP} = 2 \i F^{5d} ( \rho_{\alpha} \otimes 1 )$. Using \eqref{eq: Gamma_F_product_intermediate}, \eqref{eq: KSE_AdS_3_2} reduces to
\begin{equation} \label{eq: KSE_AdS_3_3}
     \left[ \nabla_{\alpha} -\frac{\i}{6}\Gamma_{\alpha} \Gamma^{34} F^{5d} -\frac{1}{2 {\color{black}{L}}}\Gamma_{\alpha}  \right] \varepsilon\,.
\end{equation}
Recall that the covariant derivative of a spinor is given by
\begin{equation} \label{eq: spinor_covariant}
    \nabla_{\alpha}  \varepsilon = \left( \partial_{\alpha}  + \frac{1}{4} \omega_{\alpha \, A B} \Gamma^A \Gamma^B \right) \varepsilon\,,
\end{equation}
and thus we need to identify all components of the spin connection 1-form along the $AdS_3$ directions. We start by identifying a basis of connection 1-forms as
\begin{equation}
    e_{\alpha}^a =  \frac{2}{3} y^{1/2} \bar{e}_{\alpha}^a\,, \qquad e^3_y =  {\color{black}{L}} \left( \frac{y}{q} \right)^{1/2}\,, \qquad e^4_z =  {\color{black}{L}} \frac{q^{1/2}}{6y}\,,
\end{equation}
and extract the components of the spin connection via 
\begin{equation} \label{eq: spin_connection_structure}
    \textrm{d}e^{A} + \omega^A{}_B \wedge e^B = 0\,.
\end{equation}
The $AdS_3$ tangent space directions of \eqref{eq: spin_connection_structure} are 
\begin{equation}
    \textrm{d}e^a + \omega^a{}_B \wedge e^B =0\,,
\end{equation}
which explicitly give
\begin{equation}
    \frac{1}{3} y^{-1/2} \textrm{d}y \wedge \bar{e}^a +  \frac{2}{3} y^{1/2}  \textrm{d} \bar{e}^a + \frac{2}{3} y^{1/2} \omega^a{}_b \wedge \bar{e}^b + {\color{black}L} \left( \frac{y}{q} \right)^{1/2} \omega^a{}_3 \wedge \textrm{d}y + {\color{black}L} \frac{q^{1/2}}{6y} \omega^a{}_4 \wedge \textrm{d}z = 0\,,
\end{equation}
from which we can directly read off 
\begin{equation} \label{eq: spin_conn_AdS}
    \omega^a{}_b = \bar{\omega}^a{}_b\,, \qquad  \omega^a{}_3 = {\color{black}\frac{1}{L}} \frac{q^{1/2}}{3y} \bar{e}^a\,, \qquad   \omega^a{}_4 =0\,. 
\end{equation}
Next we stu\textrm{d}y the components of \eqref{eq: spin_connection_structure} along the directions tangent to $\mathbb{\Sigma}$. First we stu\textrm{d}y the 3-direction 
\begin{equation}
    \textrm{d}e^3 + \omega^3{}_B \wedge e^B =0\,,
\end{equation}
which explicitly gives 
\begin{equation}
   \omega^3{}_4 \wedge \textrm{d}z = 0 \implies \omega^3{}_4 = f_0 \textrm{d}z\,, 
\end{equation}
and finally the function $f_0$ can be obtained by stu\textrm{d}ying the component of \eqref{eq: spin_connection_structure} along the 4-direction
\begin{equation}
      \textrm{d}e^4 + \omega^4{}_B \wedge e^B =0\,,
\end{equation}
which gives
\begin{equation}
    f_0  = -  \left( \frac{q}{y} \right)^{1/2} \left( \frac{q^{1/2}}{6y} \right)' = - \frac{a^2 - 3a y + 2 y^3}{6 y^{5/2}}\,,
\end{equation}
in agreement with \eqref{eq: spin_conn_spindle} in the main text. In summary, the non-vanishing components of the spin connection are 
\begin{equation}
     \omega^a{}_b = \bar{\omega}^a{}_b\,, \qquad \omega^a{}_3 = {\color{black}\frac{1}{L}} \frac{q^{1/2}}{3y} \bar{e}^a\,, \qquad \omega^3{}_4 = -  \left( \frac{q}{y} \right)^{1/2} \left( \frac{q^{1/2}}{6y} \right)' \textrm{d}z\,.
\end{equation}

Returning to the examination of the covariant derivative in \eqref{eq: spinor_covariant}, we only need to make use of the components of the spin connection along the $AdS_3$ directions as given in \eqref{eq: spin_conn_AdS}. We find
\begin{equation}
    \nabla_{\alpha}  \varepsilon  = \left( \partial_{\alpha}  + \frac{1}{4} \bar{\omega}_{\alpha \, a b} \Gamma^a \Gamma^b + \frac{1}{2} \omega_{\alpha \, a 3} \Gamma^a \Gamma^3  \right) \varepsilon\,, 
\end{equation}
and thus the relevant components of \eqref{eq: KSE_AdS_3_3} are
\begin{equation}
     \left( \partial_{\alpha}  + \frac{1}{4} \bar{\omega}_{\alpha \, a b} \Gamma^a \Gamma^b + \frac{1}{2} \omega_{\alpha \, a 3} \Gamma^a \Gamma^3 -\frac{\i}{6}\Gamma_{\alpha} \Gamma^{34} F^{5d} -\frac{1}{2 {\color{black}{L}}}\Gamma_{\alpha}   \right) \varepsilon\,.
\end{equation}
and we decompose the $\Gamma$-matrices using \eqref{GammaSplit} and Killing spinor via \eqref{eq: spinor_decomposition} to obtain 
\begin{equation}
    \left( \partial_{\alpha}  + \frac{1}{4} \bar{\omega}_{\alpha \, a b} (\rho^a \rho^b \otimes 1) - \frac{1}{2 {\color{black}{L}}} \bar{\rho}_{\alpha} \otimes \left( -\i \frac{q^{1/2}}{3y} \gamma^1 - \frac{2}{9} F y^{1/2} \cdot \mathbb{1} + \frac{2}{3}y^{1/2}  \gamma^3 \right) \right) (\vartheta \otimes \chi)\, , 
\end{equation}
now using the integrability condition \eqref{eq: integrability_explicit_tilde}
\begin{equation}
    \left( -\i \frac{q^{1/2}}{3y} \gamma^1 - \frac{2}{9} F y^{1/2} \cdot \mathbb{1} + \frac{2}{3}y^{1/2}  \gamma^3 \right)\chi = -\chi\,,
\end{equation}
we directly find
\begin{equation}
     \left( \partial_{\alpha}  + \frac{1}{4} \bar{\omega}_{\alpha \, a b} (\rho^a \rho^b \otimes 1) + \frac{1}{2 {\color{black}{L}}} ( \bar{\rho}_{\alpha} \otimes 1) \right) (\vartheta \otimes \chi)\,,
\end{equation}
 the $\vartheta$ part of which gives 
\begin{equation}
     \left( \partial_{\alpha}  + \frac{1}{4} \bar{\omega}_{\alpha \, a b} \rho^a \rho^b  + \frac{1}{2 {\color{black}{L}}} \bar{\rho}_{\alpha} \right) \vartheta = \left( \bar{\nabla}_{\alpha} + \frac{1}{2 {\color{black}{L}}} \bar{\rho}_{\alpha} \right) \vartheta  = 0\,,
\end{equation}
using the Killing spinor equation for $AdS_3$ \eqref{eq: ads_3_kse} which we assumed at the outset. We have demonstrated that all components of the full 5d Killing spinor equation \eqref{5dkse} are solved when $\vartheta$ is an $AdS_3$ Killing spinor and $\chi$ is a $\mathbb{\Sigma}$ Killing spinor. We note that this derivation appeared to use the integrability condition \eqref{eq: integrability_explicit_tilde} in order to arrive at the $AdS_3$ Killing spinor equation, although one can equivalently reverse this argument by using the assumption \eqref{eq: ads_3_kse} in order to derive the integrability condition \eqref{eq: integrability_explicit_tilde} as a necessary condition for \eqref{eq: KSE_AdS_3} to hold. This in fact was the logic imposed in the original derivation of these Killing spinors, see Appendix C of \cite{Ferrero:2021etw} for more details.

\section{Twisted superpotential} \label{sec: twisted_superpotential}
As we consider an abelian vector multiplet on $\mathbb{\Sigma}$ we are able to consider supersymmetric actions generated by a \textit{twisted superpotential} \cite{Benini:2012ui, Closset:2014pda}. We illuminate the general structure of such a term on $\mathbb{\Sigma}$, demonstrating that our choice of \eqref{eq: S_FI+top} in the main text arises as a particular choice of twisted superpotential. In order to construct a general twisted superpotential, we first write the vector multiplet in terms of a twisted chiral multiplet $\Sigma$, whose components are \cite{Closset:2014pda}
\begin{equation} \label{eq: Sigma_multiplet}
\Sigma = \left(\sigma + \i \rho, -\widetilde{\lambda}_+, -\lambda_-, -\widehat{D}- \i \mathcal{F} + (G+ \i H)(\sigma + \i \rho)\right)\,,  
\end{equation}
and similarly, we have the twisted anti-chiral multiplet
\begin{equation}
\widetilde{\Sigma} = \left(\sigma - \i \rho, \lambda_+ , \widetilde{\lambda}_- , \widehat{D}- \i \mathcal{F} + (G- \i H)(\sigma - \i \rho)\right)\,.
\end{equation}

Following \cite{Benini:2012ui, Closset:2014pda}, we can write down the general form of the Lagrangian from a twisted superpotential $\widehat{W}(\Sigma)$ which is a holomorphic function of $\Sigma$ and an anti-holomorphic function $\widetilde{\widehat{W}}(\widetilde{\Sigma})$.\footnote{The general expression is given in equation (6.68) of \cite{Closset:2014pda} and in order to match this to our notation we use the dictionary given in footnote 3 of \cite{GonzalezLezcano:2023cuh}. In particular, we note the dictionary between our twisted superpotential $\widetilde{W}$ and the twisted superpotential $\widetilde{\widehat{W}}$ of \cite{Closset:2014pda} 
\begin{equation} \label{eq: W_conventions}
    \widetilde{W}(-\sigma + \i \rho) =  \i \widetilde{\widehat{W}} (\sigma - \i \rho)\,, \qquad \widetilde{W}^*(-\sigma - \i \rho) = - \i \widehat{W} (\sigma + \i \rho)\,,
\end{equation}where the opposite signs in the arguments are important in order to get the derivatives to match.} We note that the general form of the action is
\begin{equation} \label{eq: superpotential_lagrangian}
   S_{W} = \int \textrm{d}^2x \, \sqrt{g_{\mathbb{\Sigma}}} \mathcal{L}_{W} = \int_{\mathbb{\Sigma}} \textrm{d}^2 x  \, \sqrt{g_{\mathbb{\Sigma}}} \left(\mathcal{L}_{\widehat{W}} + \mathcal{L}_{\widetilde{\widehat{W}}} \right)\,,
\end{equation}
where we follow the conventions of \cite{Closset:2014pda} in writing
\begin{equation} \label{eq: Lagrangian_W}
    \mathcal{L}_{\widehat{W}} = \left( -\widehat{D}- \i \mathcal{F} + (G+ \i H)(\sigma + \i \rho) \right) \partial \widehat{W} - \i (H - \i G) \widehat{W}  - \frac{\i}{2} \widetilde{\lambda} (1+\gamma_3) \lambda \partial^2 \widehat{W}\,, 
\end{equation}
and 
\begin{equation} \label{eq: Lagrangian_W_tilde}
    \mathcal{L}_{\widetilde{\widehat{W}}} = \left( \widehat{D}- \i \mathcal{F} + (G - \i H)(\sigma - \i \rho) \right) \widetilde{\partial} \widetilde{\widehat{W}} + \i (H + \i G) \widetilde{\widehat{W}}+\frac{\i}{2} \widetilde{\lambda}(1-\gamma_3) \lambda\widetilde{\partial}^2 \widetilde{\widehat{W}}\,, 
\end{equation}
where we used $\widetilde{\lambda}_+ \lambda_- = - \frac{\i}{2} \widetilde{\lambda} (1+\gamma_3) \lambda$ together with $\lambda_+ \widetilde{\lambda}_- = -\frac{\i}{2} \widetilde{\lambda}(1-\gamma_3) \lambda$. We note that $\widehat{W}$ is a holomorphic function of $(\sigma + \i \rho)$ and $\widetilde{\widehat{W}}$ is a holomorphic function of $(\sigma - \i \rho)$.

In the our localisation calculation we make use of a particularly simple choice of twisted superpotential, namely that of the Fayet-Iliopoulos Lagrangian \eqref{eq: S_FI+top}. This supersymmetric Lagrangian is obtained from the following choice of twisted superpotential
\begin{equation} \label{eq: FI_choices}
    \widehat{W} = \frac{1}{2} \tau \Sigma\,, \qquad \tau = \i \xi - \frac{\theta}{2\pi}\,, \qquad \widetilde{\widehat{W}} = \bar{\widehat{W}}\,,
\end{equation}
where the bar denotes complex conjugation.

\section{Chiral multiplet BPS locus} \label{sec: CM_BPS}

In this appendix we show that for a generic vector multiplet solution \eqref{eq: BPS_locus_real}, the only regular solution for the chiral BPS locus is that given in \eqref{eq: Chiral_BPS_locus}.  We solve the equations of motion for the action \eqref{eq: CM_deformation_action}, recalling that the sum of squares structure means that each term must vanish individually. Using this logic, we immediately note that the BPS value for the auxiliary scalar is
\begin{equation}
    \mathfrak{F} = \widetilde{\mathfrak{F}} = 0\,,
\end{equation}
and the remaining differential equation for the complex scalar $\phi$ has to be solved. Note that this will immediately give the solution for $\widetilde{\phi}$ from the reality condition \eqref{eq: chiral_bosonic_reality}. We start with the equation 
\begin{equation} \label{eq: phi_bps_equation}
    \left(\i \gamma^{\mu} D_{\mu} \phi - \i \left( \sigma + \frac{r}{2} H \right) \phi + \gamma_3 \left( \rho + \frac{r}{2} G \right) \phi \right) \epsilon = 0,
\end{equation}
and multiply by the Killing spinor $\widetilde{\epsilon}$. Upon doing this we obtain 
\begin{equation} \label{eq: phi_bps_epsilon_tilde_mult}
   \frac{1}{L} D_z \phi - \frac{\sqrt{y}}{3}\left(\sigma + \frac{r}{2} H \right) \phi - \i \frac{3y -a}{6y}\left( \rho + \frac{r}{2} G \right) \phi = 0\,,
\end{equation}
which can be manipulated into the form 
\begin{equation}
    \frac{1}{L}\left(\partial_z  -  \i r A_{z} - \i \mathcal{A}_z \right) \phi  - \frac{\sqrt{y}}{3}\left(\sigma + \frac{r}{2} H \right) \phi - \i \frac{3y -a}{6y}\left( \rho + \frac{r}{2} G \right) \phi = 0\,.
\end{equation}
We now plug in the values of the fields $A, \mathcal{A}, \sigma, H, \rho, G$ and we find 
\begin{equation}
\partial_z \phi + \left(\frac{\i \mathfrak{m}(a-3y_1)}{2y_1(n_1-n_2)} - \frac{2 \pi \i \mathfrak{m}_1}{\Delta z} - \frac{2\pi}{\Delta z}  {\color{black}{L}} \sigma_0 \right) \phi  = 0\,,
\end{equation}
an equation which immediately allows us to extract the $z$-dependence of $\phi$, notably we obtain 
\begin{equation} \label{eq: phi_z_part}
    \phi = \Phi(y) \exp\left( -  \left(\frac{\i \mathfrak{m} (a-3y_1)}{2y_1(n_1-n_2)} - \i \frac{2 \pi}{\Delta z}  \frac{\mathfrak{m}_1}{n_1} - \frac{2\pi}{\Delta z}  {\color{black}{L}} \sigma_0 \right)z \right)\,,
\end{equation}
where $\Phi(y)$ remains to be determined from another necessary condition. The second necessary condition is obtained via multiplication of \eqref{eq: phi_bps_equation} with the Killing spinor $\epsilon$, obtaining 
\begin{equation}
    \i \epsilon \gamma^{\mu} \epsilon D_{\mu} \phi + \epsilon \gamma_3 \epsilon  \left( \rho + \frac{r}{2} G \right) \phi = 0\,,
\end{equation}
where we made use of $\epsilon \epsilon =0$. We first manipulate this equation to obtain
\begin{equation}
   \frac{1}{L} \partial_y \phi=  \left[  3\i \frac{a-3y}{q} \left( \frac{\sqrt{y}}{3}\left(\sigma + \frac{r}{2} H \right) + \i \frac{3y -a}{6y}\left( \rho + \frac{r}{2} G \right)  \right) + \frac{1}{2y} \left( \rho + \frac{r}{2}G \right) \right] \phi\,,
\end{equation}
now using \eqref{eq: phi_z_part} we can solve this equation for $\Phi(y)$, 
\begin{equation} \label{eq: phi_y_part}
    \Phi(y) = C y^{c_0} (y-y_1)^{c_1} (y_2-y)^{c_2} (y_3-y)^{c_3}\,,
\end{equation}
where $C$ is an integration constant and the exponents are 
   \begin{align}
\begin{split}
c_0 &= \frac{1}{16} \left( -4 r +\frac{a^2(3r-2 {\color{black}{L}}\rho_0)}{y_1 y_2 y_3} + 8  {\color{black}{L}} \rho_0 \right)\,,\\
c_1 & = - \frac{(a-3y_1)(a \Delta z(3r-2 {\color{black}{L}}\rho_0)-3y_1 (r \Delta z - 2 \Delta z  {\color{black}{L}} \rho_0 + 8 \i \pi  {\color{black}{L}} \sigma_0))}{16 y_1 (y_1 -y_2)(y_1-y_3) \Delta z}\,, \\
c_2 & = \frac{(a-3y_2)(a \Delta z(3r-2 {\color{black}{L}}\rho_0)-3y_2 (r \Delta z - 2 \Delta z  {\color{black}{L}} \rho_0 + 8 \i \pi  {\color{black}{L}} \sigma_0))}{16 y_2 (y_1 -y_2)(y_2-y_3) \Delta z}\,, \\
c_3 & = -\frac{(a-3y_3)(a \Delta z(3r-2 {\color{black}{L}}\rho_0)-3y_3 (r \Delta z - 2 \Delta z  {\color{black}{L}}\rho_0 + 8 \i \pi  {\color{black}{L}} \sigma_0))}{16 y_3 (y_1 -y_3)(y_2-y_3) \Delta z}\,. 
\end{split}
\end{align}

Recall we are interested in the region $y \in [y_1, y_2]$ and thus we want the solution \eqref{eq: phi_y_part} to be regular at both poles of the spindle. For regularity near $y = y_1$ we require
\begin{equation}
    c_1 \geq 0 \iff a \Delta z(3r-2 {\color{black}{L}} \rho_0)-3y_1 (r \Delta z - 2 \Delta z  {\color{black}{L}} \rho_0 + 8 \i \pi  {\color{black}{L}} \sigma_0) \leq 0\,,
\end{equation}
which immediately tells us that $\sigma_0 = 0$ as otherwise the inequality would acquire an imaginary part and would not make sense as an equation. The remaining part can be manipulated as 
\begin{equation}
    a(3r-2 {\color{black}{L}}\rho_0) - 3y_1 (r- 2  {\color{black}{L}}\rho_0) \leq 0\,,
\end{equation}
which leads to
\begin{equation} \label{eq: m_bound_y_1}
    \mathfrak{m} \leq - \frac{3n_2 (n_1 + n_2)}{2(2n_1 +n_2)} r\,.
\end{equation}
We can perform a similar analysis at the south pole $y=y_2$, where regularity is 
\begin{equation}
    c_2 \geq 0 \iff  a(3r-2 {\color{black}{L}}\rho_0) - 3y_2 (r- 2 {\color{black}{L}}\rho_0) \leq 0\,,
\end{equation}
which leads to 
\begin{equation} \label{eq: m_bound_y_2}
    \mathfrak{m} \geq \frac{3n_1(n_1 + n_2)}{2(n_1 + 2n_2)} r\,. 
\end{equation}
Both inequalities \eqref{eq: m_bound_y_1} and \eqref{eq: m_bound_y_2} are clearly not satisfied for all $\mathfrak{m} \in \mathbb{Z}$ and in any case we have alrea\textrm{d}y shown that $\sigma_0 = 0$ for regularity. We thus conclude for a generic solution to the vector multiplet BPS equations (generic $\sigma_0 \in \mathbb{R}$ and $\mathfrak{m} \in \mathbb{Z}$) the only smooth chiral multiplet solution for $\phi$ is $C = 0$.

\section{Useful limits and summations} \label{sec: sums}
Here we prove the identity 
\begin{eqnarray} \label{eq:identity}
	\frac{1}{k} \sum^{k-1}_{l=0} \frac{\omega_k^{-\alpha l }}{1-\omega^l_k u } = \frac{u^{\alpha - k \lfloor \frac{\alpha}{k}\rfloor}  }{1-u^k}\, , 
\end{eqnarray}
where $\alpha$ and $k$ are integers, $u$ has unit modulus and $\omega^x _y$ is defined via $\omega^x _y = e^{\frac{2 \pi i x}{y} }$. This result is important in evaluating the one-loop determinant via the orbifold index approach, see \eqref{eq: orbifold_index_north}, where for example the first summation corresponds to $k=n_1, \, \omega_k = u_1, \, \alpha = -\mathfrak{p}_1, \, u = q_1$.

First, let us investigate the behaviour of $f_{N,k} \equiv \frac{1}{k} \frac{\exp(2\pi i N) - 1}{ \left(\exp\left(2\pi i \frac{N}{k}\right) - 1\right)}$. For $N$ and $k$ integers and $\frac{N}{k}$ not an integer, $f_{N,k}$ vanishes  as the numerator is zero with the denominator being finite. To evaluate the expression when both $N$ and $\frac{N}{k}$ are integers, let us assume $N=m k + \epsilon $ for some $m$ and investigate the behaviour of the ratio in the limit $\epsilon \rightarrow 0$. Using l'H\^{o}pital's rule
\begin{equation}
f_{N,k} =\lim_{\epsilon\rightarrow 0}f_{N,k} = \frac{1}{k}\frac{\partial_m \left( \exp(2\pi i N) - 1\right)}{\partial_m (  \left(\exp\left(2\pi i \frac{N}{k}\right) - 1)\right) } = \frac{\exp\left(2\pi i k m \right)}{\exp\left(2\pi i m\right)}\, . 
\end{equation} When $m$ is an integer, i.e., $N$ is an integral multiple of $k$, the above analysis gives $f_{N,k} = 1$. Therefore, we have  the identity
\beqa \label{eq:identity1}
\frac{1}{k} \frac{\exp(2\pi i N) - 1}{ \left(\exp\left(2\pi i \frac{N}{k}\right) - 1\right)} = \begin{cases}
	 & 1 \quad \textrm{ if } N= k \lfloor \frac{N}{k}\rfloor \in \mathbb{Z}\,,\\
	& 0 \quad \textrm{ otherwise}\, .  
\end{cases}
\eeqa 

With this intermediate result in hand, we now turn to \eqref{eq:identity}. Using notations we have defined in the main bo\textrm{d}y of the text, we can express any integer $\alpha$ as  $\alpha = k \lfloor \frac{\alpha}{k}\rfloor  + \llbracket \alpha \rrbracket_k$ with $ \llbracket \alpha \rrbracket_k\in \{0,1, \ldots , k-1\}$. Since the denominator admits a series expansion,
\beqa
\lhs = \frac{1}{k} \sum^{k-1}_{l=0} \sum^\infty_{m=0} e^{-  \frac{2 \pi  i \llbracket \alpha \rrbracket_k}{k} l} e^{  \frac{2 \pi  i m}{k} l} u^m = \frac{1}{k} \sum^{k-1}_{l=0} \sum^\infty_{m=0}  e^{  \frac{2 \pi  i l}{k} (m - \llbracket \alpha \rrbracket_k)} u^m\, . 
\eeqa Interchanging the order of the  summations above, the coefficient of $u^m$ reads
\beqa
\frac{1}{k} \sum^{k-1}_{l=0} e^{  \frac{2 \pi  i l}{k} (m- \llbracket \alpha \rrbracket_k)} = \frac{1}{k}\frac{e^{2\pi i (m - \llbracket \alpha \rrbracket_k)} - 1 }{e^{\frac{2\pi i}{k} (m - \llbracket \alpha \rrbracket_k)} - 1} =  \begin{cases}
	& 1 \quad \textrm{ if } \frac{m-\llbracket \alpha \rrbracket_k}{k}=  \lfloor \frac{ m-\llbracket \alpha \rrbracket_k}{k}\rfloor \equiv p \in \mathbb{Z}\\
	& 0 \quad \textrm{ otherwise}\, .  \qquad\qquad
\end{cases}
\eeqa where we have used \eqref{eq:identity1}. 
The above equality can be succinctly re-written as 
\begin{eqnarray}
    \frac{1}{k} \sum^{k-1}_{l=0} e^{  \frac{2 \pi  i l}{k} (m- \llbracket \alpha \rrbracket_k)} = \delta_{m,k p +\llbracket \alpha \rrbracket_k}\,.
\end{eqnarray}
As the above term was the coefficient of $u^m$, we have for the full series
\beqa
\lhs = \sum^\infty_{m=0 } \delta_{m,k p +\llbracket \alpha \rrbracket_k} ~u^m = \sum^\infty_{p=0} u^{k p + \llbracket \alpha \rrbracket_k} = \sum^\infty_{p=0} u^{k p + \alpha - k \lfloor \frac{\alpha}{k}\rfloor}  = \frac{u^{\alpha - k \lfloor \frac{\alpha}{k}\rfloor}}{1-u^k}\,. \,\, \, \quad 
\eeqa

\bibliographystyle{JHEP}
\bibliography{spindle.bib}

\end{document}